\documentclass[traditabstract]{aa} 
\usepackage{graphicx}
\usepackage{natbib}
\usepackage{txfonts}
\usepackage{color}
\begin{document}
   \title{Seismic sensitivity to sub-surface solar activity\\
  from 18 years of GOLF/SoHO observations}

   \author{D. Salabert, R. A. Garc\'ia and S. Turck-Chi\`eze}
        
   \institute{Laboratoire AIM, CEA/DSM-CNRS, Universit\'e Paris 7 Diderot, IRFU/SAp, Centre de Saclay, 91191 Gif-sur-Yvette, France\\
              \email{david.salabert@cea.fr}
 }

   \date{Received xx xx xx; accepted xx xx xx}

  \abstract
{Solar activity has significantly changed over the last two Schwabe cycles. After a long and deep minimum at the end of Cycle~23,  the weaker activity of Cycle~24 contrasts with the previous cycles.
In this work, the response of the solar acoustic oscillations to solar activity is used in order to provide insights on the structural and magnetic changes in the sub-surface layers of the Sun during this on-going unusual period of low activity.
   We analyze 18 years of continuous observations of the solar acoustic oscillations collected by  the Sun-as-a-star GOLF instrument onboard the SoHO spacecraft. From the fitted mode frequencies,
 the temporal variability of the frequency shifts of the radial, dipolar, and quadrupolar modes are studied for different frequency ranges which are sensitive to different layers in the solar sub-surface interior.  
   The low-frequency modes show  nearly unchanged frequency shifts between Cycles~23 and 24, with a time evolving signature of the quasi-biennial oscillation, which is particularly visible for the quadrupole component revealing the presence of a complex magnetic structure. The modes at higher frequencies show frequency shifts 30\% smaller during Cycle~24, which is in agreement with the decrease observed in the surface activity between Cycles~23 and 24.
The analysis of 18 years of GOLF oscillations indicates that the structural and magnetic changes responsible for the frequency shifts remained comparable between Cycle~23 and Cycle~24 in the deeper sub-surface layers below 1400~km as revealed by the low-frequency modes. The frequency shifts of the higher-frequency modes, sensitive to shallower regions, show that Cycle~24 is magnetically weaker in the upper layers of Sun.}

   \keywords{Methods: data analysis --
                Sun: helioseismology -- Sun: activity }

 \titlerunning{GOLF/SoHO seismic sensitivity to sub-surface solar activity}
   \maketitle
%

\section{Introduction}
The unexpected long activity minimum of the solar Cycle 23 in 2007--2008 puzzled over the complexity and the variability of the emerging solar magnetic fields generated in the convective zone. It revealed even more the difficulty to perform reliable prediction of the 11-year solar cycle. The current  Cycle 24 is showing one of the weakest level of activity since a century, with a decreased surface activity by about 30\% compared to Cycle 23. Moreover, \citet{livingston12} observed a continue drop in sunspot magnetic field strength since the last 30 years arguing that sunspots could disappear from the Sun's surface during Cycle 25 as the 1500~G threshold to form dark sunspots would be reached. Deeper inside the Sun, \citet{howe13a} have used helioseismic data to follow the zonal flow pattern, called torsional oscillation, during Cycles 23 and 24. They showed in particular that the high-latitudinal pattern, one of the signatures of a new Cycle, is actually absent since the beginning of Cycle 24, probably due to the lower polar magnetic field strength.

The solar acoustic (p) oscillation frequencies were revealed to respond to changes in the surface activity, and to vary in correlation with solar proxies such the sunspot number or the 10.7-cm radio emission. The temporal variations of the solar p-mode frequencies with solar activity were first reported by \citet{wood85}, shortly confirmed by \citet{fossat87} and \citet{libbrecht90}. They were quickly attributed to changes taking place in the sub-surface layers. It is also known for more than a decade that the extraction of the  sound-speed profile in the core must be done carefully. It is why this information was obtained in using only low-degree modes with frequencies smaller than 2.4 mHz \citep{turck2001,turcklopes2012}.

The solar frequencies are today identified as a unique proxy of solar activity, as they might reveal inferences on sub-surface changes with solar activity not visible at the surface. They could also contribute to build varying topologies of magnetic field below the surface.
As longer and higher-quality helioseismic observations became available, the temporal variability of the p-mode frequencies  was studied in more details. These frequency shifts were actually observed to be frequency dependent, the shifts being larger at higher frequencies, and to be angular-degree ($l$) dependent, i.e., mode-inertia dependent, although the  $l$ dependence is  small for low-degree modes \citep[see e.g.][]{gelly02,howe02,chano04,salabert04}. The shifts can be then interpreted to arise from structural changes in the sub-surface layers, such as changes in temperature \citep{kuhn88} or in the size of the acoustic cavity \citep{dziem05}. 
Moreover, \citet{howe02} showed that the temporal and  latitudinal distribution of the frequency shifts is correlated with the spatial distribution of the surface magnetic field.
Thus, the frequency shifts cannot be purely explained by structural changes 
 and the determination of the dipolar and the toroidal magnetic fields of the sub-surface layers appear as one of the  important outputs from these observables \citep{Baldner2009}.

However, while the frequency shifts varied closely to solar surface activity with very high levels of correlation over the last three complete solar Cycles (21, 22, and 23), spanning thirty years of solar activity \citep{chaplin07}, significant differences were observed between acoustic frequencies and solar activity during the unusual minimum of Cycle 23 as reported in \citet{salabert09}, \citet{broomhall09}, and \citet{tripathy10}. Moreover, \citet{basu12} showed that the structure of the solar sub-surface layers was very different between Cycle 22 and Cycle 23, with deeper changes during Cycle 22 compared to Cycle 23.

In this study we investigate in more details the differences in the  variability of the  radial, dipolar, and quadrupolar low-degree modes. We show how the temporal evolution of the magnetic configuration imprints differently these modes during Cycle~23 and the weak on-going Cycle~24 by analyzing the low-degree oscillation frequencies extracted from the observations collected  by the space-based, Sun-as-a-star Global Oscillations at Low Frequency  instrument \citep[GOLF;][]{gabriel95} onboard the {\it Solar and Heliospheric Observatory}  \citep[SoHO;][]{domingo95} spacecraft.  
SoHO was launched in December 2, 1995, and since then is collecting continuous data of the Sun from a stable orbit around the L1 Lagrangian point. With more than 18 years of space observations available today, free from any perturbations related to the Earth atmosphere as it occurs from ground-based observations, the GOLF/SoHO observations provide an exquisite and unique dataset for global helioseismic studies of the low-degree oscillation modes.
In Section~2, we describe the data used in this paper and how the seismic parameters were extracted. The frequency shifts thus estimated are presented and discussed in Section~3. In Section~4, we provide possible explanations and interpretations of the related structural and magnetic changes of the sub-surface layers during Cycle 24. The main results are summarized in Section 5. Finally, the tables containing the oscillation frequencies of the GOLF low-degree modes extracted from the 365-day spectra analyzed in this study are made electronically available as described in Appendix~A. In Appendix~B, 18 years of frequency shifts at individual angular degree $l$ are presented in more details. \\

\section{Data and analysis}
\subsection{Data sets}
We analyzed 18 years of continuous observations collected by the GOLF instrument,  a resonant scattering spectrophotometer measuring the Doppler wavelength shift -- integrated over the solar surface -- in the D1 and D2 Fraunhofer sodium lines at 589.6 and 589.0~nm respectively. 
It has been designed to measure the Doppler shift by switching from one side of the wing to the other side, an additional coil allowing two near point measurements along each side of the wings in the more stable part of the lines \citep{gabriel95}. 
However, the GOLF instrument has not been used in its nominal configuration due to a malfunction in its polarization switching mechanism identified shortly after the launch \citep{garcia04}. As a consequence, observations from only  one side of the sodium doublet are collected, from which a proxy for the Doppler velocity signal is formed. Moreover, observations have been collected from each side of the doublet as follows:  in the blue-wing configuration from 1996 April 11 until 1998 June 26, and later on from 2002 November 19 until now (the so-called {\it blue periods}); in the red-wing configuration of the sodium doublet between 1998 October 30 until 2002 November 18 (the so-called {\it red period}). These two configurations imply different spatial weightings of the solar disk \citep{garcia98}.

In addition, two extended gaps are present in the analyzed time series because of the temporary loss of the SoHO spacecraft. The first gap of about 100 days happened during
the summer of 1998 after a bad maneuver of the SoHO rotation
spacecraft. The second one of a period around 1 month in
January 1999 occurred while new software was being uploaded
to the spacecraft due to a failure in the SoHO gyroscopes. 

The observational blue-wing configuration of the GOLF instrument has not  been modified since November 2002, except for readjusting the photomultiplier high voltages in order to compensate its aging. It corresponds today to almost 12 years of continuous measurements, with a duty cycle close to 100\%. 
The response function of the GOLF sodium Fraunhofer line is different between the blue- and the red-wing configurations \citep{lefebvre08}, with averaged heights in the solar atmosphere of 322~km and 480~km respectively \citep{chano07}. Then, the measurements obtained in the red-wing configuration originate higher up in the solar atmosphere with larger contributions from the chromosphere.  

A total of 6538 days of GOLF observations with a duty cycle of 96.0\% covering 18 years between April 11, 1996 and March 5, 2014  were analyzed. 
The GOLF velocity time series were obtained following \citet{garcia05} and the amplitudes of the blue- and red-wing signals normalized as in \citet{chano03}. 
To obtain relations between mode frequencies and solar activity, this dataset was split into 70 contiguous 365-day sub series, with a four-time overlap of 91.25 days. \\

\subsection{Extraction of the p-mode parameters}
\label{sec:extraction}
The power spectrum of each time series was fitted to estimate the mode parameters of the $l=0$, 1, 2, and 3 modes as described in \citet{salabert07}. Each mode component of radial order $n$, angular degree $l$, and azimuthal order $m$ was parameterized with an asymmetric Lorentzian profile \citep{nigam98}, as:
\begin{equation}
{\cal L}_{n,l,m}(\nu) = H_{n,l} \frac{(1+b_{n,l} x_{n,l})^2+b_{n,l}^2}{1+ x_{n,l}^2},
\label{eq:mlemodel}
\end{equation} 
\noindent
where $x_{n,l} = 2(\nu-\nu_{n,l})/\Gamma_{n,l}$, and $\nu_{n,l}$, $\Gamma_{n,l}$, and $H_{n,l}$ represent the mode frequency, linewidth, and  height of the spectral density, respectively. The peak asymmetry is described by the parameter $b_{n,l}$.
Because of their close proximity in frequency, modes are fitted in pairs (i.e., $l=2$ with 0, and $l=3$ with 1). While each mode parameter within a pair of modes are free, the peak asymmetry is set to be the same within pairs of modes. When being present inside the fitting window, the $l = 4$ and 5 modes were included in the fitted model. An additive value $B$ is added to the fitted profile representing a constant background noise in the fitted window. Since SoHO observes the Sun equatorwards, only the $l+|m|$ even components are visible in Sun-as-a-star observations from the GOLF instrument. The amplitude ratios between the $l=0,1,2$, and 3 modes and the $m$-height ratios of the $l=2$ and 3 multiplets calculated in \citet{salabert11a}  for the GOLF measurements were used.  
Finally, the mode parameters were extracted by maximizing the likelihood function, the power spectrum statistics being by a $\chi^2$ with two degrees of freedom distribution.
The natural logarithms of the mode height, linewidth, and background noise were varied resulting in normal distributions. The formal uncertainties in each parameter were then derived from the inverse Hessian matrix. \\

\begin{figure*}[tbp]
\begin{center} 
\includegraphics[width=0.24\textwidth]{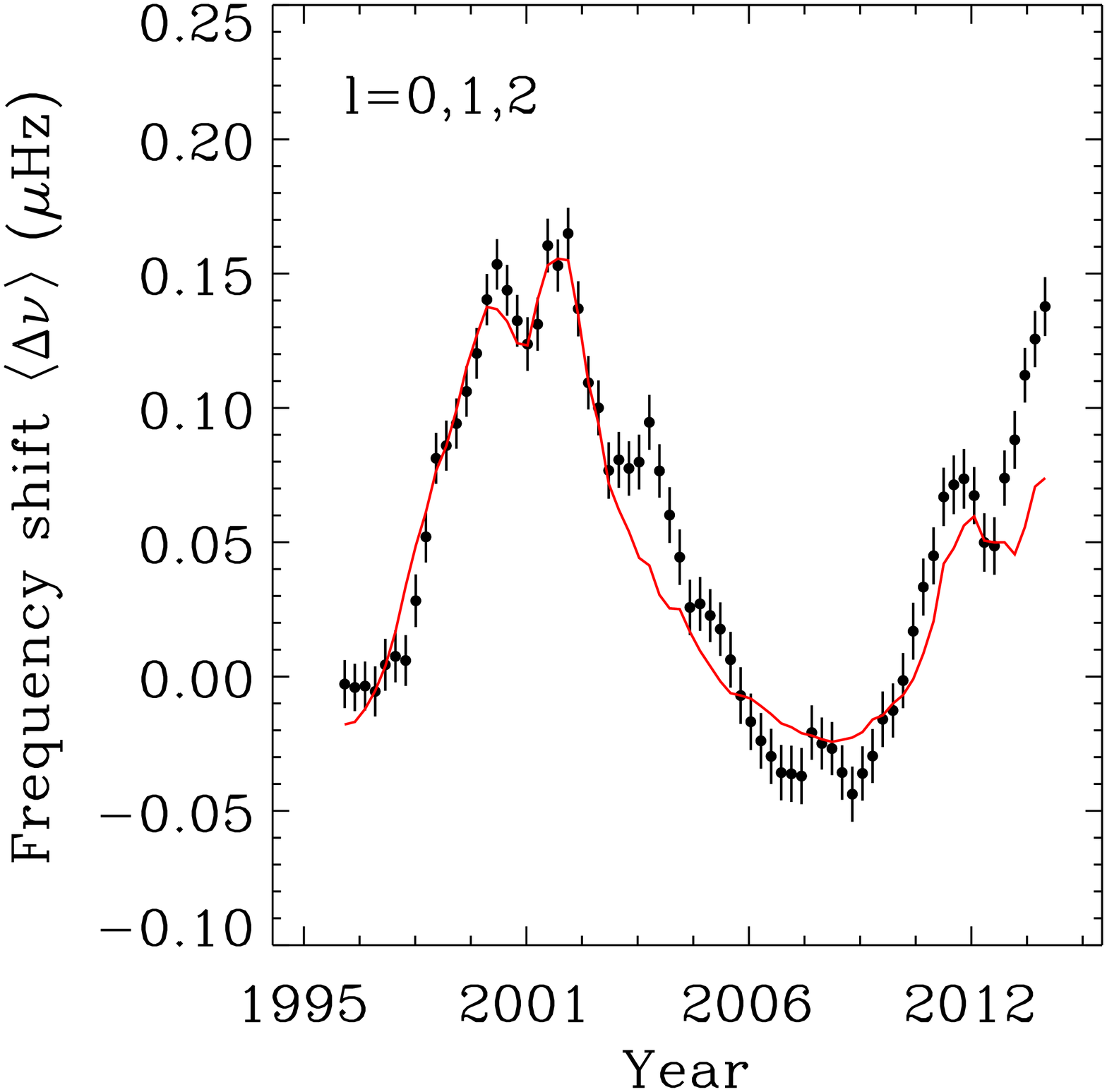}
\includegraphics[width=0.24\textwidth]{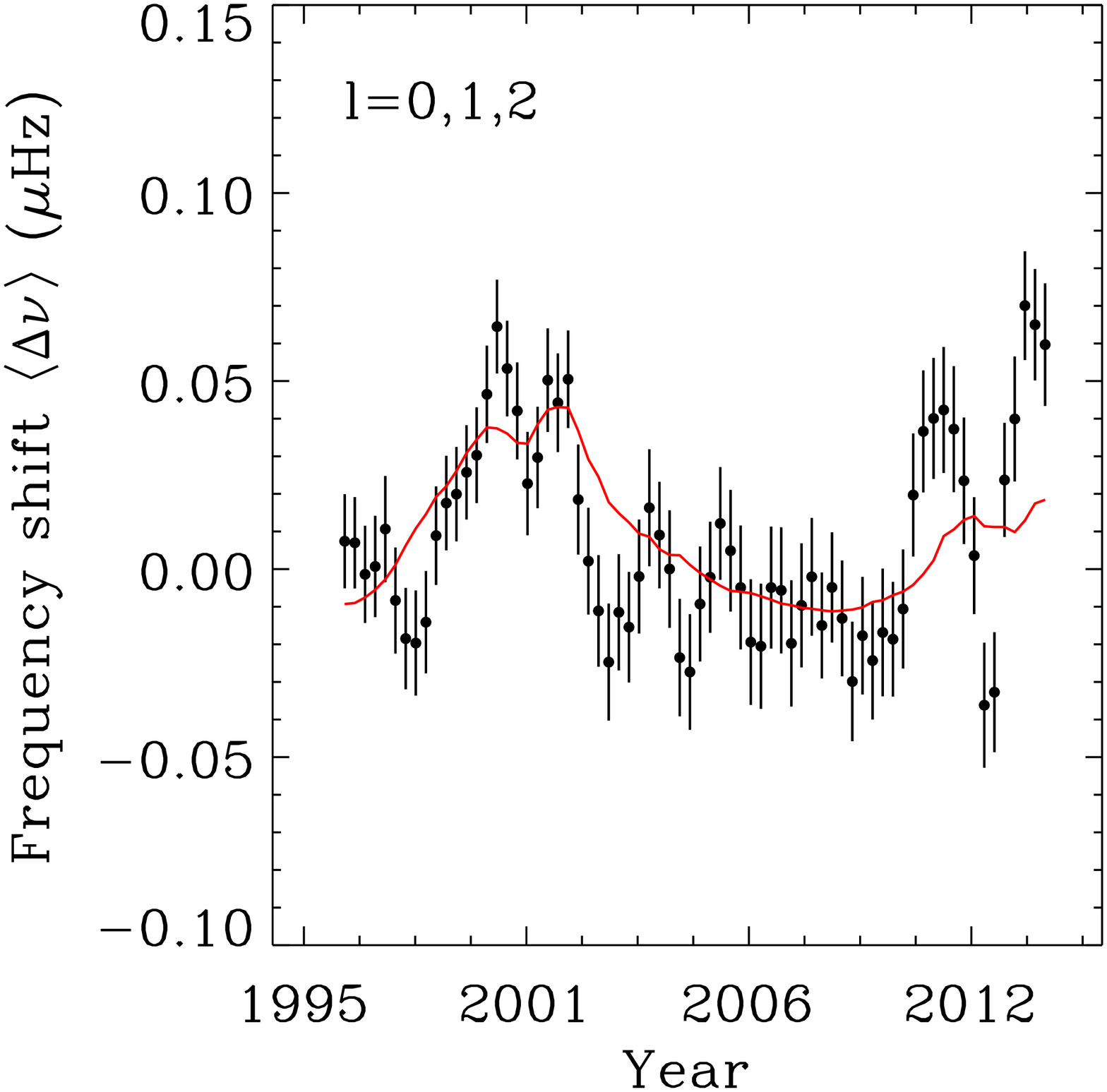}
\includegraphics[width=0.24\textwidth]{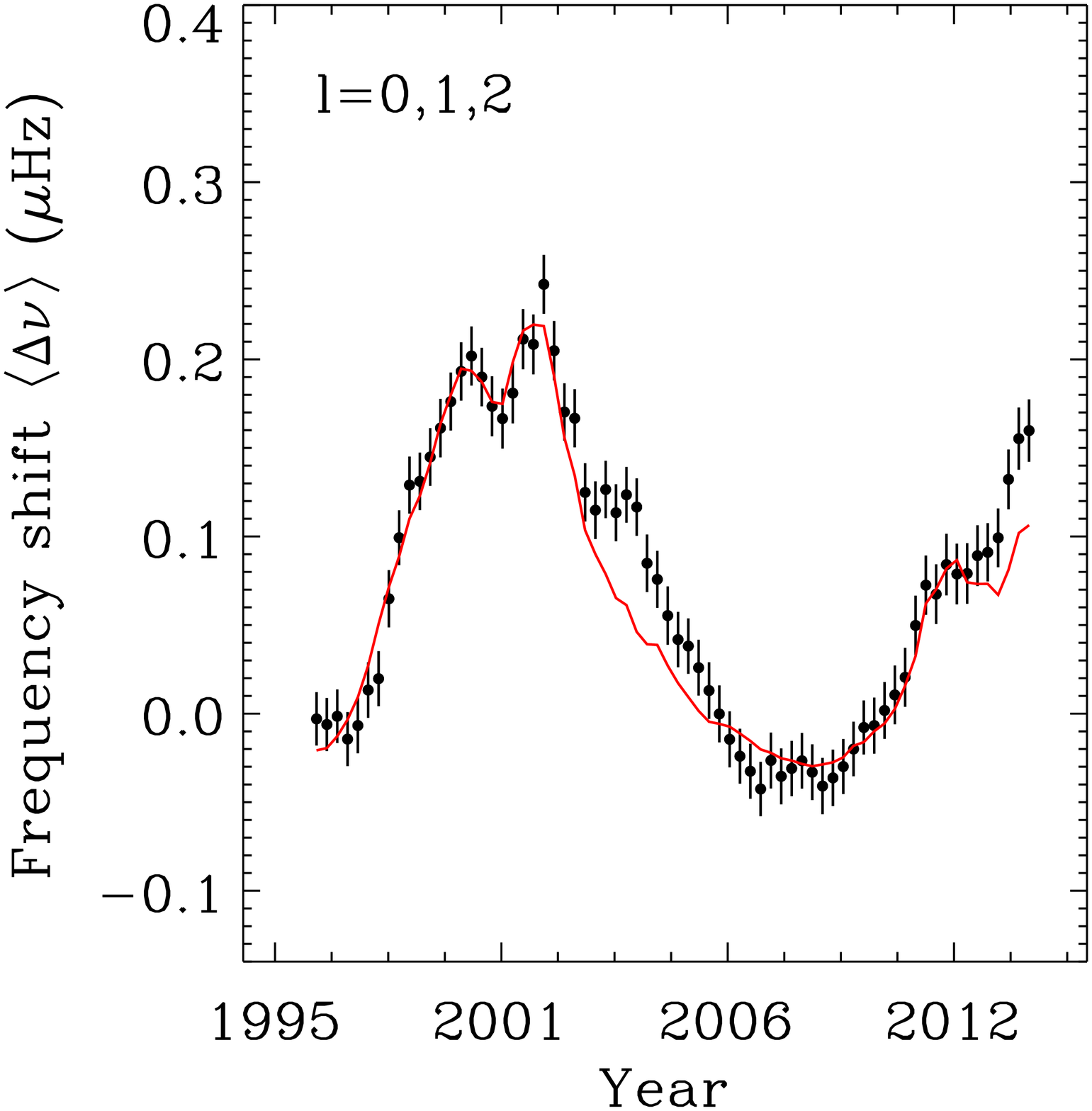}
\includegraphics[width=0.24\textwidth]{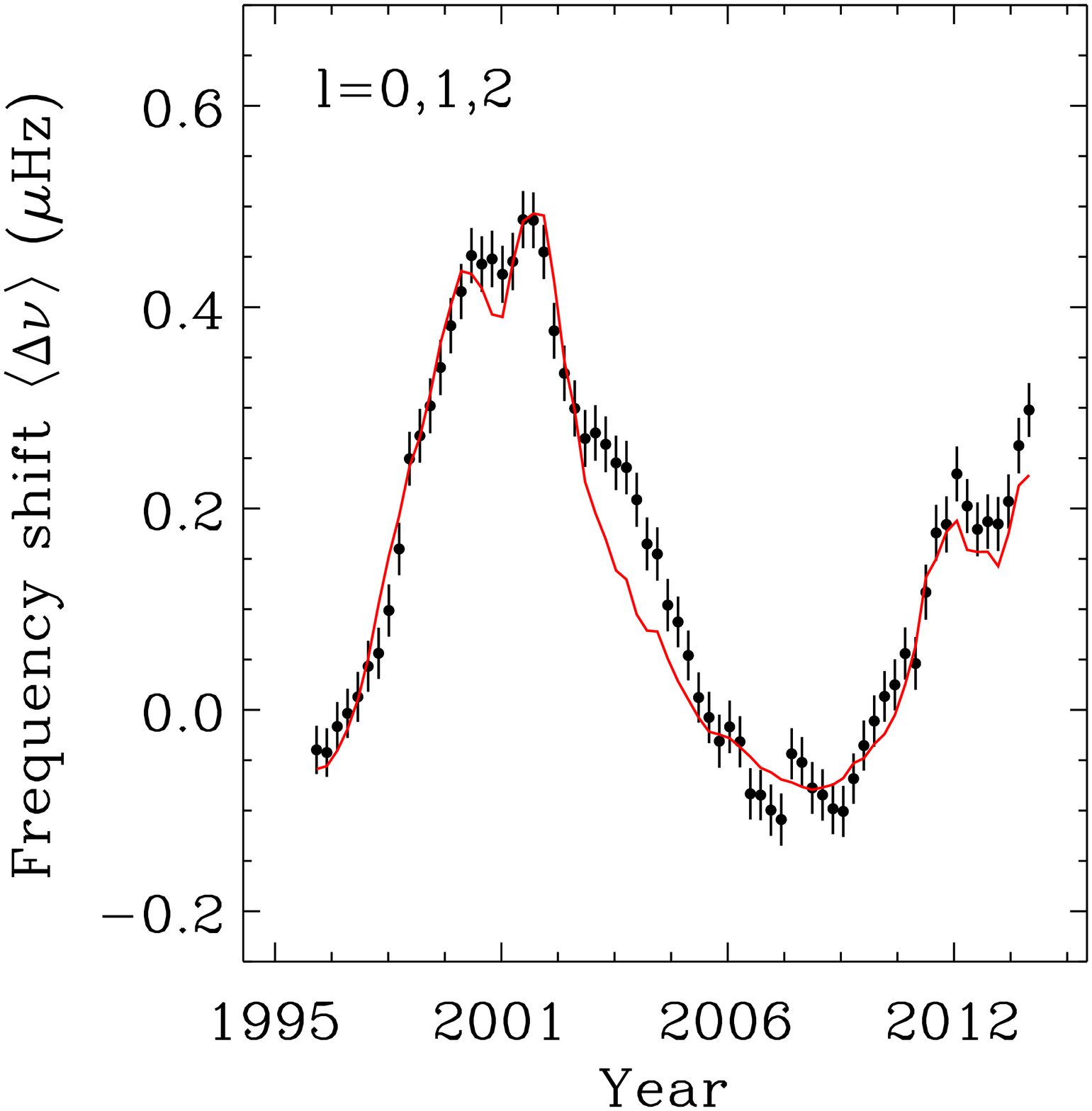}
\end{center}
\caption{\label{fig:fig1} 
Temporal variations of the frequency shifts in $\mu$Hz averaged over the modes $l=0,1$, and 2, $\langle  \Delta\nu_{n,l=0,1,2} \rangle$, and calculated for four different frequency ranges (black dots).  From left to right, the frequency ranges are the following: a) 1800~$\mu$Hz  $\leq \nu <$ 3790~$\mu$Hz; b) 1800~$\mu$Hz  $\leq \nu <$ 2450~$\mu$Hz; c) 2450~$\mu$Hz  $\leq \nu <$ 3110~$\mu$Hz; and d) 3110~$\mu$Hz  $\leq \nu <$ 3790~$\mu$Hz. The 10.7-cm radio flux, $F_{10.7}$, averaged over the same 365-day timespan and scaled to match the rising phase and maximum of Cycle 23 is shown as a proxy of the solar surface activity (solid line).}
\end{figure*} 
\begin{figure*}[tbp]
\begin{center} 
\includegraphics[width=0.24\textwidth]{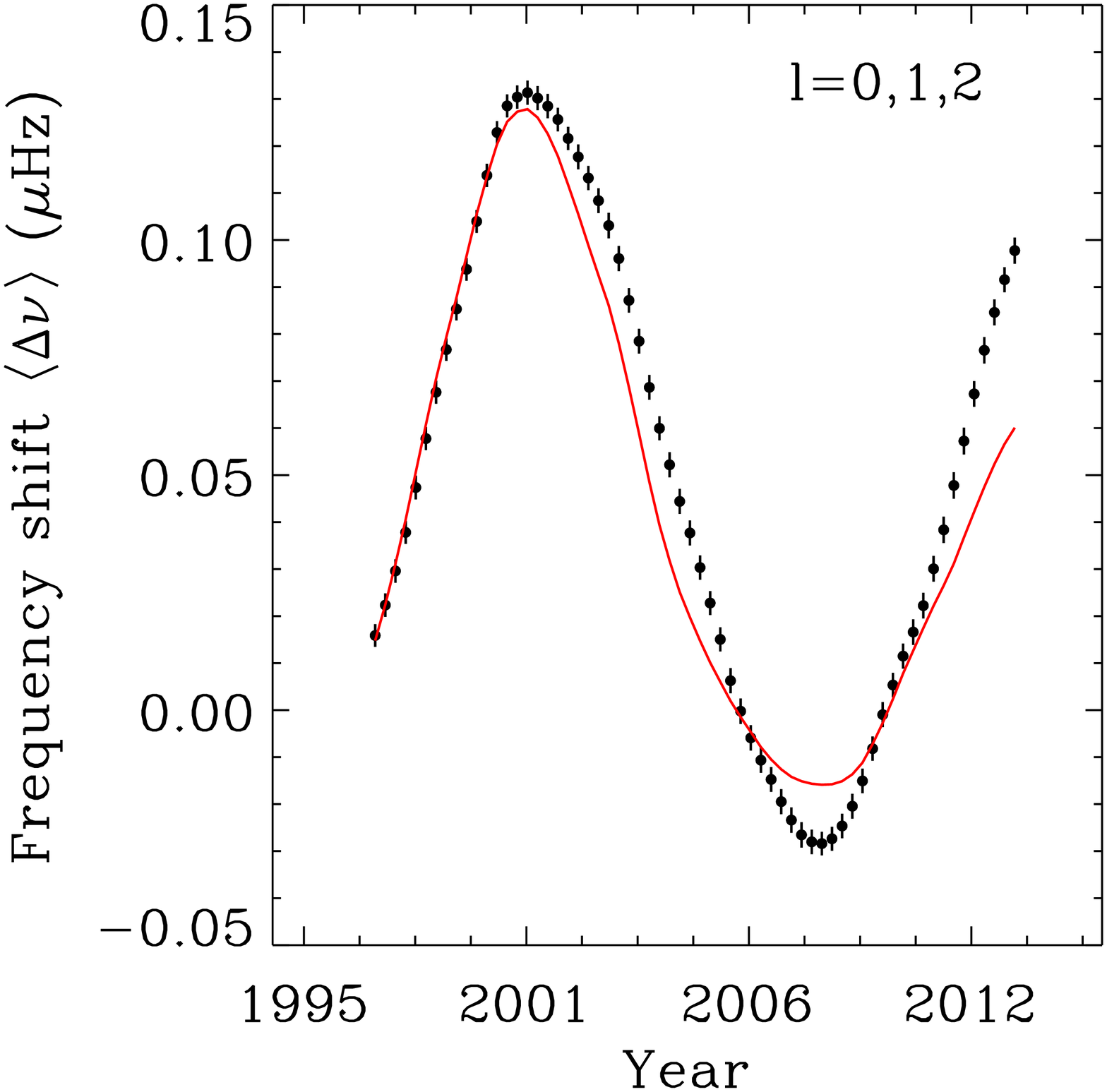}
\includegraphics[width=0.24\textwidth]{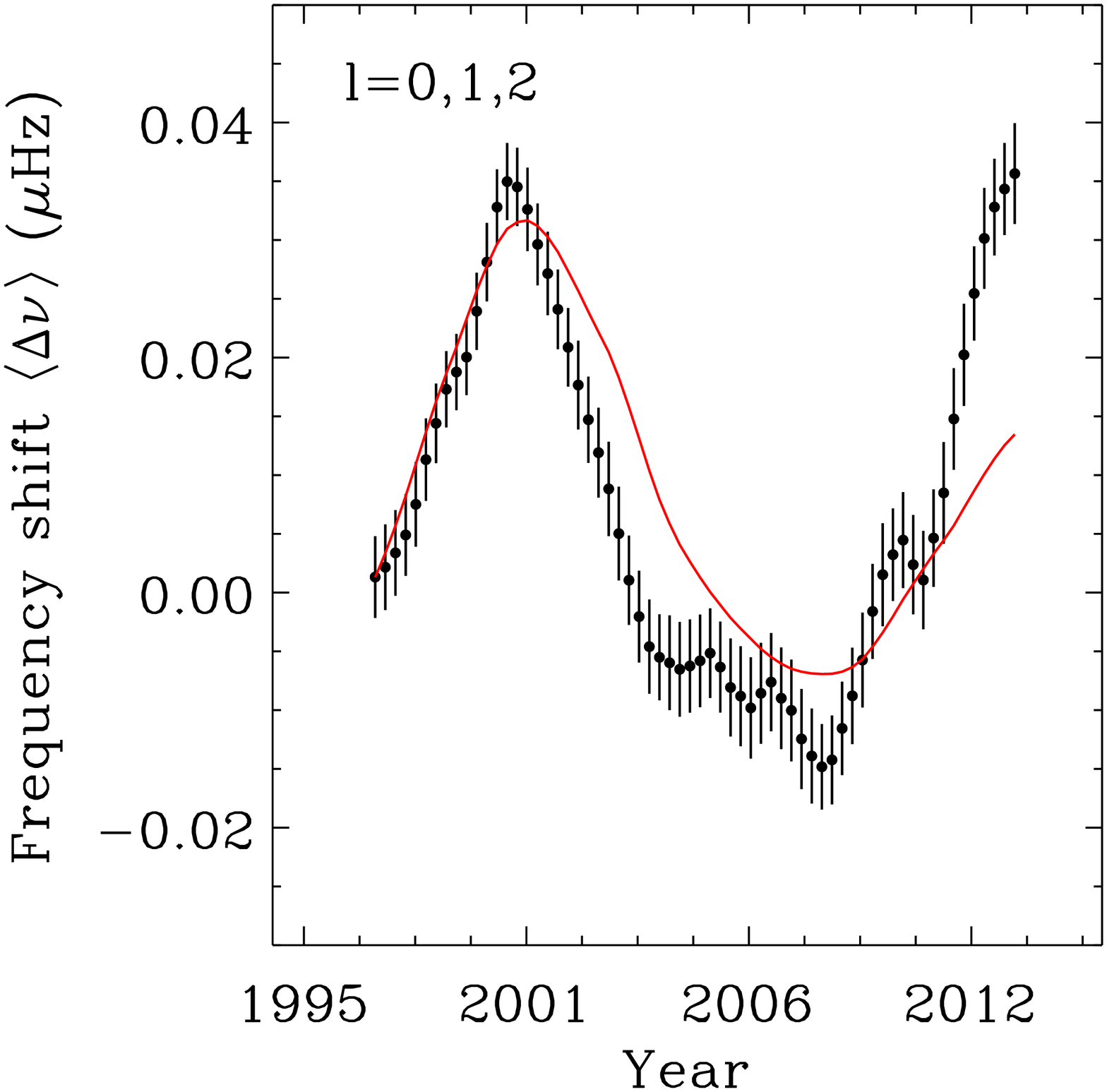}
\includegraphics[width=0.24\textwidth]{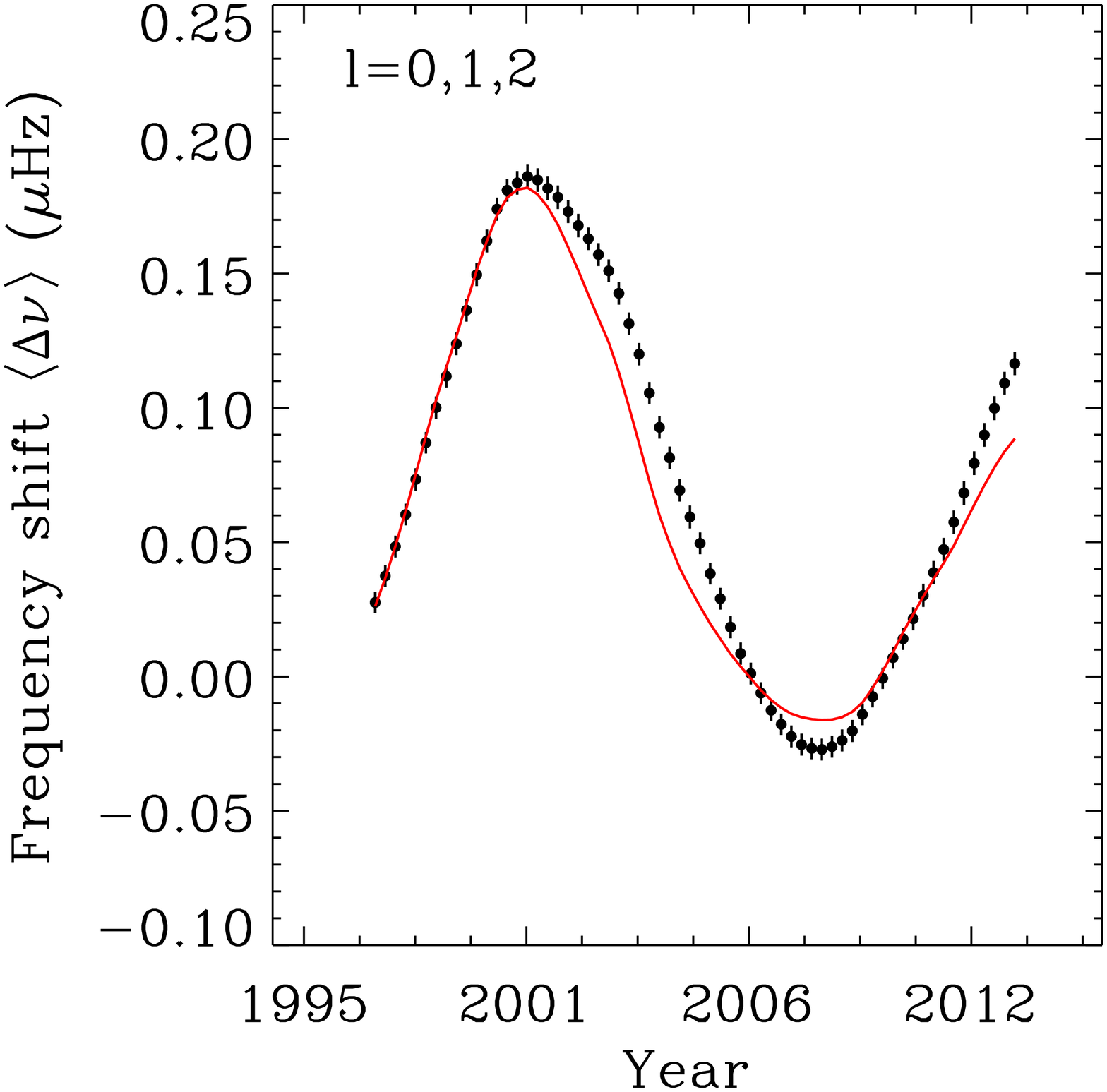}
\includegraphics[width=0.24\textwidth]{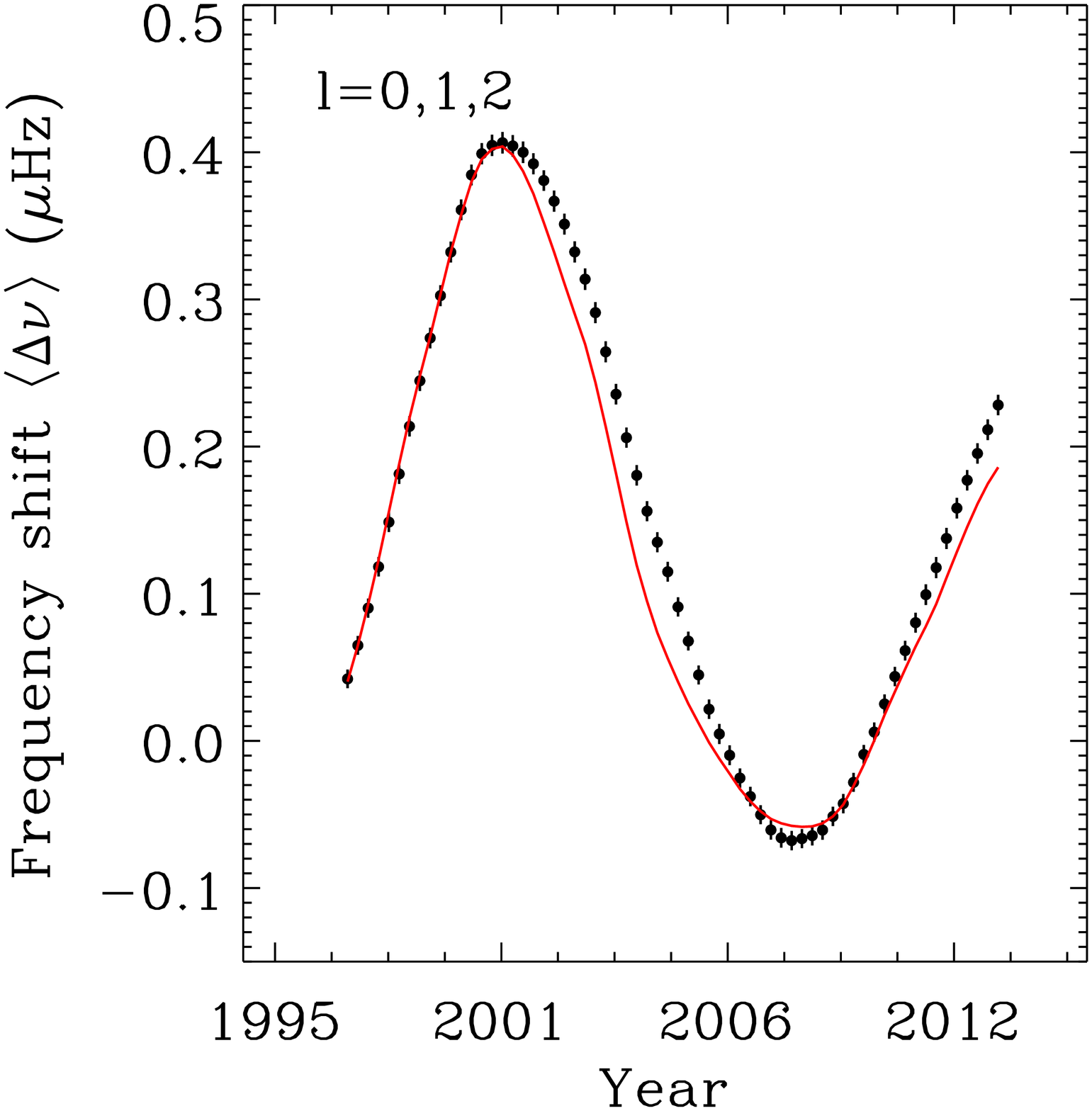}
\end{center}
\caption{\label{fig:fig5} Same as Fig.~\ref{fig:fig1} but after removing the QBO's signature from the temporal variations of the frequency shifts, $\langle  \Delta\nu_{n,l=0,1,2} \rangle$ (black dots).  The 10.7-cm radio flux, $F_{10.7}$,  was smoothed in the same way and scaled to match the rising phase and maximum of Cycle 23 (solid line).}
\end{figure*} 

\section{Results}
The temporal variations of the mode frequencies, $\langle  \Delta \nu_{n,l} \rangle$,  were defined as the differences between the frequencies observed at different dates and reference values of the corresponding modes.  The set of reference frequencies was taken as the average over the years 1996--1997, corresponding to the minimum of the activity Cycle 22. The formal uncertainties resulting from the peak-fitting analysis were used as weights in the average computation. 
The frequency shifts thus obtained were then weigthed averaged over 15 consecutive radial orders between 1800~$\mu$Hz and 3790~$\mu$Hz. The lower frequency was defined by the lowest modes which are accurately fitted in all the analyzed 365-day spectra, i.e., $n=12$ for $l=0$ and 1, and $n=11$ for $l=2$. Note that we did not use the $l = 3$ mode in the following analysis because of its lower signal-to-noise ratio. In addition, mean values of daily measurements of the 10.7-cm radio flux, $F_{10.7}$, were used as a proxy of the solar surface activity\footnote{The 10.7-cm radio flux data are available from the National Geophysical Data Center at http://www.ngdc.noaa.gov/stp/solar/solardataservices.html.}. 

\subsection{At different depths below the surface}
As demonstrated, for example, by \citet{chaplin98}, \citet{howe99}, \citet{chano01}, and \citet{salabert04} for both low- and medium-angular degrees,  the frequency shifts associated to the solar Schwabe cycle present a strong frequency dependence, going from small shifts at low frequency to much larger shifts at high frequency. As modes with different frequencies are sensitive to different layers in the sub-surface outer parts of the Sun, this information can be used to examine differences between  solar Cycles 23 and 24 at different depths. 

We thus defined three  distinct frequency ranges containing each five consecutive radial orders covering different reflecting points in the solar atmosphere. 
The three frequency ranges were defined as follows: (1) the low-frequency range where 1800~$\mu$Hz $\leq \nu <$  2450~$\mu$Hz; (2) the medium-frequency range where 2450~$\mu$Hz  $\leq \nu <$ 3110~$\mu$Hz; and (3) the high-frequency range where 3110~$\mu$Hz $\leq \nu <$ 3790~$\mu$Hz. These three frequency ranges allow us  to study the response of the p-mode oscillations to solar activity at different depths below the surface, and thus to infer localizations of the layers where the magnetic fields affect the oscillation frequencies. Indeed, \citet{basu12}  calculated that the averaged kernels of the mid- and the high-frequency ranges have their largest sensitivities at $0.9981~R_{\sun}$ (i.e., 1300~km) and $0.9989~R_{\sun}$ (i.e., 760~km) respectively, while the low-frequency modes peak deeper in the solar interior at $0.9965~R_{\sun}$ (i.e., 2400~km). 
Nevertheless, while the mid- and the high-frequency modes correspond to a thin layer below the Sun's surface, the averaged kernel of the low-frequency modes covers a much wider region ranging from 1400~km down to 4000~km deep. Finally, four frequency ranges were thus used to calculate the frequency shifts, when we include the one between 1800~$\mu$Hz and 3790~$\mu$Hz.

\begin{table*}
\begin{minipage}{\textwidth}
\caption{Frequency shifts per unit of change in the 10.7-cm solar radio flux  and linear correlations extracted from the analysis of independent 365-day GOLF spectra for different phases of the solar Cycles 23 and 24.  The values were obtained from the 15 consecutive radial orders between 1800~$\mu$Hz $\leq \nu <$ 3790~$\mu$Hz, and for 5 consecutive radial orders  in the low-frequency range 1800~$\mu$Hz $\leq \nu <$ 2450~$\mu$Hz, 5 consecutive radial orders in the mid-frequency range 2450~$\mu$Hz $\leq \nu <$ 3110~$\mu$Hz, and 5 consecutive radial orders in the high-frequency range 3110~$\mu$Hz $\leq \nu <$ 3790~$\mu$Hz.}
\label{table:correl_1800-3790}      
\centering               
\renewcommand{\footnoterule}{}  
\begin{tabular}{c c c c c  c c c c  c c c  }        
\hline\hline            
& \multicolumn{2}{c}{All data}   &  \multicolumn{2}{c}{Rising phase Cycle 23}   &   \multicolumn{2}{c}{Declining phase Cycle 23}   &   \multicolumn{2}{c}{Rising phase Cycle 24}        \\   

 & \multicolumn{2}{c}{(Apr. 1996 / Apr. 2014)}   & \multicolumn{2}{c}{(Apr.  1996 / Oct. 2001)}   &   \multicolumn{2}{c}{(Oct. 2001 / Jan. 2009)}   &   \multicolumn{2}{c}{(Jan. 2009 / Apr. 2014)}     \\ 
 
$l$ & \multicolumn{1}{r}{Gradient\footnote{Gradient against the 10.7-cm radio flux in units of nHz RF$^{-1}$ (with 1~RF =  10$^{-22}$~J~s$^{-1}$~m$^{-2}$~Hz$^{-1}$).}} & $r_p$ & Gradient$^a$ & $r_p$ & Gradient$^a$ & $r_p$  &Gradient$^a$ & $r_p$ \\    
\hline                  
         &  \multicolumn{8}{c}{1800~$\mu$Hz $\leq \nu <$ 3790~$\mu$Hz}        \\  
$\langle 0,1,2 \rangle$ & 1.7~$\pm$~0.1 &  0.95 &  1.4~$\pm$~0.1 &  0.98 &  1.6~$\pm$~0.1 & 0.95  &  2.1~$\pm$~0.1 & 0.95   \\
0  & 1.7~$\pm$~0.1 & 0.95 &  1.2~$\pm$~0.1 & 0.97 &  1.7~$\pm$~0.1 & 0.97 &  2.1~$\pm$~0.2 & 0.99     \\
1  & 2.1~$\pm$~0.1 & 0.95 & 1.7~$\pm$~0.1 & 0.97 &  1.7~$\pm$~0.1 & 0.95 &  2.3~$\pm$~0.2 & 0.86 \\
2  & 1.4~$\pm$~0.1 & 0.82 &  1.4~$\pm$~0.1 & 0.94 &  1.3~$\pm$~0.1 & 0.81 & {\bf 2.0~$\pm$~0.2} & 0.75   \\
\\
&  \multicolumn{8}{c}{1800~$\mu$Hz $\leq \nu <$ 2450~$\mu$Hz}        \\  
$\langle 0,1,2 \rangle$ & 0.4~$\pm$~0.1 &  0.49 &  0.4~$\pm$~0.1 &  0.77 &  0.3~$\pm$~0.1 & 0.50  & 0.8~$\pm$~0.3 & 0.56   \\
0  & 0.5~$\pm$~0.2 & 0.38 & 0.6~$\pm$~0.1 & 0.74 &  0.5~$\pm$~0.2 & 0.58&  0.5~$\pm$~0.3 & 0.33     \\
1  & 0.4~$\pm$~0.2 & 0.46 & 0.3~$\pm$~0.1 & 0.59 &  0.3~$\pm$~0.1 & 0.26 &  0.9~$\pm$~0.3 & 0.48  \\
2  & 0.4~$\pm$~0.1 & 0.33 &  0.4~$\pm$~0.1 & 0.54 &  0.1~$\pm$~0.1 & 0.13 &  1.0~$\pm$~0.3 & 0.40    \\
\\
&  \multicolumn{8}{c}{2450~$\mu$Hz $\leq \nu <$ 3110~$\mu$Hz}        \\  
$\langle 0,1,2 \rangle$ & 2.1~$\pm$~0.1 &  0.97   & 2.0~$\pm$~0.1 &  0.96   & 2.3~$\pm$~0.1 & 0.95    & 2.4~$\pm$~0.2 & 0.96   \\
0  & 1.5~$\pm$~0.2 & 0.92   & 1.1~$\pm$~0.1 & 0.95  & 1.9~$\pm$~0.1 & 0.95  & 2.0~$\pm$~0.2 & 0.94     \\
1  & 2.5~$\pm$~0.2 & 0.97  & 2.7~$\pm$~0.1 & 0.98 & 2.5~$\pm$~0.1 & 0.97  & 2.4~$\pm$~0.3 & 0.87  \\
2  & 2.3~$\pm$~0.2 & 0.89  & 2.3~$\pm$~0.2 & 0.96 & 2.8~$\pm$~0.2 & 0.89 & 3.2~$\pm$~0.3 & 0.91    \\
\\
&  \multicolumn{8}{c}{ 3110~$\mu$Hz $\leq \nu <$ 3790~$\mu$Hz }     \\  
$\langle 0,1,2 \rangle$ & 4.6~$\pm$~0.2 &  0.98 &  4.6~$\pm$~0.1 &  0.99  & 4.6~$\pm$~0.1 & 0.95  & 5.2~$\pm$~0.3 & 0.99   \\
0  & 4.1~$\pm$~0.3 & 0.96   & 3.8~$\pm$~0.2 & 0.97  & 4.2~$\pm$~0.2 & 0.96 & 5.5~$\pm$~0.4 & 0.96  \\
1  & 5.2~$\pm$~0.3 & 0.97  & 5.3~$\pm$~0.2 & 0.99 & 5.2~$\pm$~0.3 & 0.93 &  5.7~$\pm$~0.4 & 0.96  \\
2  & 4.7~$\pm$~0.3 & 0.94 & 4.9~$\pm$~0.3 & 0.95 & 4.6~$\pm$~0.3 & 0.95 &  4.1~$\pm$~0.5 & 0.93   \\
 \hline                                
\end{tabular}
\end{minipage}
\end{table*}

Figure~\ref{fig:fig1} shows the temporal variations of the frequency shifts, $\langle  \Delta\nu_{n,l=0,1,2} \rangle$, averaged over the modes $l=0,1$, and 2 and calculated over the four defined frequency ranges. 
The corresponding 10.7-cm radio flux averaged over the same 365-day sub series  is also shown for comparison. It was scaled to match the rising phase and the maximum of  the solar Cycle 23, by performing a linear least-square regression with the frequency shifts. In Table~\ref{table:correl_1800-3790}, we present the frequency shifts, $\langle  \Delta \nu_{n,l=0,1,2} \rangle$, calculated per unit of change in the 10.7-cm radio flux obtained by minimizing weighted linear regressions between these two quantities, as well as the associated  linear correlation coefficients, derived for the different frequency ranges. Note also that the values given in Table~\ref{table:correl_1800-3790} correspond to one set of independent spectra (one over four points) and that they are consistent within $1\sigma$ with any other of the chosen sets. 
As already widely established (see e.g. the references given in Section~1), the temporal variabilities of the p-mode oscillation frequencies extracted from 18 years of GOLF data are observed to be closely correlated with surface activity following the 11-year periodicity of the solar cycle. The correlation coefficients are over 0.95 for the mid- and the high-frequency ranges, but the low-frequency modes show a smaller correlation with surface activity of about 0.5. 
While the higher frequency modes show the largest variations along the solar cycle with frequency shifts being more than 10 times larger than the ones at low frequency, the low-frequency p modes still show significant variations between periods of low and high surface activity (Fig.~\ref{fig:fig1}).
Moreover,  the minimum of Cycle 23 in the frequency shifts of the mid- and high-frequency modes is observed to be lower than the minimum of Cycle 22 in May-June 1996, which is in agreement with the weaker surface activity since the long and deep minimum of Cycle 23 compared to Cycle~22. The quasi-biennial oscillation (QBO) \citep{fletcher10} with an amplitude modulated by the 11-year solar cycle is also visible on Fig.~\ref{fig:fig1} in all the frequency ranges.

\subsection{At different periods of the cycles}
Significant differences in the relationship between acoustic  oscillation frequencies and activity proxies were observed during the unusual extended minimum of Cycle 23 compared to other phases of activity \citep{salabert09,broomhall09,tripathy10}. For these reasons, we examined three separate periods:
 a) the rising phase of Cycle 23 going from April 1996 to October 2001 with a mean 10.7-cm radio flux of 129~RF\footnote{The 10.7-cm radio flux has for units  1~RF =  10$^{-22}$~J~s$^{-1}$~m$^{-2}$~Hz$^{-1}$.}; b) the declining phase of Cycle 23 going from October 2001 to January 2009 with a mean 10.7-cm radio flux of 114~RF; and c) the rising phase of Cycle 24 going from January 2009 to April 2014 with a mean 10.7-cm radio flux of 99~RF. The starting and ending dates of each of the three activity phases were defined using the 10.7-cm radio flux properly smoothed to remove the signature of the QBO as described in \citet{fletcher10}. Table~\ref{table:correl_1800-3790} gives the gradients and the correlation coefficients between frequency shifts and radio flux during the three phases of activity for the four frequency ranges. It shows that the frequency shifts averaged over 1800~$\mu$Hz~$\leq \nu <$~3790~$\mu$Hz are observed to be about 50\% larger per unit of change in radio flux during the rise of Cycle 24 than during the rise of Cycle 23, whereas the radio emission decreased by about 30\%   over the same period of time. In addition, during the declining phase of Cycle 23, the gradients are observed to be about 15\% larger than during the rising phase of Cycle 23. Moreover, the discrepancy in the relationship between frequencies and radio flux is more important for the low-frequency modes, i.e., for the modes going into deeper layers of the Sun, than for the higher-frequency modes confined in shallower layers, and it is accentuated during the rise of Cycle ~24 (Table~\ref{table:correl_1800-3790}). The correlation coefficients with solar activity indicate as well a decrease of almost 40\% between Cycle 23 and Cycle 24 for the low-frequency modes, while for the mid- and the high-frequency modes they remained similar.

\begin{figure*}[tbp]
\begin{center} 
\includegraphics[width=0.3\textwidth]{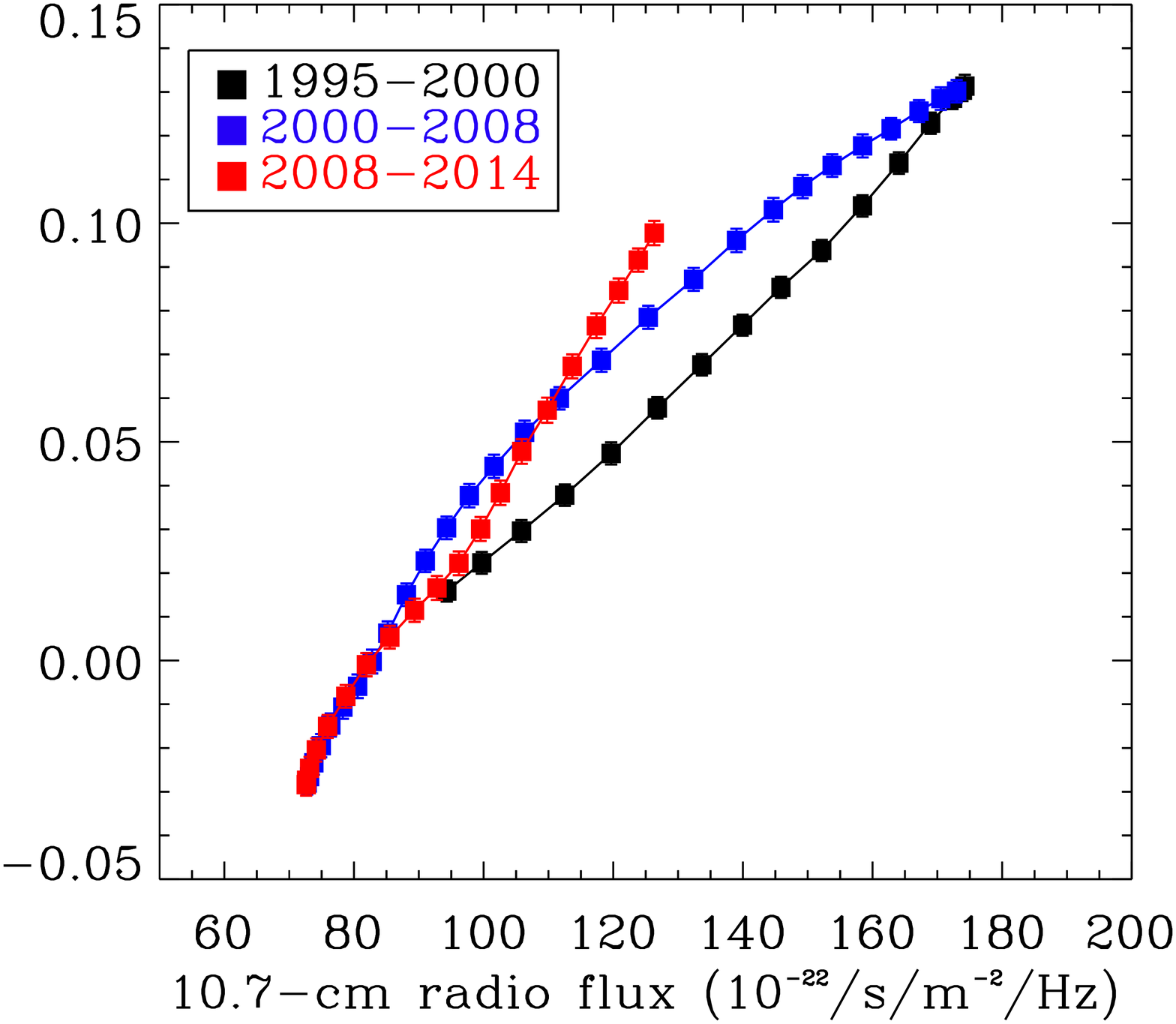}
\includegraphics[width=0.3\textwidth]{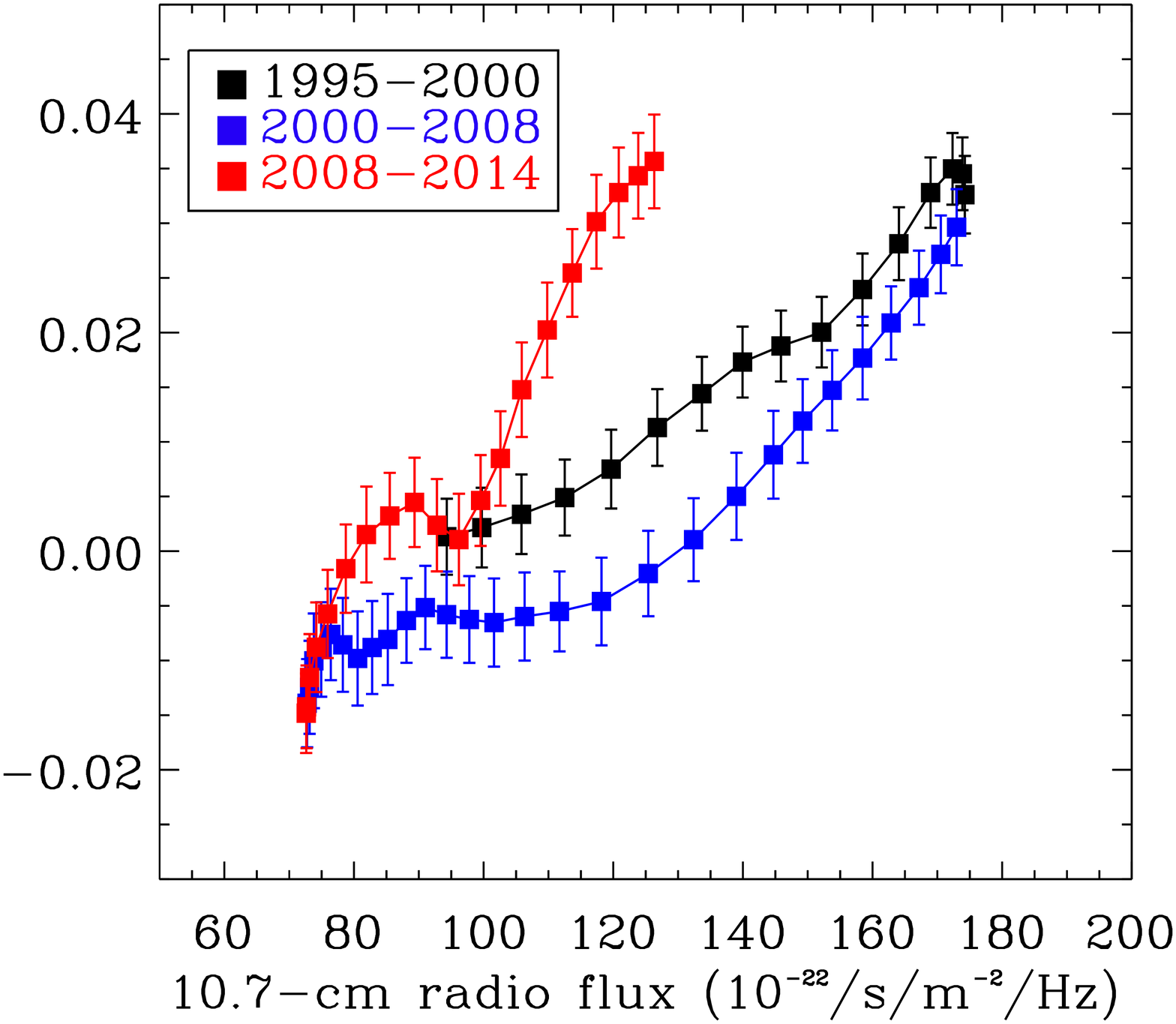}
\includegraphics[width=0.3\textwidth]{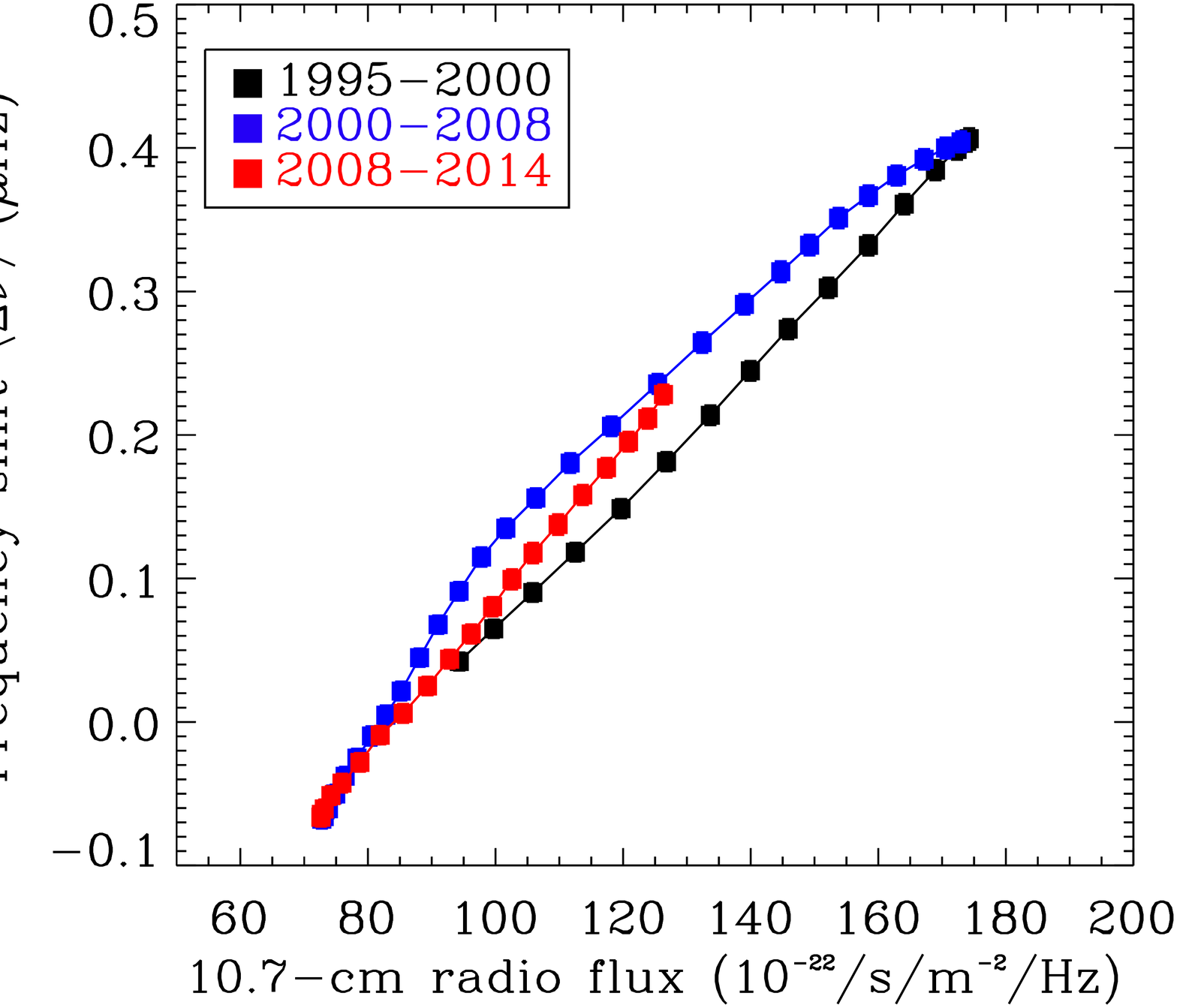}
\end{center} 
\caption{\label{fig:fig2} (Left panel) Frequency shifts, $\langle  \Delta\nu_{n,l=0,1,2} \rangle$, in $\mu$Hz averaged over 15 radial orders over 1800~$\mu$Hz~$\leq \nu <$ ~3790~$\mu$Hz as a function of the corresponding 10.7-cm radio flux, $F_{10.7}$, once the QBO's signature was removed. The rising (black dots) and declining (blue dots) phases of Cycle 23, and the rising phase (red dots) of Cycle 24 are indicated by different colors. (Middle panel) Same as left panel, but for the low-frequency modes averaged between 1800~$\mu$Hz~$\leq \nu <$~2450~$\mu$Hz. (Right panel) Same as the left and middle panels, but for the high-frequency modes between 3110~$\mu$Hz~$\leq \nu <$~3790~$\mu$Hz.}
\end{figure*} 

\begin{table*}[ht]
\begin{minipage}{\textwidth}
\caption{Peak-to-peak variations of the frequency shifts, $\langle  \Delta\nu_{n,l=0,1,2} \rangle$,  averaged over the modes $l=0$, 1, and 2 (in nHz)  calculated between different activity phases of Cycles 23 and 24 over four different frequency ranges.}
\label{table:absshift}      
\centering               
\renewcommand{\footnoterule}{}  
\begin{tabular}{c c c c c }        
\hline\hline   
 Frequency range  & (\textsc{Max$_{23}-$Min$_{22}$}) & (\textsc{Min$_{23}-$Max$_{23}$}) & (\textsc{Max$_{24}-$Min$_{23}$})  & (\textsc{Max$_{24} - $Min$_{23}$)/(Max$_{23}-$Min$_{22}$})\\    
 & nHz& nHz & nHz & \\    
\hline                           
 1800~$\mu$Hz $\leq \nu <$  3790~$\mu$Hz & 142.9~$\pm$~5.3 &  $-$168.1~$\pm$~3.9 & 114.1~$\pm$~4.3  & 0.80~$\pm$~0.04  \\
 \\
1800~$\mu$Hz $\leq \nu <$  2450~$\mu$Hz & 36.4~$\pm$~7.4 &  $-$54.3~$\pm$~5.7 & 39.9~$\pm$~6.5  & 1.09~$\pm$~0.29  \\
2450~$\mu$Hz $\leq \nu <$  3110~$\mu$Hz  & 198.8~$\pm$~9.0 &  $-$220.6~$\pm$~6.4 & 131.7~$\pm$~6.8 & 0.66~$\pm$~0.05  \\
3110~$\mu$Hz $\leq \nu <$  3790~$\mu$Hz  & 455.6~$\pm$~14.5 &  $-$500.1~$\pm$~10.5 & 281.5~$\pm$~10.9  & 0.62~$\pm$~0.03 \\
 \hline                                
\end{tabular}
\end{minipage}
\end{table*}

Figure~\ref{fig:fig5} shows the frequency shifts, $\langle  \Delta\nu_{n,l=0,1,2} \rangle$, averaged over the modes $l=0,1$, and 2 once the QBO's signature was removed by applying a proper smoothing of 2.5 years as prescribed in \citet{fletcher10} for the four frequency ranges previously defined. The solid line on each panel corresponds to the scaled 10.7-cm radio flux, smoothed in the same manner. Note that in order to avoid misinterpretation due to possible border effects introduced by the smoothing, we do not consider the extreme points corresponding to the first three and the last three points of the smoothed data. The larger frequency shifts compared to the surface activity proxy is clearly seen on Fig.~\ref{fig:fig5} during the rise of Cycle~24 from 2008. The low-frequency modes show the largest deviation with the 10.7-cm radio flux, with much larger frequency shifts than would be expected based on the surface activity proxy. The declining phase associated to Cycle 23 is much steeper as well than the activity proxy, unlike what is observed for  the mid- and high-frequency p modes.  
Moreover, while the mid- and high-frequency ranges show temporal variations cleaned from the QBO, a longer period seems to appear in the low-frequency modes since 2006.  Although that needs further observations in order to be confirmed, it could indicate changes related to the QBO deeper than 1400~km in the Sun's sub-surface layers.
See Appendix~B for additional figures based on the analysis of the frequency shifts at individual angular degree. 

The left panel of Fig.~\ref{fig:fig2} shows the frequency shifts, $\langle  \Delta\nu_{n,l=0,1,2} \rangle$, averaged over the 15 radial orders between 1800~$\mu$Hz and 3790~$\mu$Hz as a function of the associated 10.7-cm radio flux for the three phases of solar activity defined above, both quantities being properly smoothed as in \citet{fletcher10}. A clear magnetic hysteresis appears  between oscillation frequencies and radio flux over  Cycle 23, but
 the pattern changed since the minimum of Cycle 23 with larger frequency shifts for a given level of radio flux during the rising phase of Cycle 24. The middle and right panels of Fig.~\ref{fig:fig2} show respectively the low- and the high-frequency shifts as a function of the corresponding radio flux. The different behavior between the lower and the higher frequency modes is indicating differences in the structural and magnetic changes as a function of depth in the sub-surface layers of the Sun between Cycle 23 and Cycle 24 (see also the discussion in Section~4 and Appendix~B for additional figures of hysteresis at each $l$).
The frequency shifts are actually measured to be ahead of the radio flux by about 60~days in the velocity GOLF data, while a longer delay of about 90~days is observed with the data from the three Sun photometers composing the Variability of Solar Irradiance and Gravity Oscillations  \citep[VIRGO;][]{froh95}  instrument onboard SoHO \citep{salabert13}.  
 
 Table~\ref{table:absshift} gives the  differences in the frequency shifts averaged over $l=0$, 1, and 2 modes between the maximum and minimum of each solar cycle for the different frequency ranges as illustrated in Fig.~\ref{fig:fig5}. The periods of minimum and maximum of activity were defined by selecting the data points included within 15\% from the minimum (or the maximum) of the radio flux. The mean values of the frequency shifts over these selected periods were obtained before  calculating the differences.  
  The last column in Table~\ref{table:absshift} shows the ratio of amplitudes between Cycle 24 and Cycle 23, in the sense (\textsc{Max$_{24}-$Min$_{23}$})~$-$~(\textsc{Max$_{23}-$Min$_{22}$}). This indicates how the frequency shifts respond to the weaker Cycle 24 compared to Cycle 23.
The low-frequency modes have comparable shifts between Cycle~23 and Cycle~24, while the mid- and the high-frequency modes are about 30\% smaller during Cycle~24, which
is comparable to the decrease observed in radio flux emission between Cycles 23 and 24. The differences between  different frequency ranges thus suggest that the changes associated to the weaker Cycle 24 occur in shallower layers of the Sun, while the structure and the magnetic field deeper than 1400~km must have remained unchanged between Cycle 23 and Cycle 24, as inferred from the low-frequency modes. \\

\subsection{At individual angular degrees}
Oscillation modes with different angular degrees $l$ respond differently to the solar cycle \citep{salabert09} in relation to the spatial distribution of the surface magnetic field \citep{howe02,chano04}.  Studying them individually can provide insights about the differences observed between activity cycles (see also Appendix~B).

\begin{figure*}[tbp]
\begin{center} 
\includegraphics[width=0.24\textwidth]{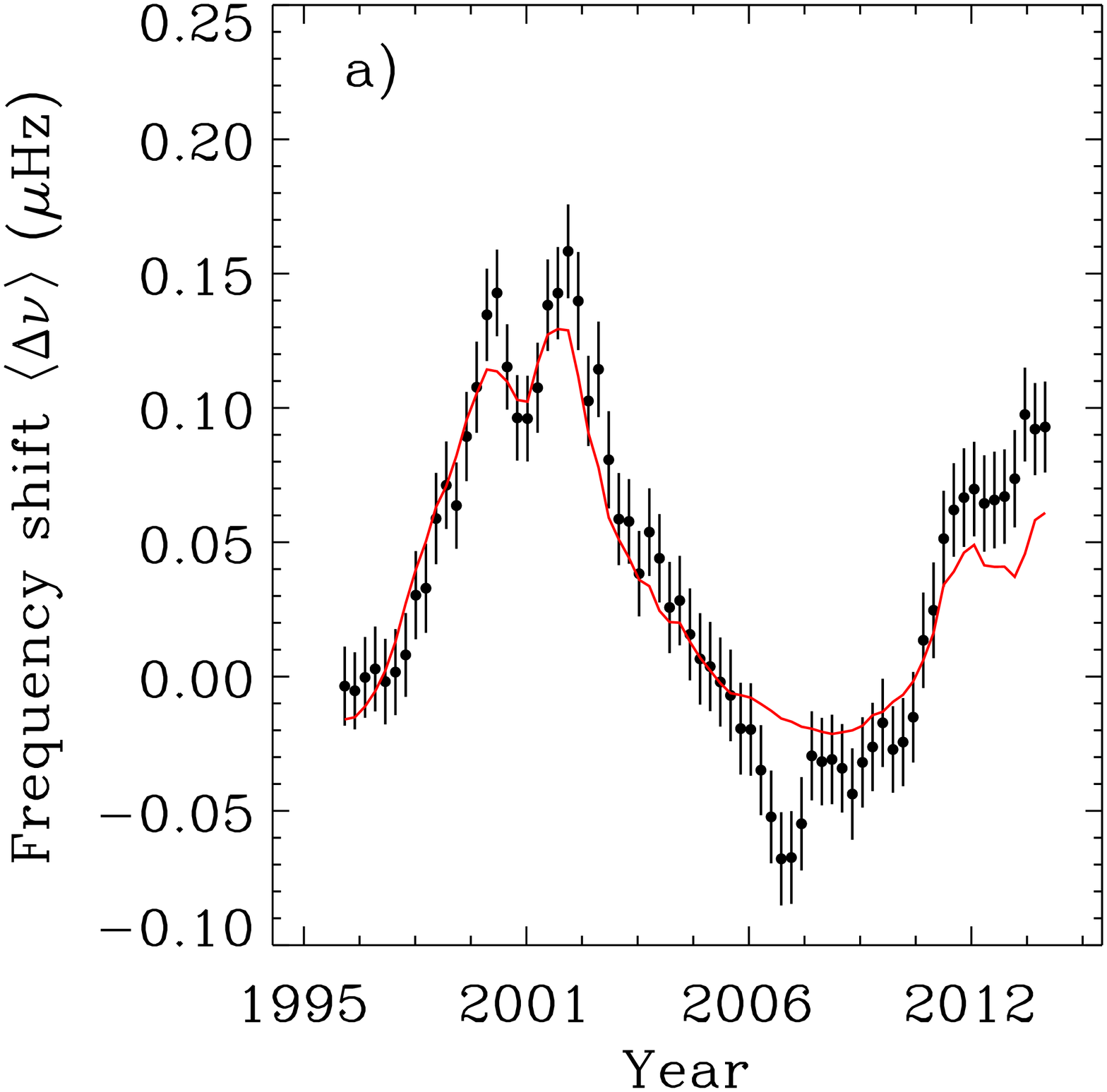}
\includegraphics[width=0.24\textwidth]{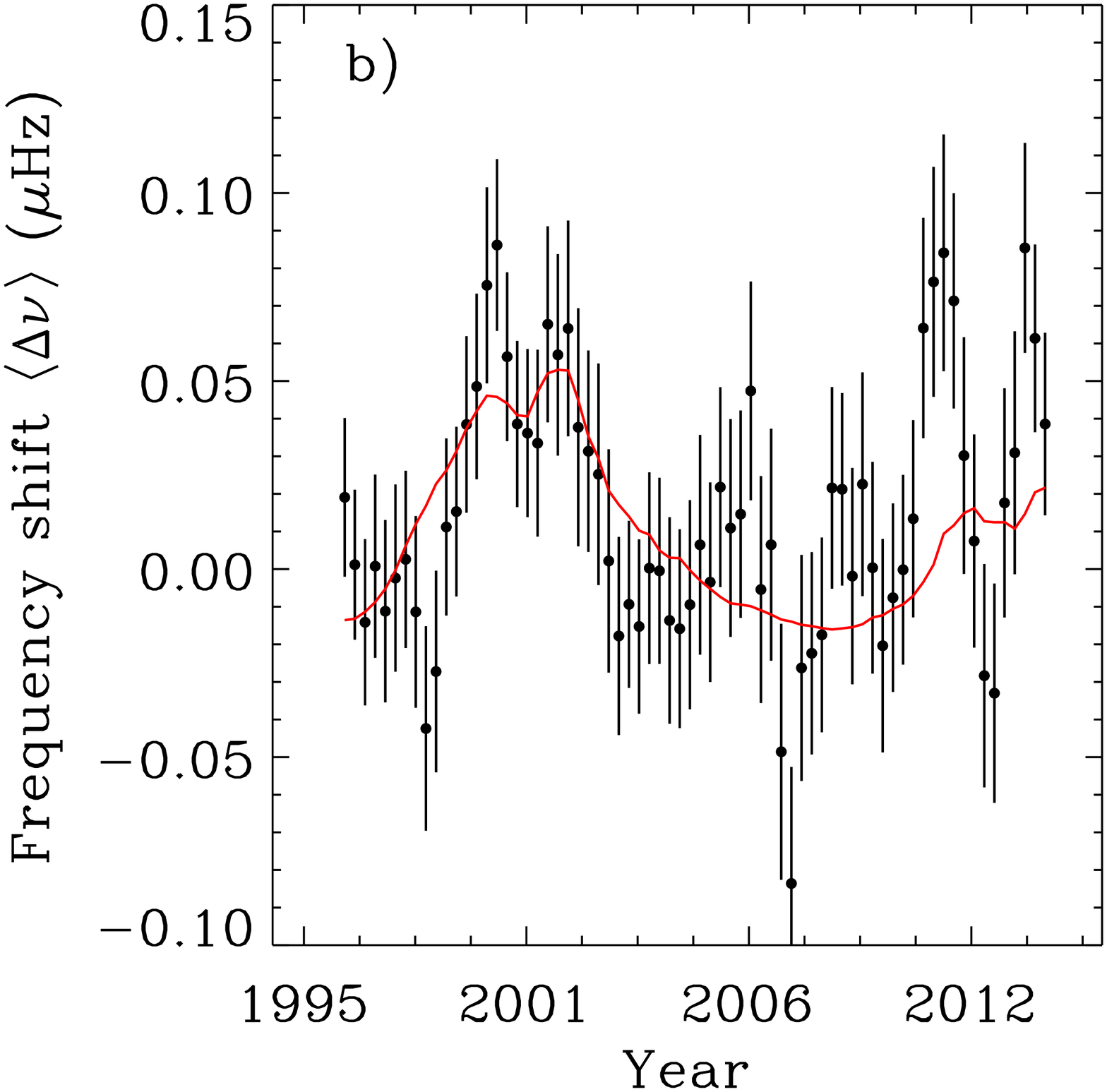}
\includegraphics[width=0.24\textwidth]{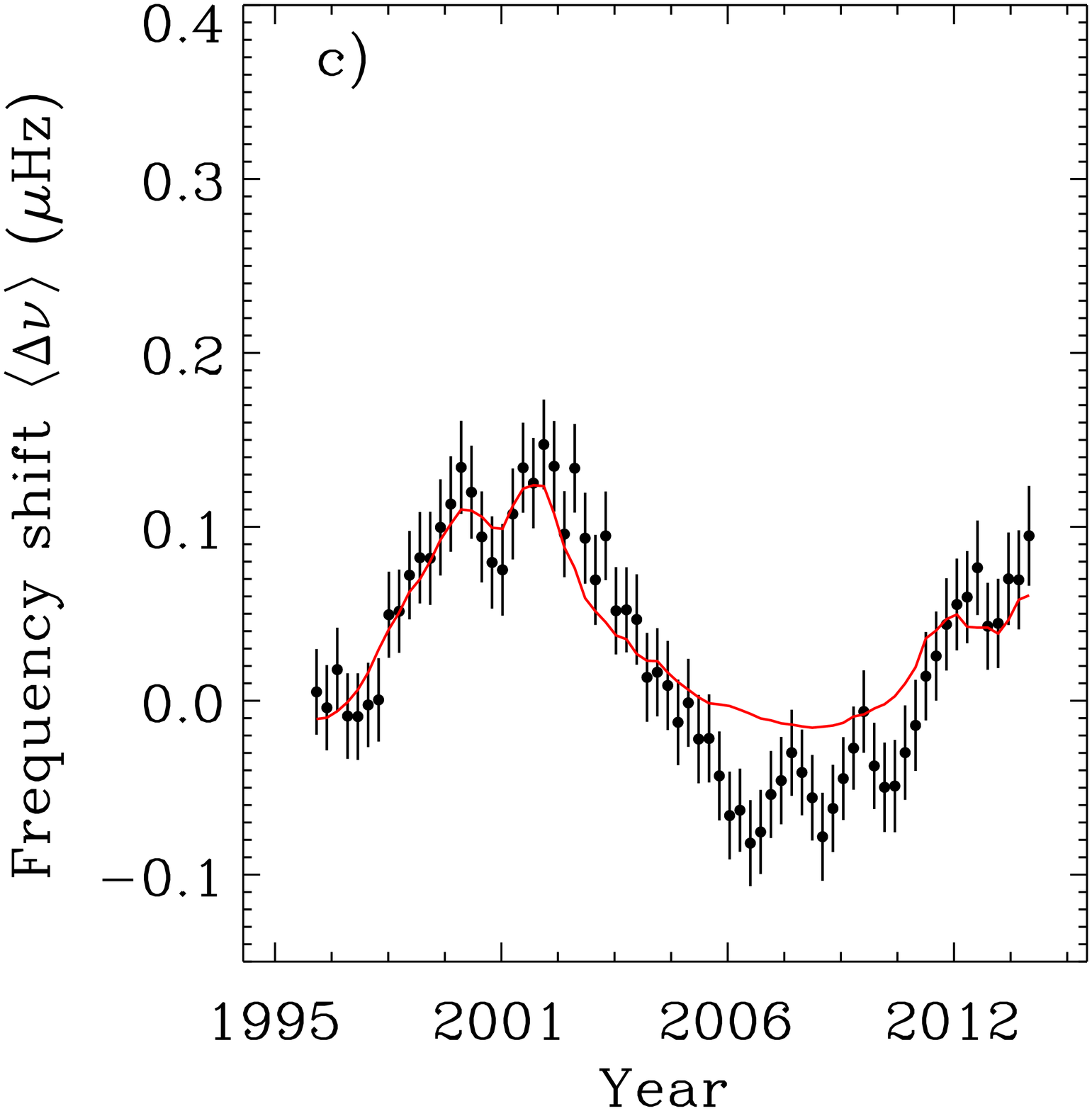}
\includegraphics[width=0.24\textwidth]{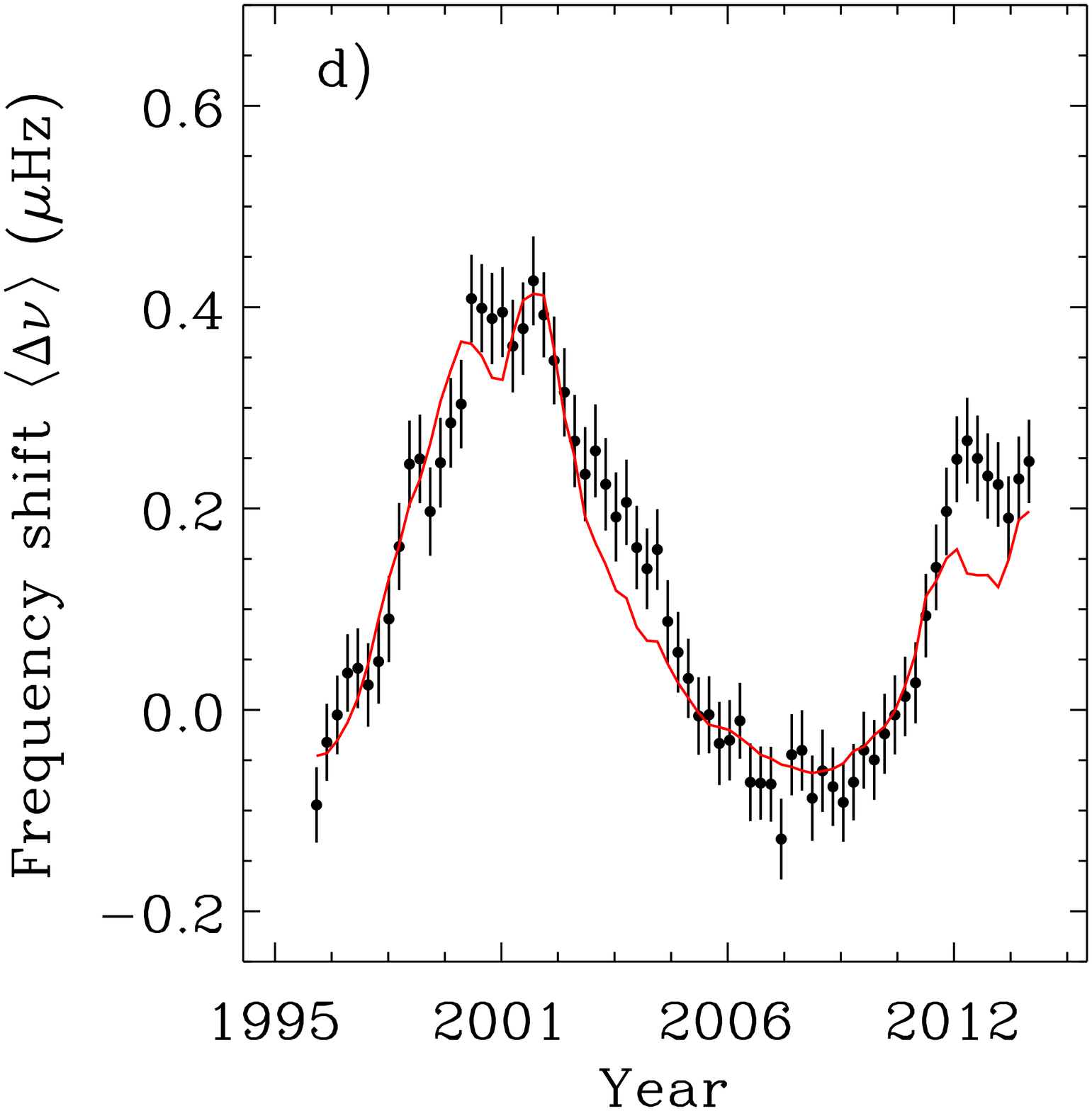}
\includegraphics[width=0.24\textwidth]{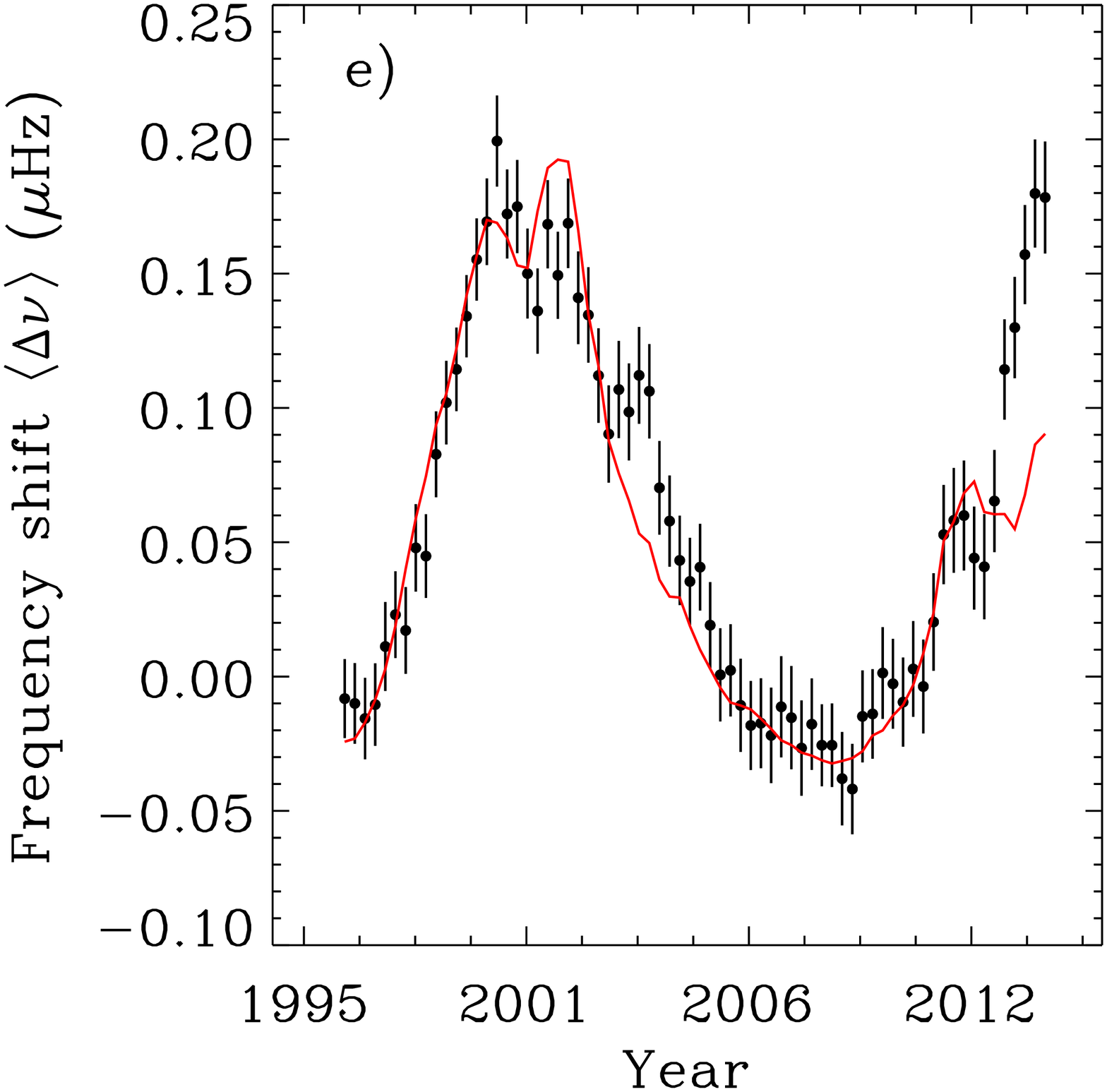}
\includegraphics[width=0.24\textwidth]{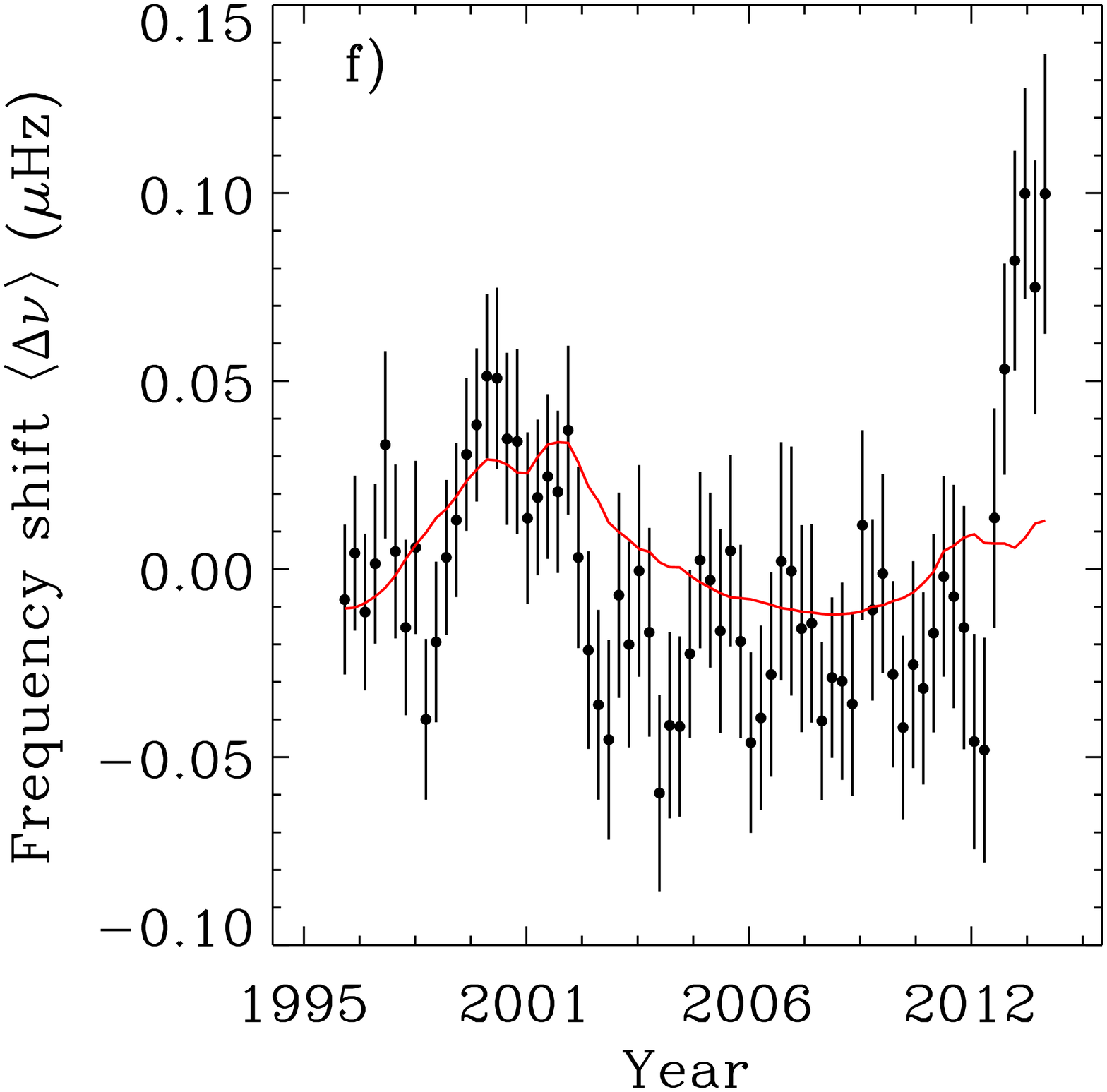}
\includegraphics[width=0.24\textwidth]{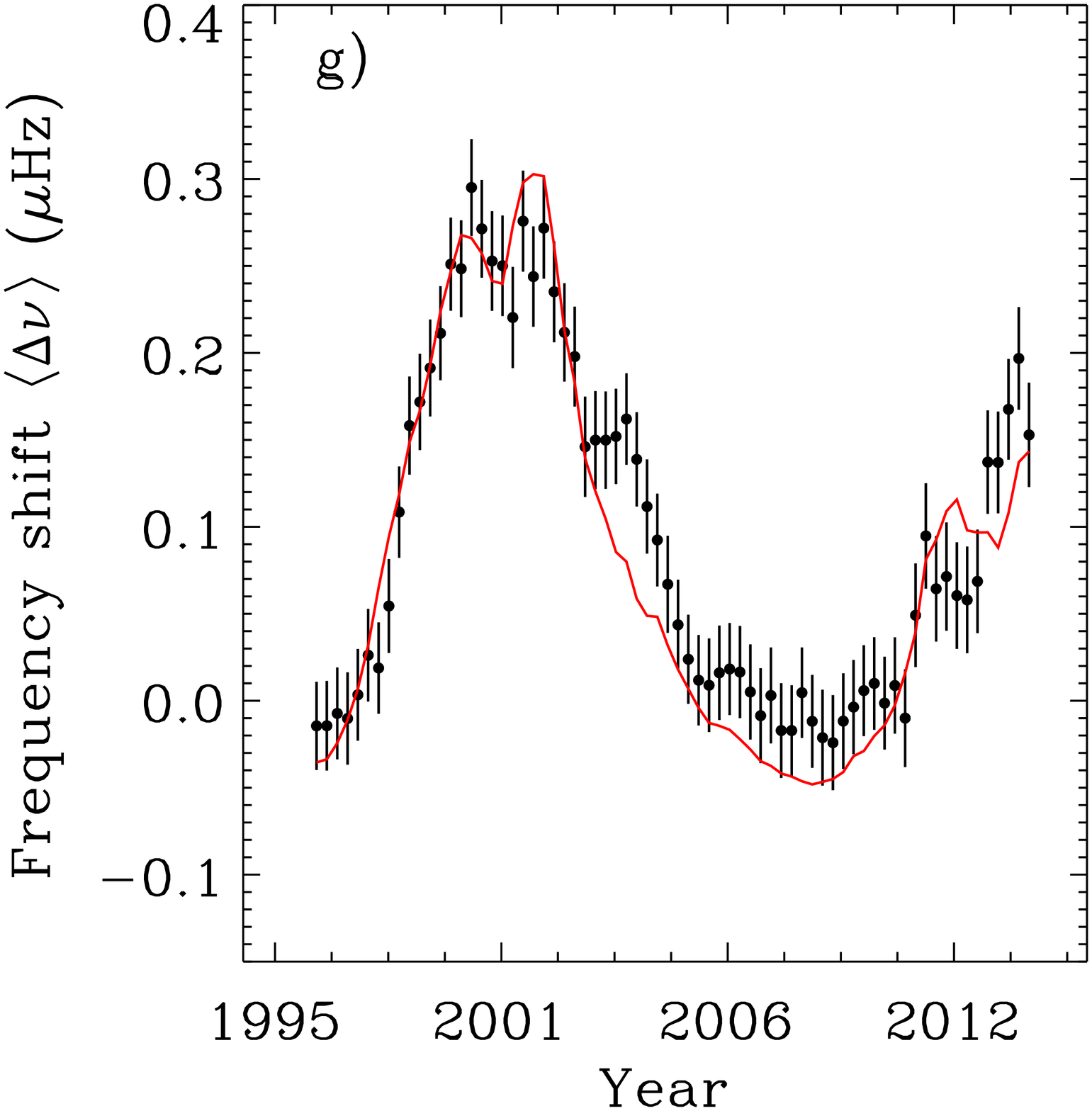}
\includegraphics[width=0.24\textwidth]{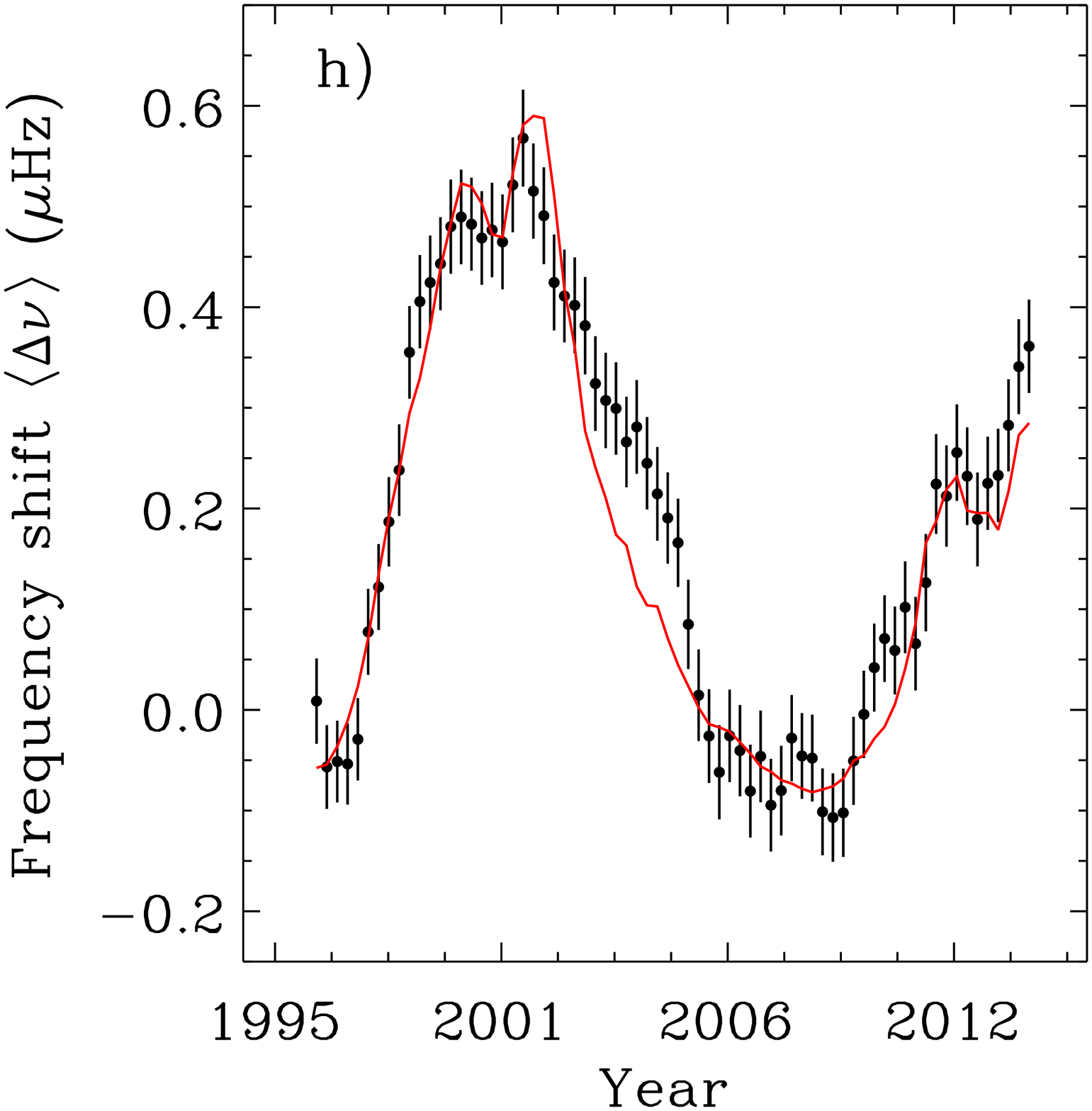}
\includegraphics[width=0.24\textwidth]{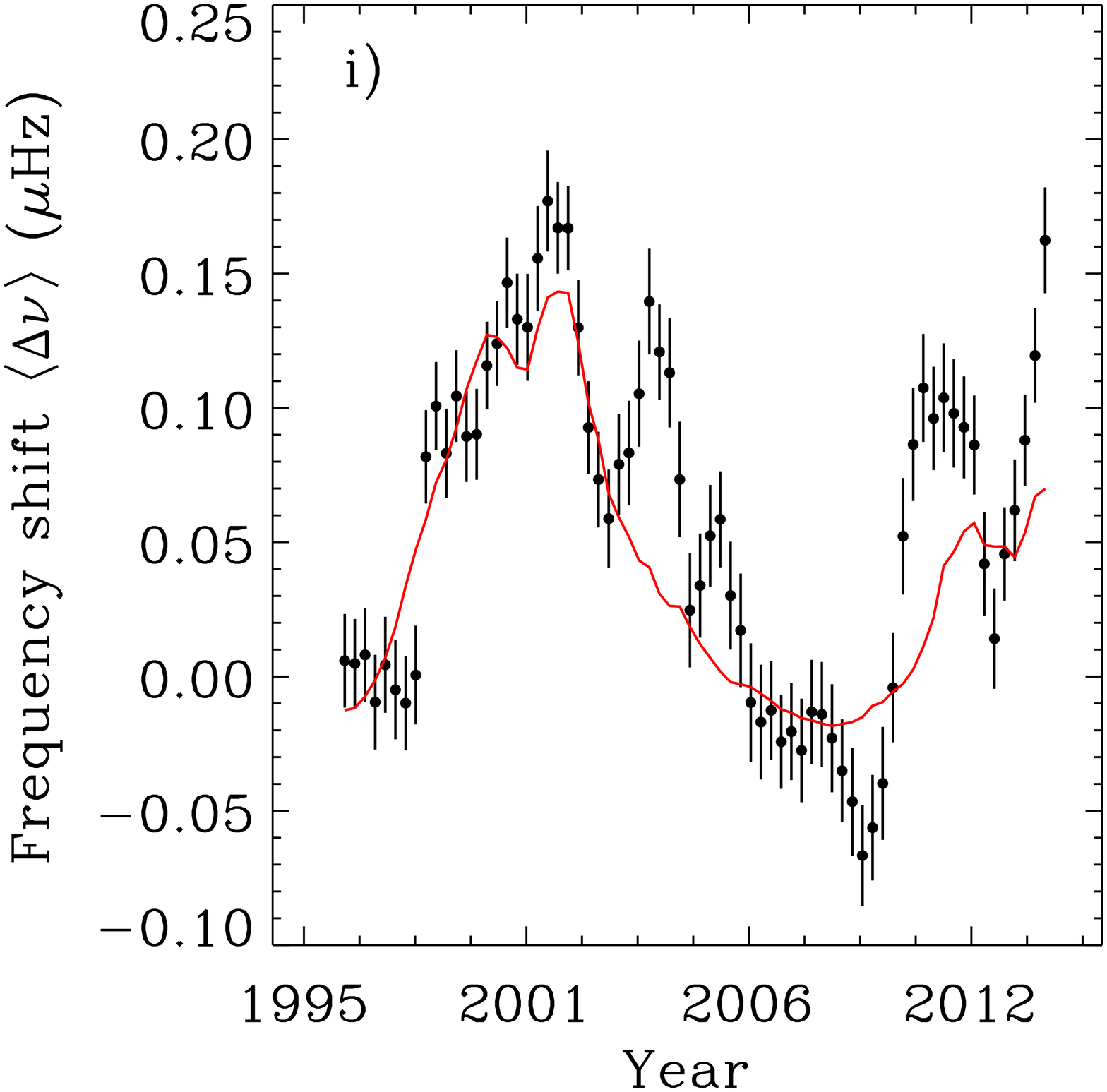}
\includegraphics[width=0.24\textwidth]{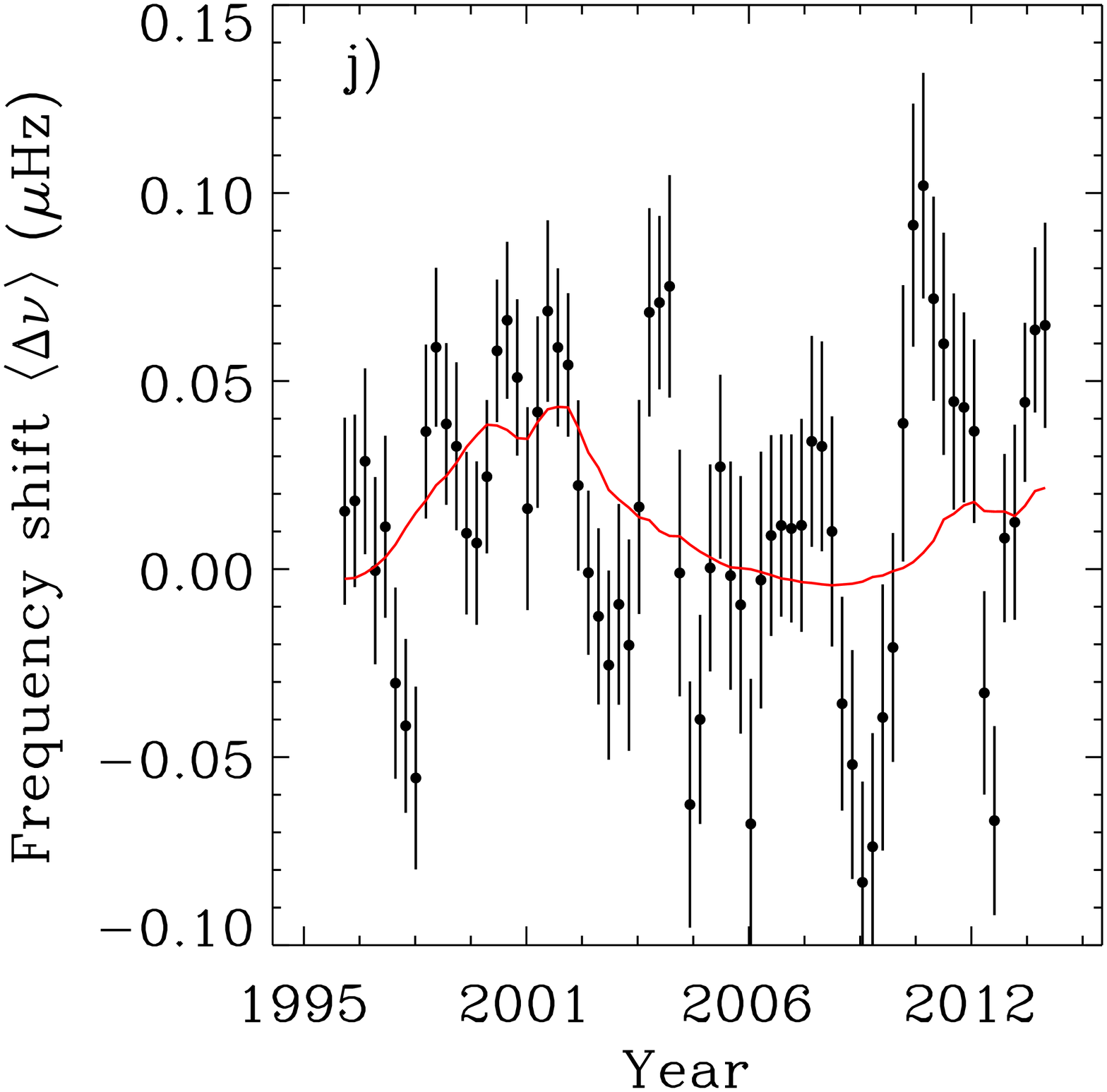}
\includegraphics[width=0.24\textwidth]{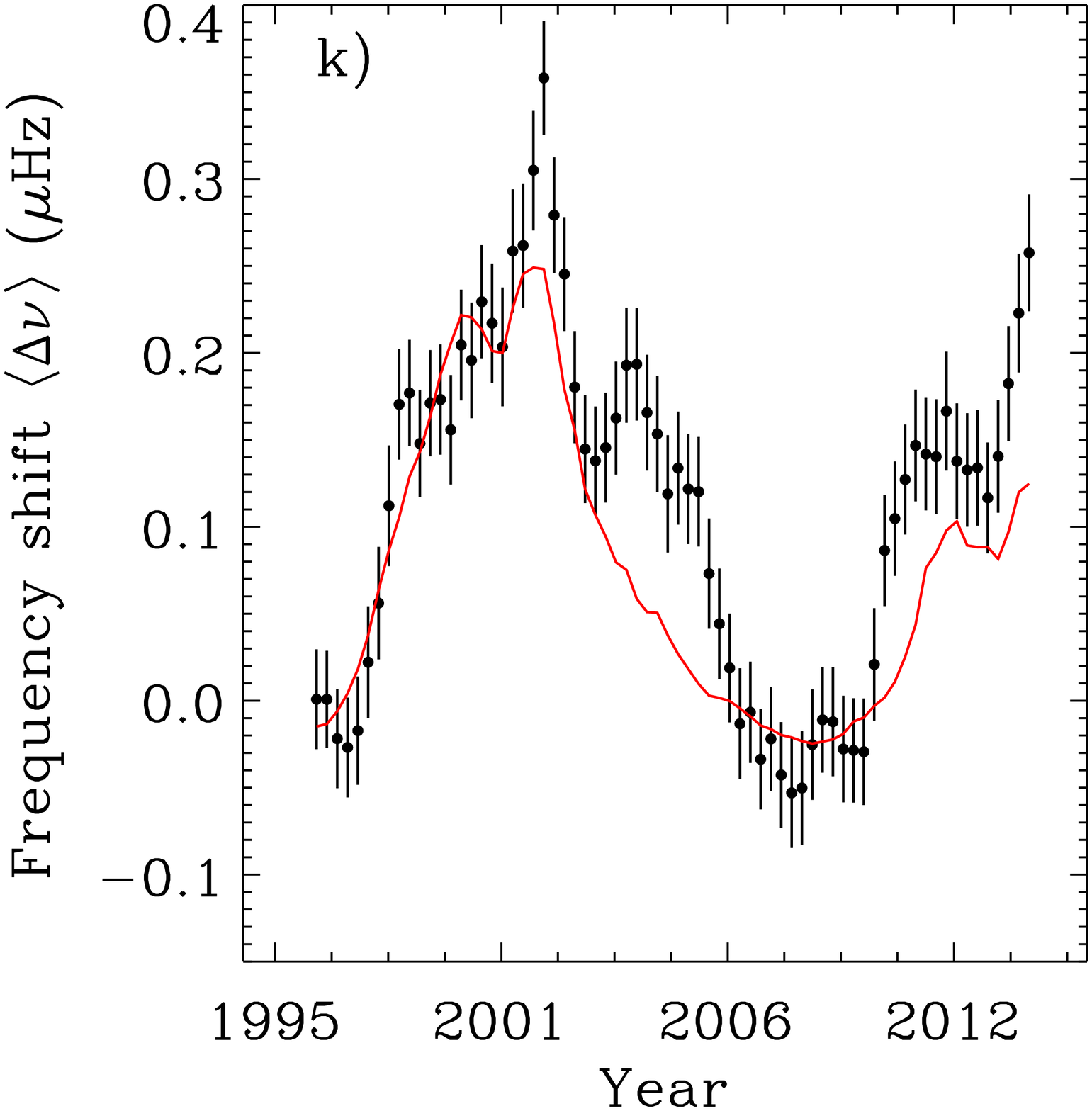}
\includegraphics[width=0.24\textwidth]{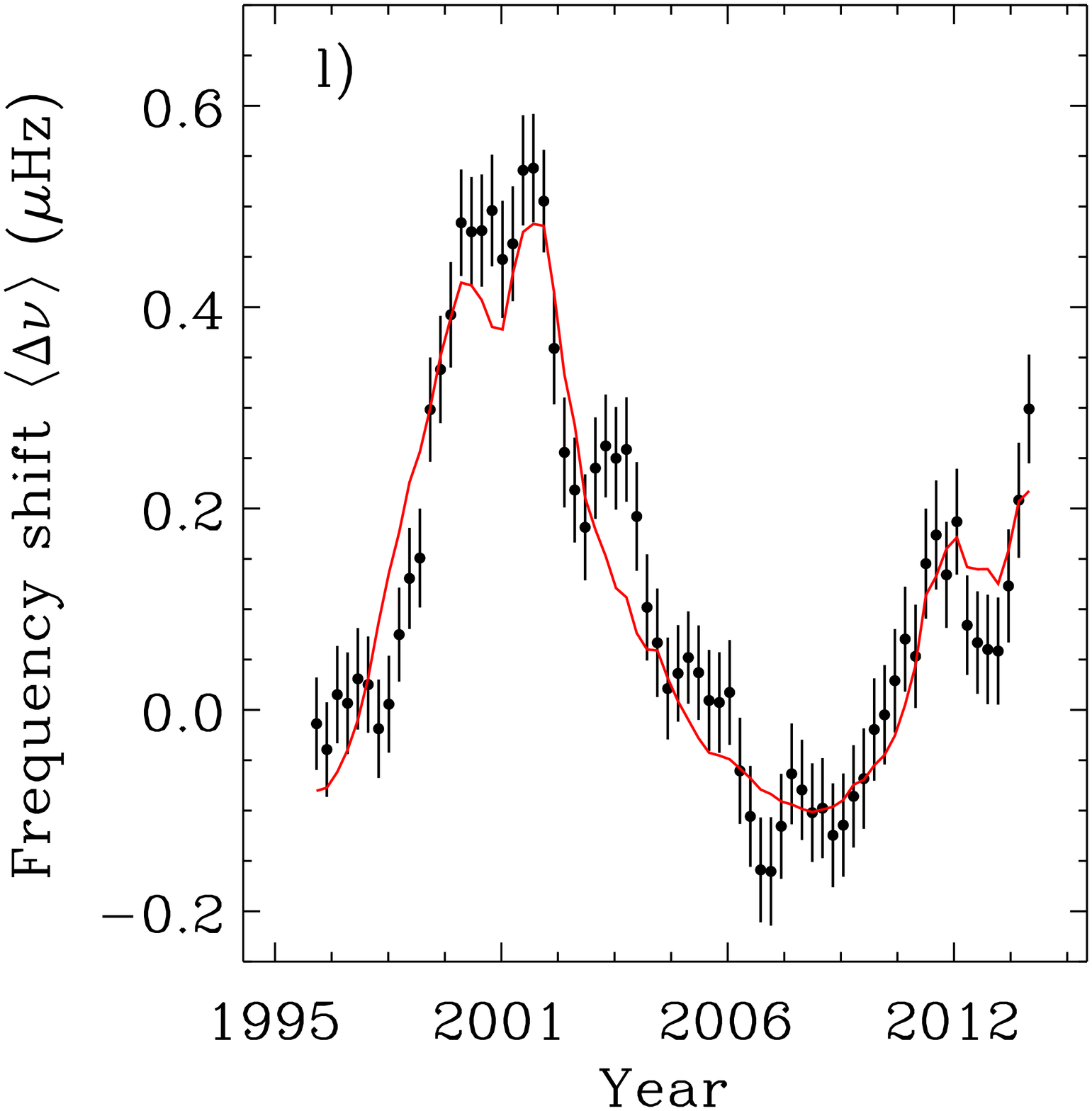}
\end{center}
\caption{\label{fig:fig4} Same as Fig.~\ref{fig:fig1}, but, from top to bottom, for the temporal variations  in $\mu$Hz of the frequency shifts, $\langle  \Delta\nu_{n,l=0} \rangle$ , $\langle  \Delta\nu_{n,l=1} \rangle$ , $\langle  \Delta\nu_{n,l=2} \rangle$, at each individual angular degree $l=0$, 1, and 2  (black dots).  The associated 10.7-cm radio flux, $F_{10.7}$, scaled to match the rising phase and the maximum of Cycle 23 is shown as a proxy of the solar surface activity (solid line).}
\end{figure*}

The twelve panels of Fig.~\ref{fig:fig4} show the temporal variations of the frequency shifts extracted from the GOLF data at each individual $l$ and calculated over the four frequency ranges. The individual $l=0$, 1, and 2 modes are represented from top to bottom, while the different frequency ranges are shown from left to right as illustrated in Fig.~\ref{fig:fig1}. The  10.7-cm radio flux is also represented, scaled to match the rising phase and the maximum of Cycle 23 as previously. The corresponding gradients and the correlation coefficients between frequency shifts at each $l$  and radio flux are also given in Table~\ref{table:correl_1800-3790}. Each individual $l$ modes show different responses depending on the frequency range, indicating different  temporal and latitudinal sensitivities to the structural and magnetic sub-surface changes with solar activity. The gradients between each $l$ are of the same order of magnitude over a given frequency range. Nevertheless, the detailed temporal evolution of the individual frequency shifts is different from one degree to the other, and from one frequency range to the other, providing precious insights into the structural and magnetic changes as a function of latitude and depth that could be related to the dipole and the quadrupole components of the magnetic field, as discussed in the next section.

\section{Discussion}
The temporal variability of the solar oscillation frequencies with solar activity results from the combination of structural changes in the sub-surface layers of the Sun, associated to other changes resulting from the topology of the magnetic field. In fact, the internal magnetic field does not act directly on the structure as the magnetic pressure is very low compared to the gas pressure in the solar interior, so it does not perturb it directly. This is the reason why the 1D modeling of the Sun, through the Standard Solar Model (SSM) or  the Seismic Solar Model (SeSM), do not contain any equations including the magnetic field \citep[e.g.,][]{turck11,basu14}.

\begin{figure*}[tbp]
\begin{center} 
\includegraphics[width=0.45\textwidth, angle=90]{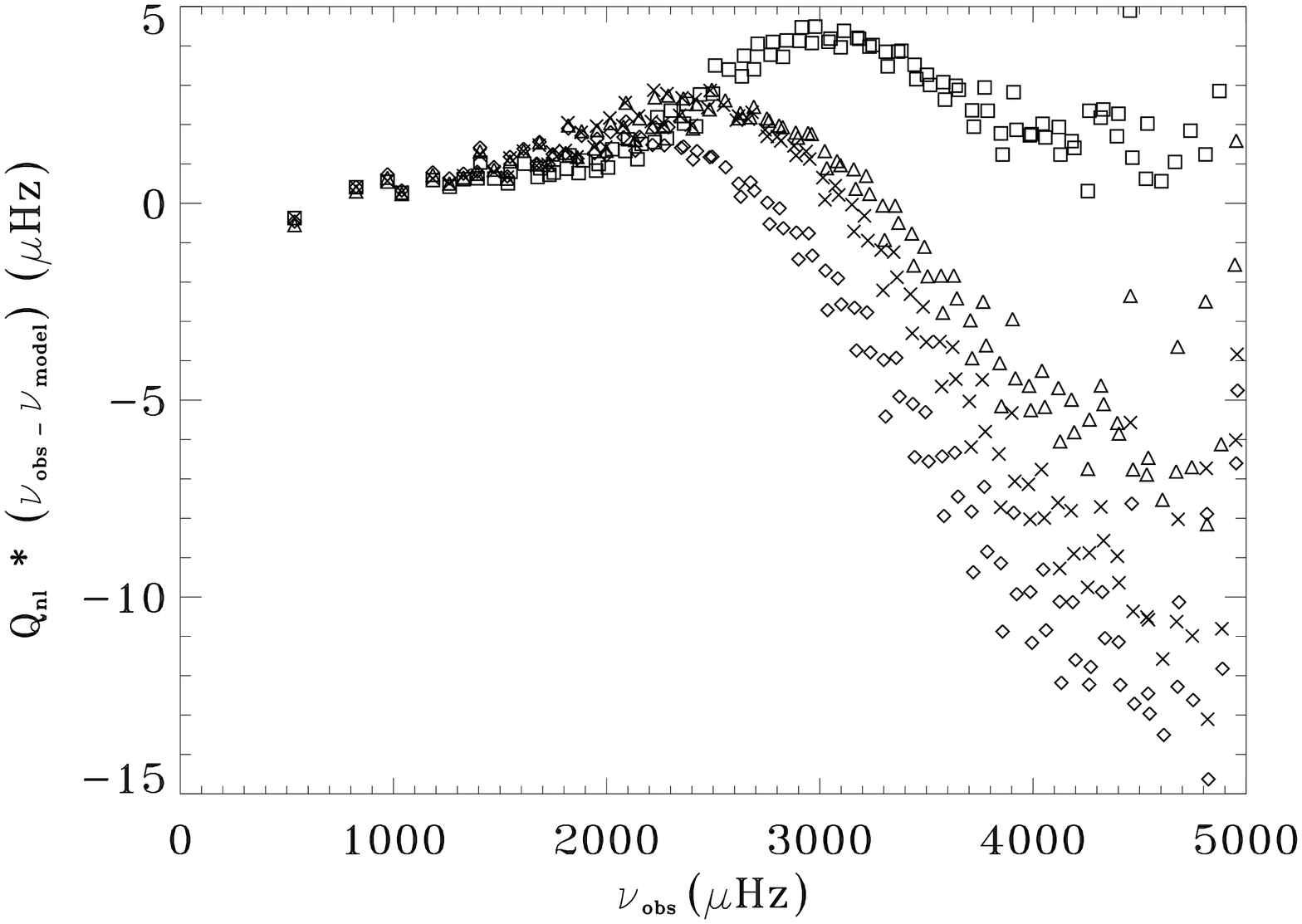}
\end{center} 
\caption{\label{fig:sylvaine} Differences in $\mu$Hz between observed GOLF frequencies used for sound speed inversion \citep{turcklopes2012} and theoretical frequencies of the SSM scaled by the inertia ratio Q$_{n, l}$ using MLT convection (diamonds); the Canuto prescription of convection (triangles) \citep{Canuto1991}; the mean thermal description (crosses); and 3D simulations from \citep{Nordlund1998} (squares). See also \citet{piau14} for the results obtained with the MDI data.}
\end{figure*}

It is now well established that the variability of the acoustic modes reveals a very small but important zone of the Sun located just below the surface called the superadiabatic region -- see Figure 1 of \cite{lefebvre09} illustrating the stratification of the upper layers of the Sun. The 3D simulations performed with STAGGER    \citep{Nordlund1998} represent one of the best ways to study the action of the magnetic field in this region as these simulations correspond to layers located between $-$2000 km and $+$900 km around the surface. STAGGER properly describes the turbulent pressure and the motions that act below the solar surface, and it reproduces remarkably well  the convective motions at the surface. The addition of a magnetic field has two effects in that region of the Sun: it partly inhibits the turbulent pressure which cannot be properly  described in a 1D solar model, and it directly acts on the gas pressure when the magnetic pressure is relatively not negligible around the solar surface. The advantage to combine a real 3D turbulent structure with a 1D model was demonstrated by \citet{piau14}, as it improves the prediction of the absolute values of the acoustic frequencies and help to interpret the observations of the sub-surface regions. The  introduction of an accurate turbulent pressure modifies the absolute value of  the theoretical acoustic mode frequencies. While the SSM or SeSM lead to theoretical frequencies greater than the observed ones by up to 10~$\mu$Hz, an hydrodynamical model reduces the differences with the seismic values within  3~$\mu$Hz because these layers are better described. Figure~\ref{fig:sylvaine} shows this result compared to different ways to treat the convection in 1D.

The introduction of a magnetic field strength of 1200~G  at the bottom of the simulation compatible with the strength of the toroidal field at 0.996~$R_{\sun}$  extracted by \citet{Baldner2009} from the   seismic observations collected by the Michelson Doppler Imager \citep[MDI;][]{scherrer95} onboard SoHO, reduces slightly the turbulence and leads to a small shrink in the solar radius of dozens of kilometers at the surface \citep{piau11}. 
The $l=0$ radial differences at the high-frequency range extracted from the GOLF observations (Fig.~\ref{fig:fig4}) of about $1.3\times 10^{-4}$ between minimum and maximum of activity  could be attributed to pure structural changes. If we follow the relation ${\Delta R \over R}= {-3 \over 2} {\Delta \nu \over  \nu}$,  that would correspond to an upper limit for a change in solar radius of about 50~km  at maximum of Cycle~23 and of about 30~km during Cycle~24.
\citet{bush2015}  extracted a radius change of  $4\times10^{-5}$ during the last 4 years using the observations from the Helioseismic and Magnetic Imager \citep[HMI;][]{scherrer12} instrument onboard the Solar Dynamics Observatory spacecraft, which is compatible with the estimate extracted from the radial mode observed by the GOLF instrument.
The $l=1$ dipole and $l=2$ quadrupole temporal variations shown in Figure 4 at the high-frequency range, show reasonably well the same order of changes combined with a clear increase of the quasi-biennal component which probably translates the beating of the dipolar and quadrupolar components of the magnetic field just below the surface \citep{Tobias1996,Simoniello2013,TobiasCattaneo2013}.\\

The temporal variations of the medium-frequency modes  are clearly less sensitive to radius changes ($0.15~\mu$Hz instead $0.4~\mu$Hz for the $l = 0$ modes).  \cite{lefebvre09} demonstrated that a change in radius of $2\times10^{-4}$ perturbs the internal structure down to 0.99 R$_{\sun}$ through opacity and equation of state variations that could induce pulsations of these layers by kappa mechanisms. The differences observed here cannot be then interpreted as a pure structural change when the magnetic field in the sub-surface layers slightly changes. 
In order to better understand the situation, we also need to consider the  observations of the emerging solar magnetic field. Three distinct results 
 are useful to that respect: 1) the observation of the solar wind collected by the Ulysses mission has shown very different minima for Cycles~22 and 23, the minimum of Cycle~23 having a rather opened field near the equator which shows the increasing importance of the toroidal component \citep{McComas2008}; 2) the dipolar and the radial magnetic fields observed at the Wilcox Solar Observatory (WSO) \citep{Hoeksema2009} since 1978; and 3) the measurements  of  the sunspot magnetic field strength which is  decreasing since Cycle~22 and the measurement of the 1500~G  threshold needed in order to dark sunspots to form \citep{livingston12}.   The variability of the  $l=1$ and $l=2$ modes is practically twice larger than the $l=0$ mode variability, with a more perturbed temporal evolution than the one observed in the 10.7-cm radio flux and  the $l=0$ modes. This is particularly true for the $l=2$ modes. This is actually in agreement with the observations of the magnetic field performed at WSO which indicate that even if both dipolar and quadrupolar components of the external fields are decreasing since at least 1985, the ratio between quadrupolar and dipolar components of the radial field shows a constant increase \citep{Inceoglu2015}.

The low-frequency range is the most interesting region to consider. Even if the effects are very small and that the maximum of Cycle~23 is visible, we observe very perturbed signals for the low-degree modes mainly for the $l=2$ modes. Table~\ref{table:correl_1800-3790} shows low  levels of correlation with the 10.7-cm radio flux since the maximum of Cycle~23,  and Table~\ref{table:absshift} indicates that the ratio of the peak-to-peak variations of the frequency shifts between Cycle 24 and Cycle 23, (\textsc{Max$_{24}-$Min$_{23}$})~$-$~(\textsc{Max$_{23}-$Min$_{22}$}), is larger for  the low-frequency modes than for the medium- and high-frequency modes.
Moreover, the toroidal field, responsible for the local partial inhibition of the turbulence and associated consequences, was observed to increase by about a factor 3 during the maximum of  Cycle~23 toward the solar interior between 0.999 R$_{\sun}$ and 0.996 R$_{\sun}$ \citep{Baldner2009}.  What is unclear is  to know if  the observed  variability of the acoustic low-frequency modes comes from oscillations due to a  kappa mechanism or from a beating between the poloidal and toroidal components of the magnetic field.
The first assumption implies  that the QBO's signature should decrease as a function of time, which is not the case. We can then  speculate about an amplification of the variations of the low-frequency shifts to an increase of the ratio between components of the magnetic field from Cycle~22 to Cycle~24.
Nevertheless, the sudden increase of the $l=1$ shifts at low frequency is unexpected in this context and justifies then more investigations (see Appendix~B). Also, as observed in sunspot number, the last solar cycles present clear hemispheric asymmetries\footnote{Hemispheric sunspot numbers are provided by SILSO, Royal Observatory of Belgium, Brussels at http://sidc.oma.be/silso/datafiles.}, with successive dominant north and south hemispheres during the maximum of activity.

 The study presented here indicates  that the deeper toroidal magnetic field seems rather stable along Cycle~23 and 24, while the magnetic field at the surface is overall decreasing.  As a consequence, a relative disconnection between acoustic frequencies and surface proxy, such as the 10.7-cm radio flux, is observed to be more and more visible in the deeper sub-surface layers of the Sun below 1400~km. Its physical reason is not yet understood.  One possibility is the existence of two dynamos, a deep one and a shallow one. Another possibility, among others,  is that some variability, larger than the 11-year timescale,  exists in the tachocline and produces the emergence at the surface of varying flux tubes in addition to the small-scale convection present in the sub-surface layers. Moreover, the chaotically modulated solar dynamo, proposed by \cite{Tobias1995}, does not appear to be in contradiction with the present observations.
Any progress supposes a good knowledge of the deep magnetic field in the quiet and active Sun. More realistic 3D spherical simulations of the global dynamo coupled to superficial cartesian 3D simulations  with different topologies of the magnetic field could provide invaluable understanding of the internal dynamical Sun.\\

\section{Conclusions}
We analyzed 18 years of the Sun-as-a-star, space-based radial velocity helioseismic GOLF observations spanning the entire Cycle 23 and the rising phase of the weak Cycle 24. We studied the temporal variations of the low-degree oscillation frequencies along this 18-year long dataset in order to propose inferences on the structural and magnetic changes in the sub-surface layers of the Sun which could explain the progression of Cycle 24. Indeed, the averaged kernels of the acoustic modes at different frequency ranges have their largest sensitivities at different depths. While the sensitivity of the low-degree, high-frequency modes (between 3110~$\mu$Hz $\leq \nu <$ 3790~$\mu$Hz) peaks at 0.9989~R$_{\sun}$ (i.e., 760~km) from the Sun's surface, the low-degree, low-frequency modes (1800~$\mu$Hz $\leq \nu <$  2450~$\mu$Hz) are mostly sensitive to deeper-lying layers at 0.9965~R$_{\sun}$ (i.e., 2400~km). 

We showed that while the frequency shifts associated to layers above 1400~km have been following the decrease of surface activity during Cycle 24 compared to Cycle 23, the frequency shifts associated to layers below 1400~km remained similar between Cycle~23 and Cycle~24 for the extrema of the Schwabe cycle. The study of the temporal variations of the individual mode oscillation frequencies observed by GOLF thus suggests that while Cycle 24 is clearly magnetically weaker in the shallower parts of the Sun, the solar Cycle 23 and 24 are magnetically and structurally  similar below 1400~km, except the quasi-biennial oscillation for the radial and quadrupole modes which is observed to be slightly amplified and to possibly have a longer temporal period in these deeper-lying parts during the rising phase of Cycle~24. The dipole modes show however  a specific and not explained behavior yet.
Inversions combining low- and  high-degree modes would provide a more precise localization of the evolution of the magnetic field from the surface to deeper parts of the Sun.

 \begin{acknowledgements}
The GOLF instrument onboard SoHO is a cooperative effort of scientists, engineers, and technicians, to whom we are indebted. SoHO is a project of international collaboration between ESA and NASA. The authors strongly acknowledge the french space agency, CNES, for its support to GOLF  since the launch of  SoHO. We are also particularly grateful to Catherine Renaud for her daily check of the GOLF data. DS acknowledges the financial support from CNES. The research leading to these results has also received funding from the European Community's Seventh Framework Programme ([FP7/2007-2013]) under grant agreement no. 312844 (SPACEINN). The 10.7-cm solar radio flux data were obtained from the National Geophysical Data Center. The authors acknowledge J. Ballot for useful discussions.
 \end{acknowledgements}

\begin{appendix} 
\section{GOLF/SoHO low-degree acoustic oscillation frequencies}
\begin{table*} 
\caption{Acoustic oscillation frequencies of the $l=0$, 1, 2, and 3 modes extracted from the 365-day GOLF/SoHO spectrum calculated between 11-04-1996 and 10-04-1997.} 
\label{table:001} 
\centering 
\begin{tabular}{c c c c c}\hline \hline 
$n$ & $l=0$ & $l=1$ & $l=2$ & $l=3$ \\
\hline
 9 &  $\pm$  &  $\pm$  & 1535.872 $\pm$   0.034 &  $\pm$   \\ 
10 & 1548.414 $\pm$   0.041 & 1612.734 $\pm$   0.028 & 1674.568 $\pm$   0.057 &  $\pm$   \\ 
11 & 1686.538 $\pm$   0.043 & 1749.297 $\pm$   0.040 & 1810.304 $\pm$   0.052 &  $\pm$   \\ 
12 & 1822.197 $\pm$   0.043 & 1885.106 $\pm$   0.037 & 1945.823 $\pm$   0.045 &  $\pm$   \\ 
13 & 1957.469 $\pm$   0.051 & 2020.840 $\pm$   0.033 & 2082.058 $\pm$   0.049 & 2137.784 $\pm$    0.064 \\ 
14 & 2093.510 $\pm$   0.038 & 2156.811 $\pm$   0.053 & 2217.819 $\pm$   0.083 & 2273.370 $\pm$    0.075 \\ 
15 & 2228.819 $\pm$   0.049 & 2291.996 $\pm$   0.064 & 2352.285 $\pm$   0.074 & 2407.785 $\pm$    0.122 \\ 
16 & 2362.799 $\pm$   0.065 & 2425.473 $\pm$   0.060 & 2485.921 $\pm$   0.083 & 2541.352 $\pm$    0.126 \\ 
17 & 2496.222 $\pm$   0.060 & 2559.231 $\pm$   0.059 & 2619.644 $\pm$   0.070 & 2676.191 $\pm$    0.098 \\ 
18 & 2629.906 $\pm$   0.055 & 2693.303 $\pm$   0.052 & 2754.474 $\pm$   0.061 & 2811.351 $\pm$    0.087 \\ 
19 & 2764.146 $\pm$   0.052 & 2828.103 $\pm$   0.056 & 2889.557 $\pm$   0.053 & 2946.983 $\pm$    0.058 \\ 
20 & 2898.947 $\pm$   0.055 & 2963.273 $\pm$   0.054 & 3024.664 $\pm$   0.064 & 3082.219 $\pm$    0.093 \\ 
21 & 3033.772 $\pm$   0.054 & 3098.152 $\pm$   0.063 & 3159.856 $\pm$   0.066 & 3217.738 $\pm$    0.088 \\ 
22 & 3168.619 $\pm$   0.055 & 3233.168 $\pm$   0.065 & 3295.107 $\pm$   0.090 & 3353.672 $\pm$    0.164 \\ 
23 & 3303.376 $\pm$   0.071 & 3368.612 $\pm$   0.081 & 3430.871 $\pm$   0.109 & 3489.737 $\pm$    0.361 \\ 
24 & 3438.884 $\pm$   0.086 & 3503.887 $\pm$   0.105 & 3566.714 $\pm$   0.200 & 3626.576 $\pm$    0.466 \\ 
25 & 3574.584 $\pm$   0.170 & 3640.344 $\pm$   0.135 & 3703.153 $\pm$   0.348 &  $\pm$   \\ 
26 & 3710.794 $\pm$   0.235 & 3776.536 $\pm$   0.208 & 3839.292 $\pm$   0.638 &  $\pm$   \\ 
27 & 3847.099 $\pm$   0.334 & 3913.674 $\pm$   0.267 &  $\pm$  &  $\pm$   \\ 
\hline 
\end{tabular} 
\end{table*}

Table~A.1 illustrates the content of the frequency tables made electronically available through this work. They contain the central frequencies of the low-degree acoustic modes of oscillations of the Sun extracted from  365-day sub series of 18 years of GOLF/SoHO observations between April 11, 1996 and March 5, 2014. As a four-time overlap of 91.25~days was used, a total of 70 frequency tables are provided. Note then that one every four frequency tables contain frequencies extracted from independent sub series.
See Section~2 for a detailed description of the methodology applied to extract the mode frequencies. 
Quality criteria were defined based on the fitted mode parameters and their associated uncertainties in order to remove outliers, such as: 1) the error of the mode frequency must be less than its mode width; 2) the signal-to-noise ratio must be larger than 1; and 3) the mode width must be larger than the frequency resolution. The entire set of tables containing the GOLF $l=0$, 1, 2, and 3 p-mode frequencies between 1500~$\mu$Hz  and 4000~$\mu$Hz, and their associated formal 1$\sigma$ uncertainties are available in a machine readable format at CDS. The GOLF velocity time series used in this work, as well as the extracted frequency tables are directly available on the GOLF website at CEA\footnote{The GOLF website at CEA is also mirrored through the SPACEINN website http://www.spaceinn.eu.} and through the SoHO data archive at NASA and ESA.

\section{Frequency shifts at individual angular degree}
\begin{figure*}[tbp]
\begin{center} 
\includegraphics[width=0.24\textwidth]{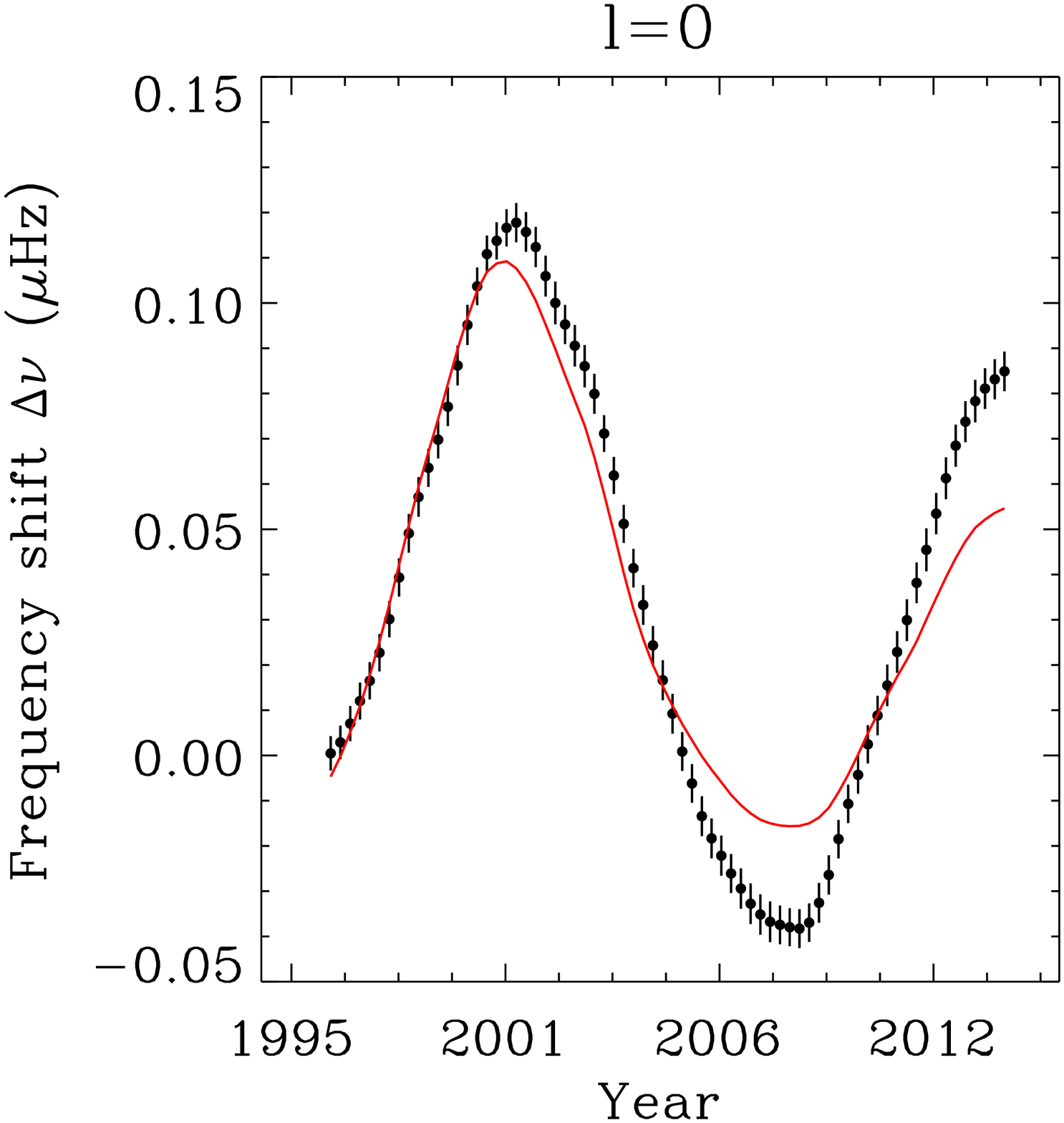}
\includegraphics[width=0.24\textwidth]{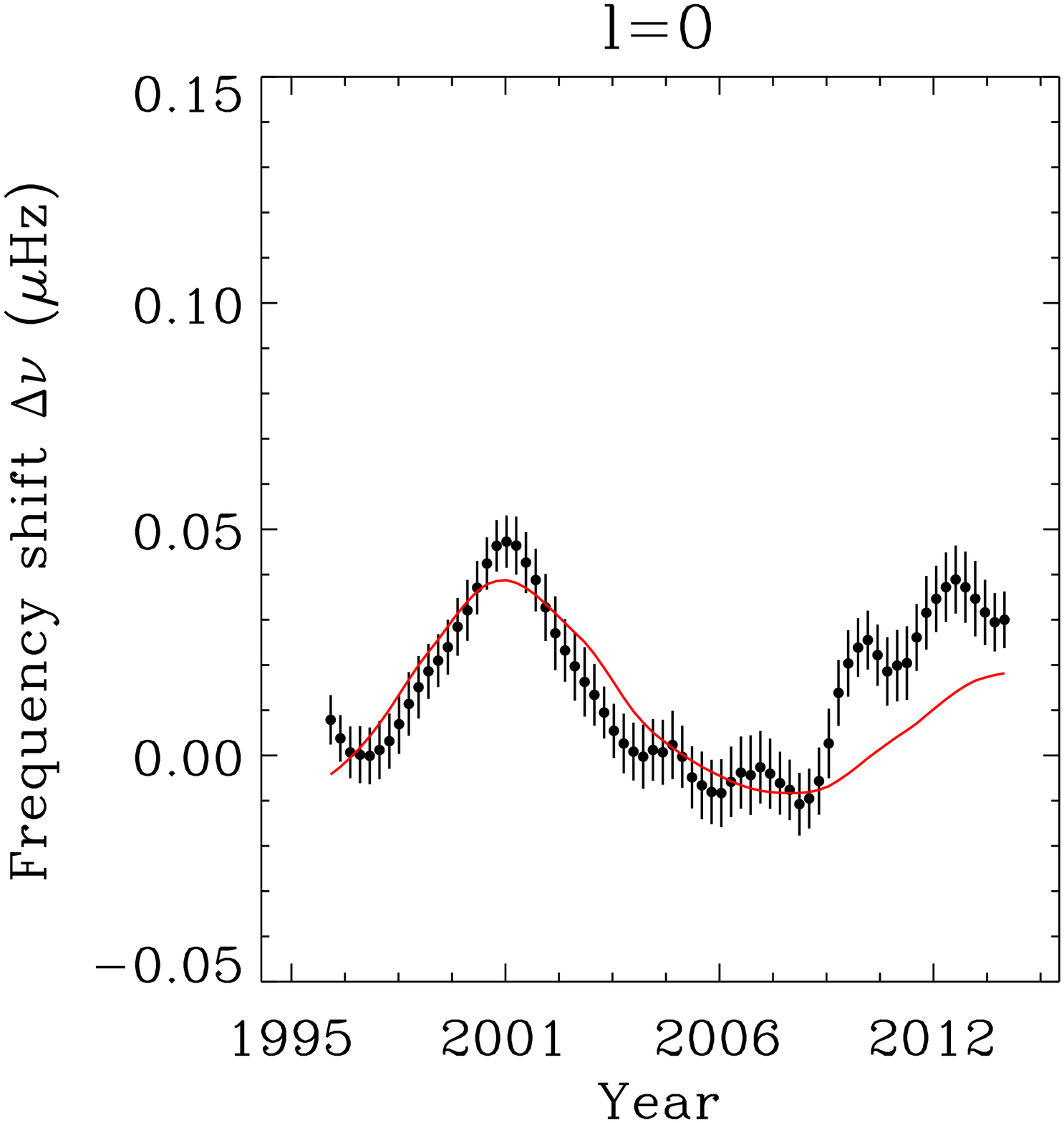}
\includegraphics[width=0.24\textwidth]{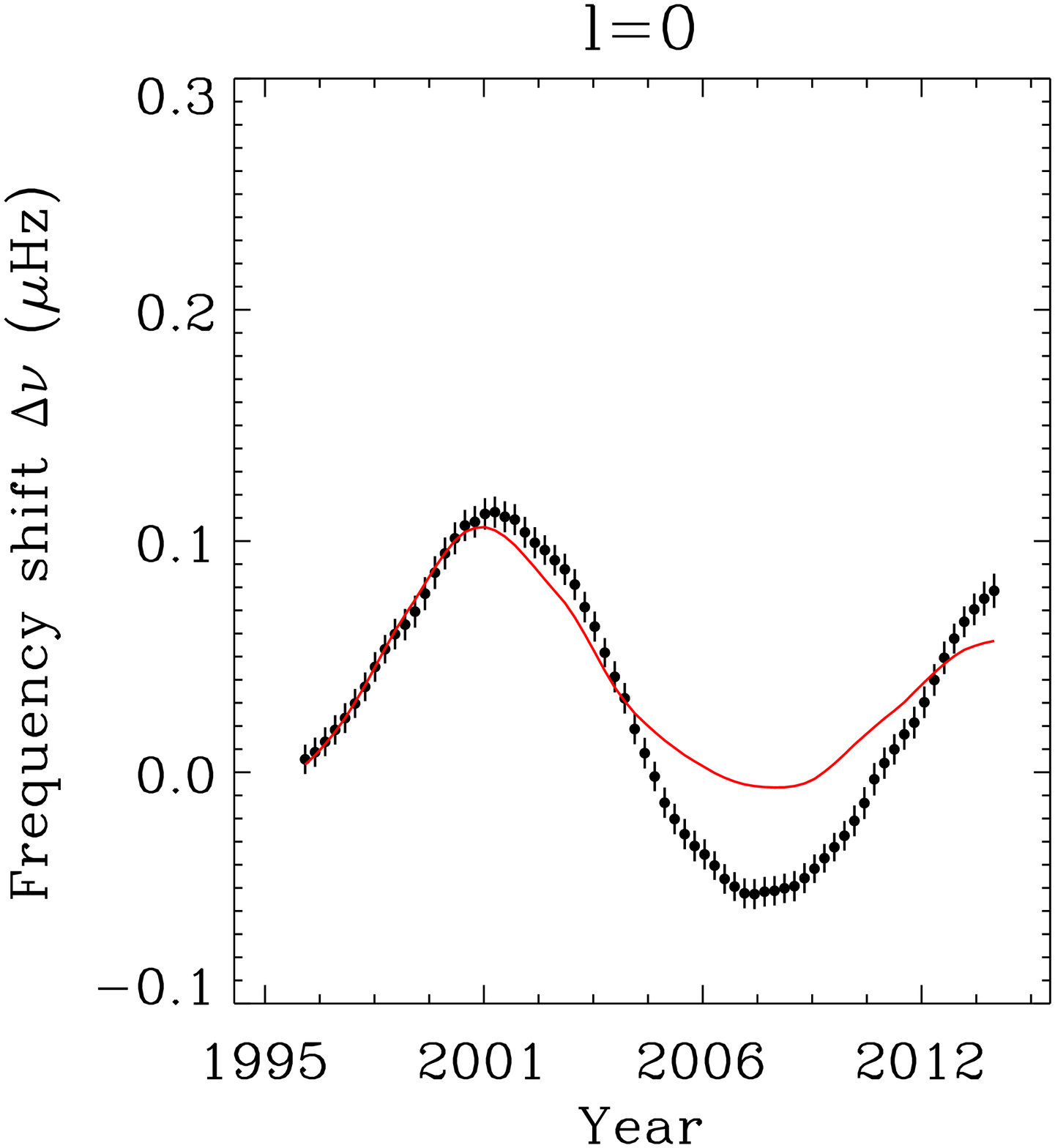}
\includegraphics[width=0.24\textwidth]{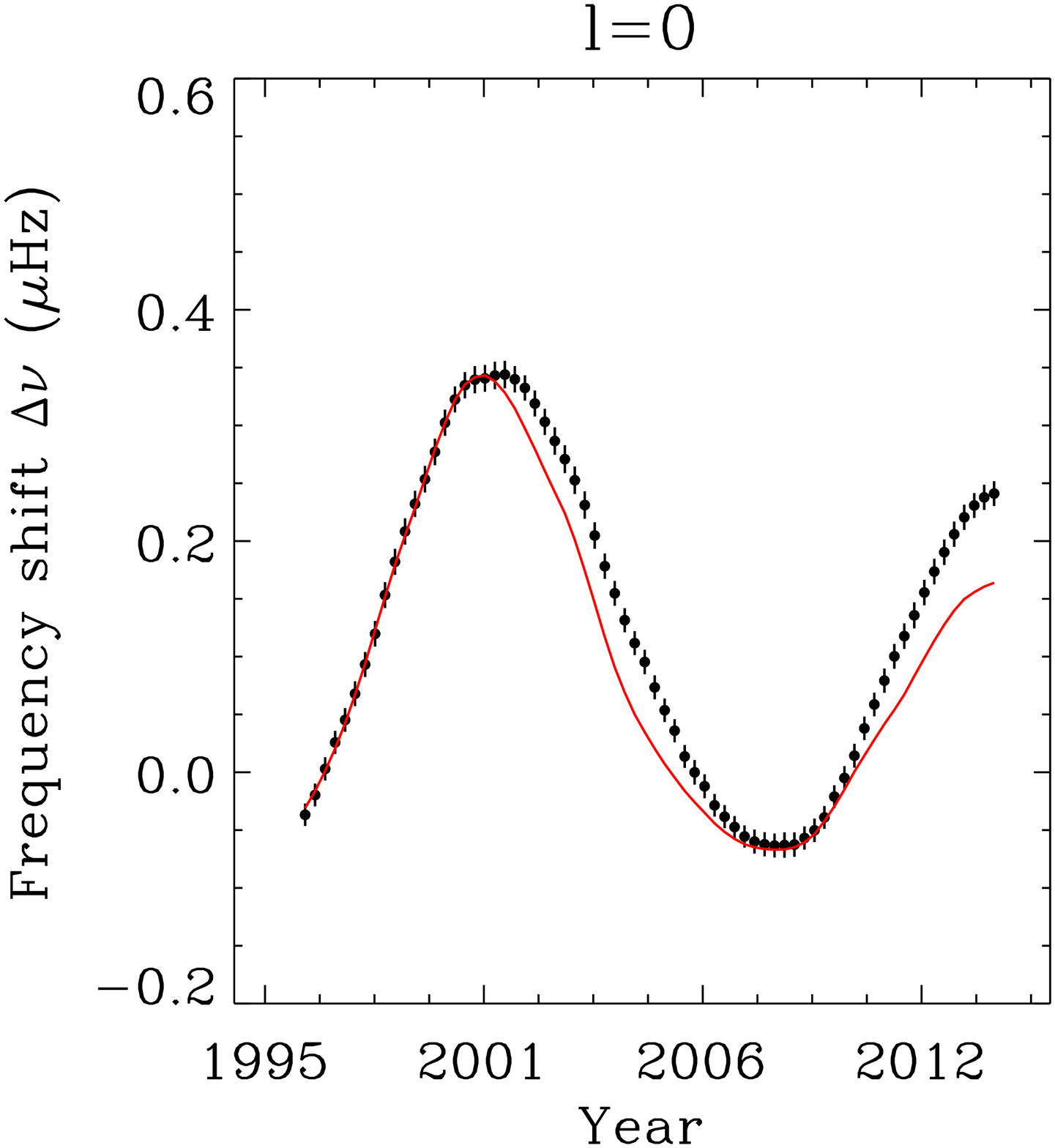}

\includegraphics[width=0.24\textwidth]{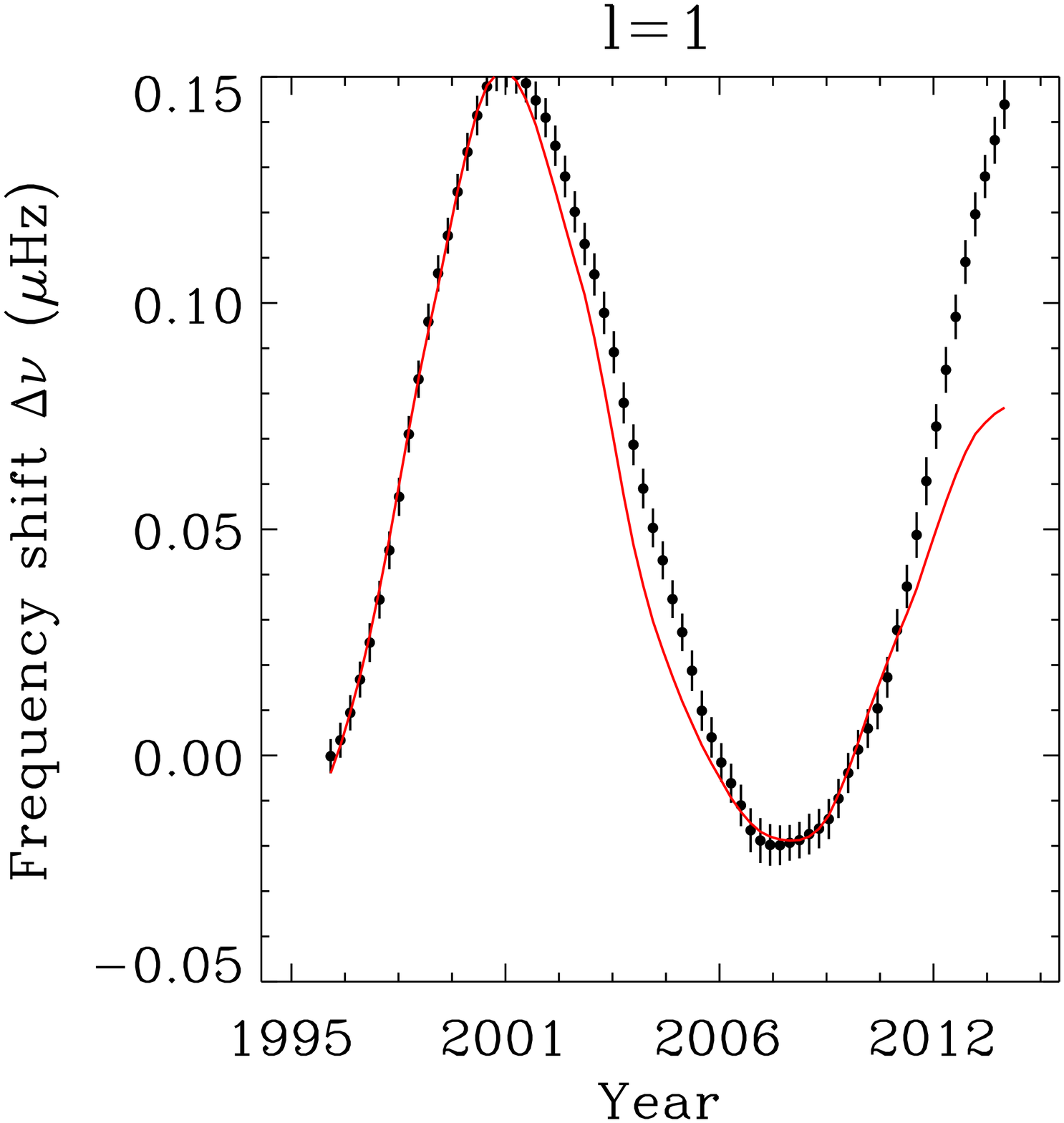}
\includegraphics[width=0.24\textwidth]{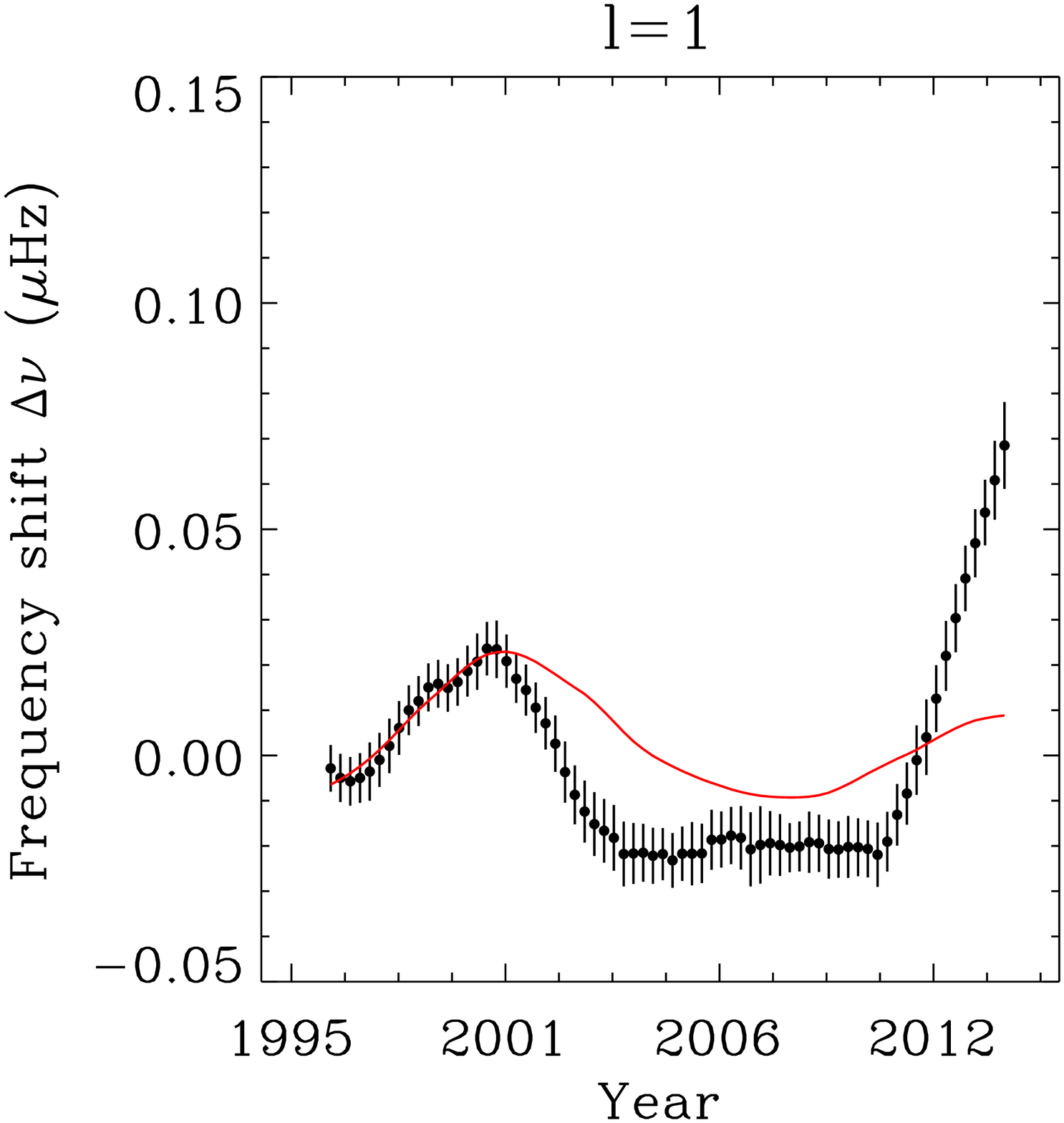}
\includegraphics[width=0.24\textwidth]{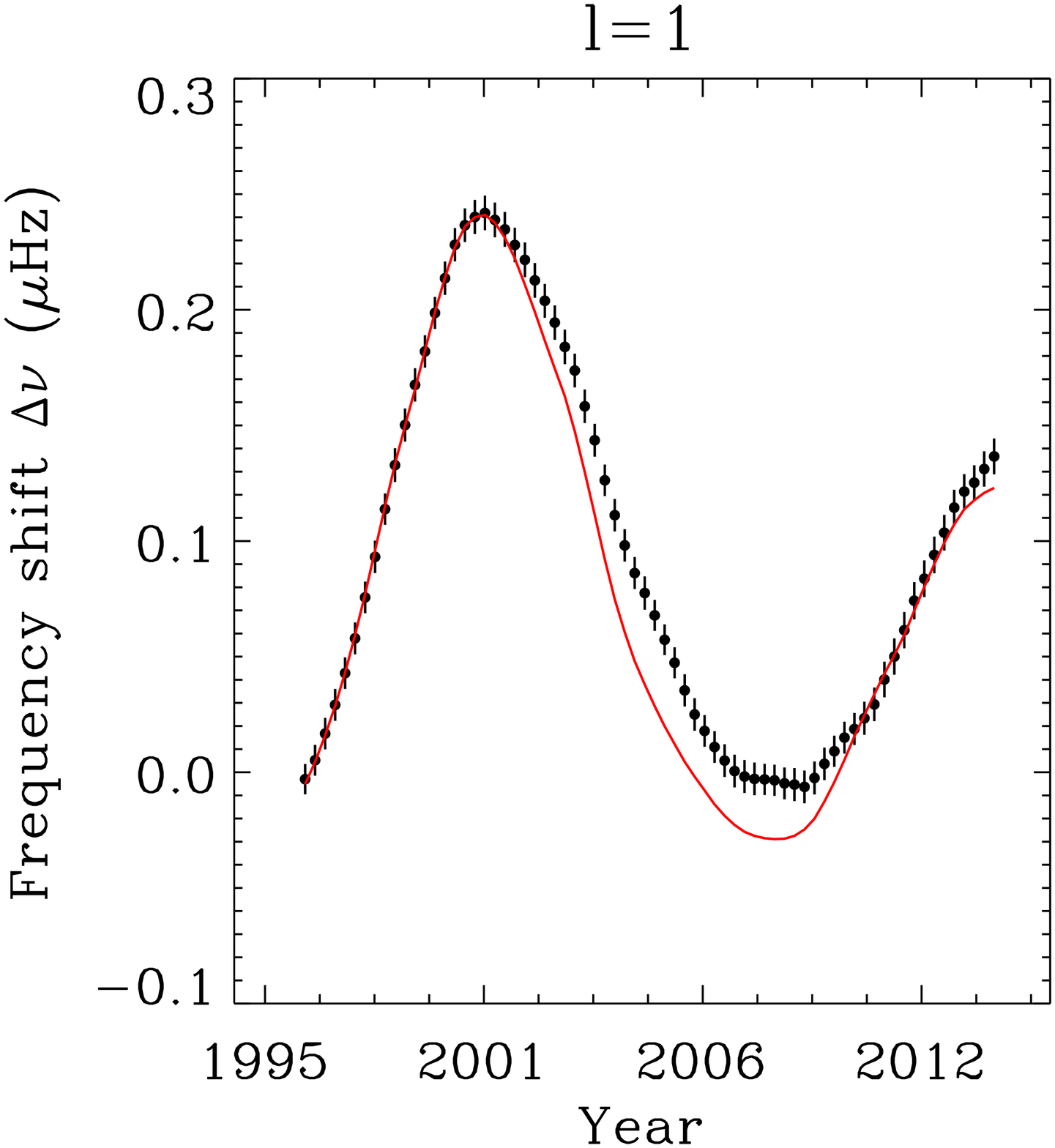}
\includegraphics[width=0.24\textwidth]{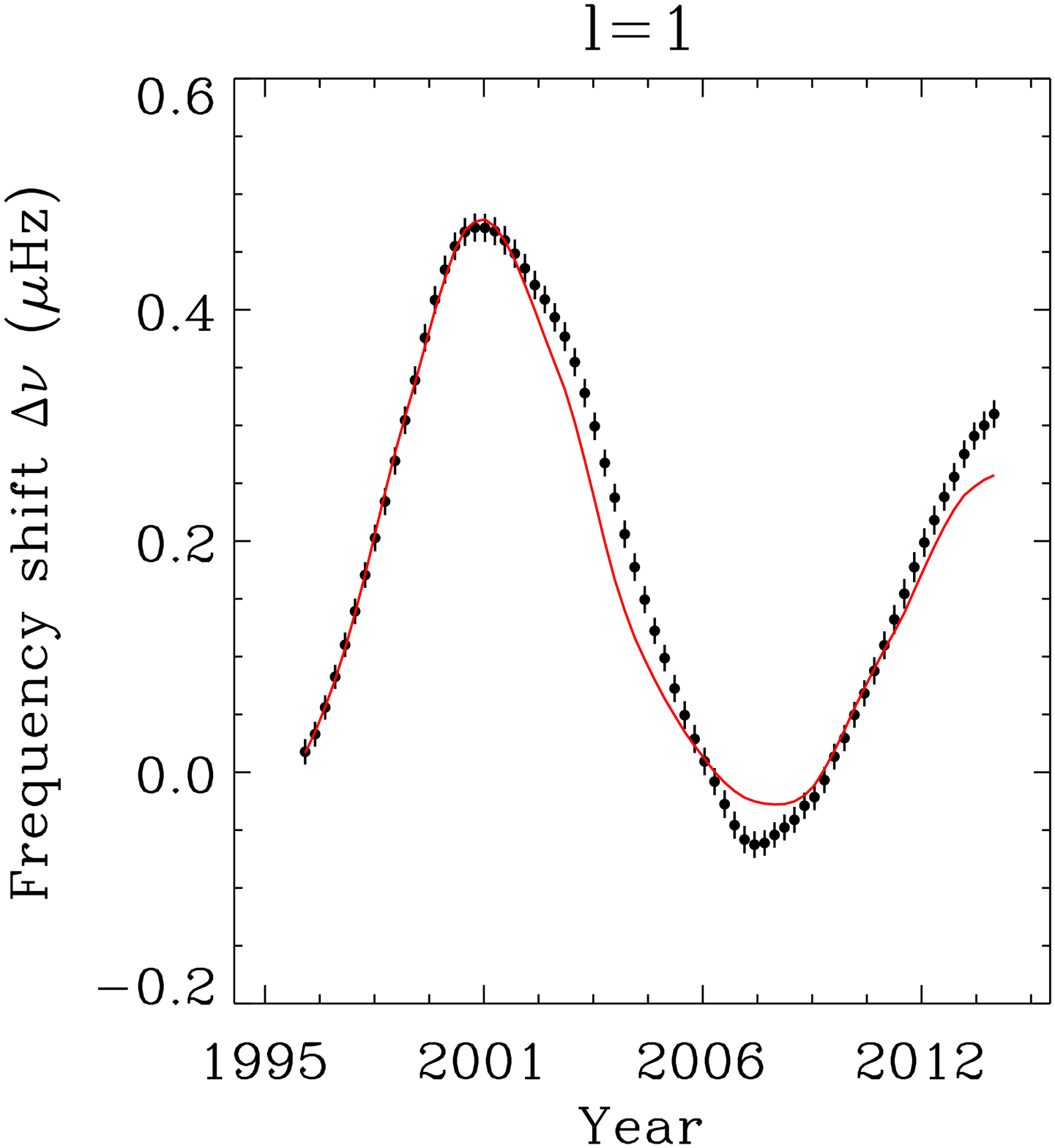}

\includegraphics[width=0.24\textwidth]{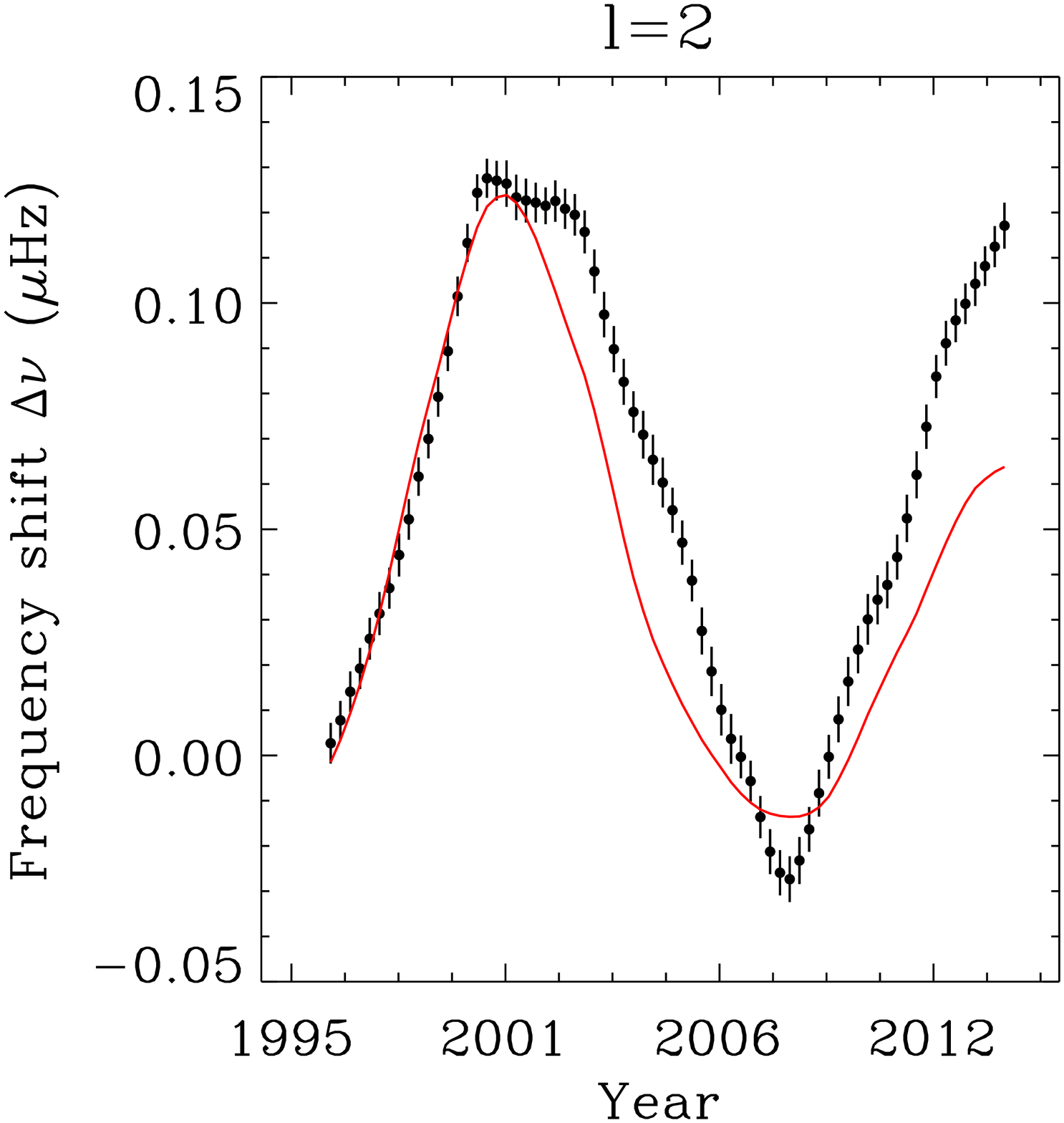}
\includegraphics[width=0.24\textwidth]{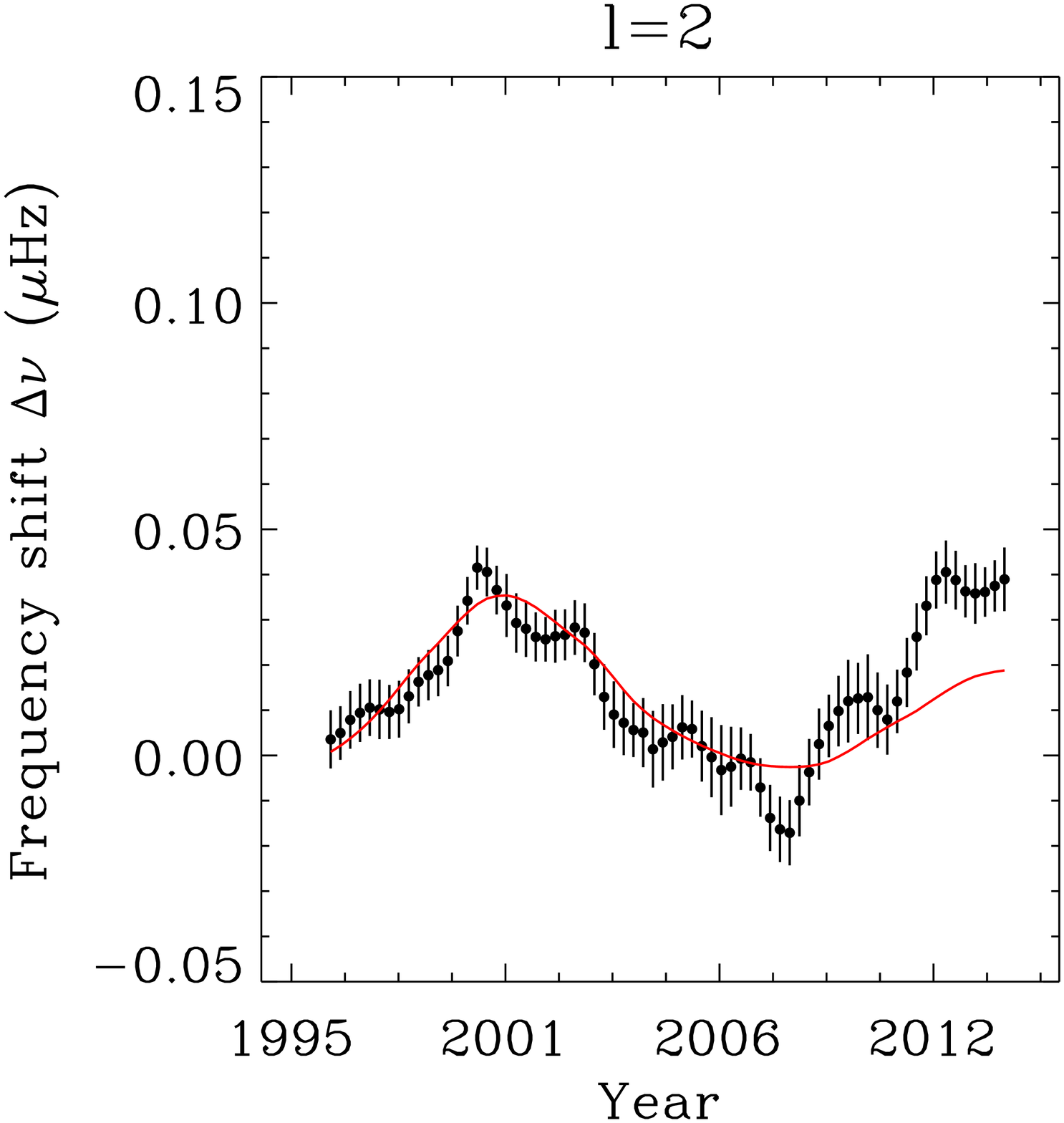}
\includegraphics[width=0.24\textwidth]{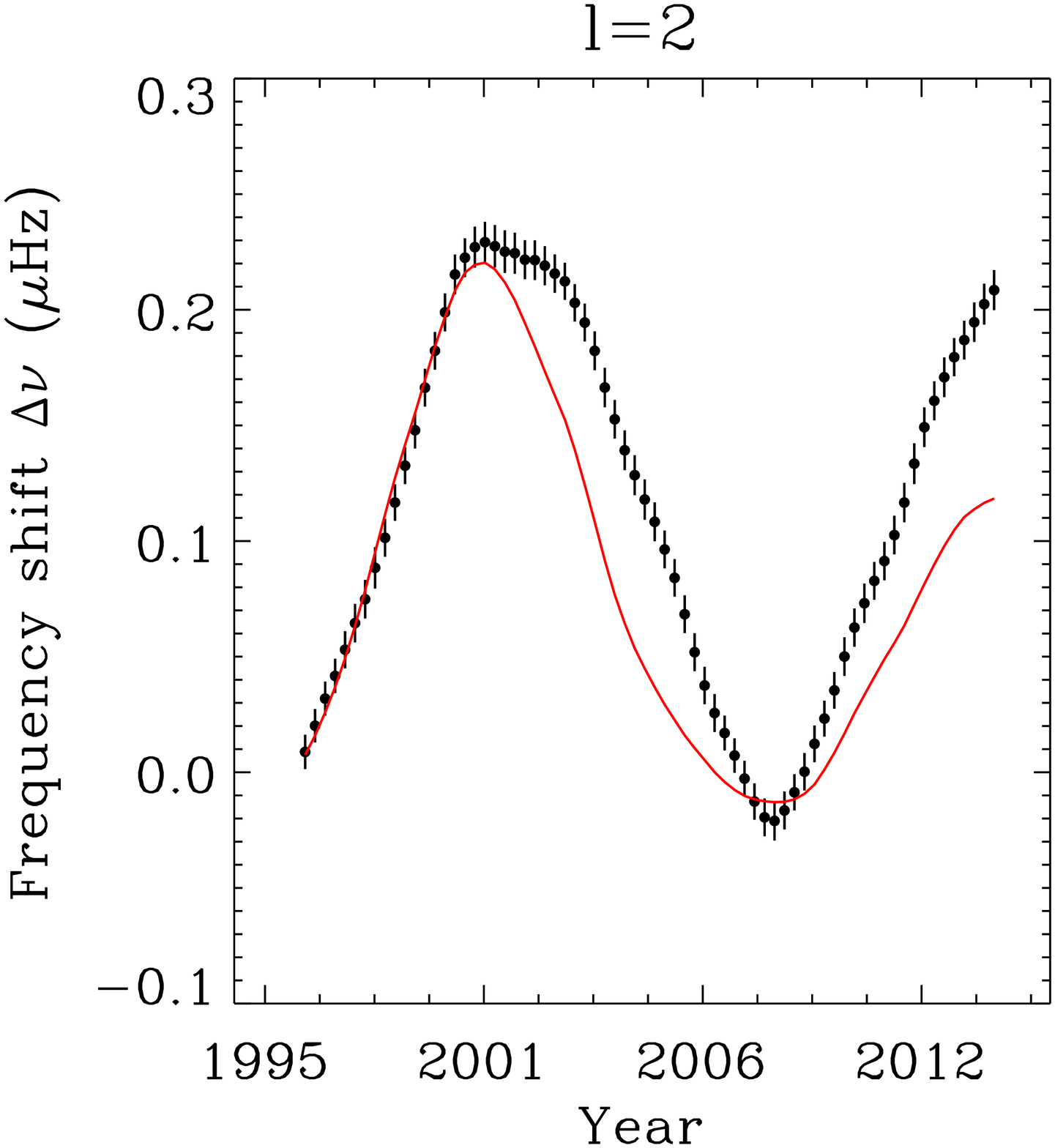}
\includegraphics[width=0.24\textwidth]{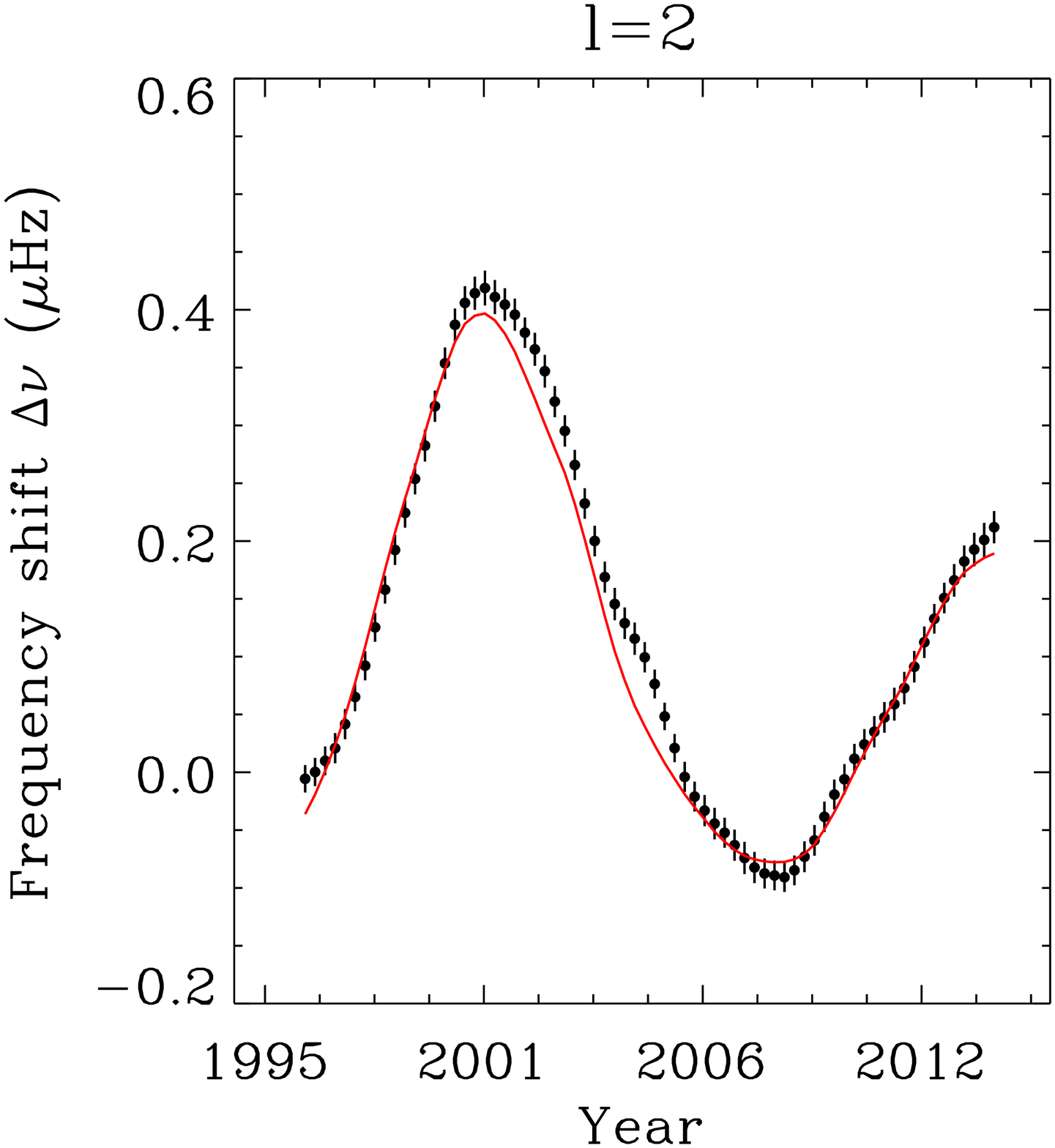}

\end{center} 
\caption{\label{fig:figappendixB1}Temporal variations in $\mu$Hz of the frequency shifts, $\langle  \Delta\nu_{n,l=0} \rangle$ , $\langle  \Delta\nu_{n,l=1} \rangle$ , $\langle  \Delta\nu_{n,l=2} \rangle$, at each individual angular degree $l$ (from top to bottom), and calculated for four different frequency ranges, the QBO's signature being removed   (black dots).  From left to right, the frequency ranges are the following: a) 1800~$\mu$Hz  $\leq \nu <~$ 3790~$\mu$Hz; b) 1800~$\mu$Hz  $\leq \nu <~$2450~$\mu$Hz; c) 2450~$\mu$Hz  $\leq \nu <~$3110~$\mu$Hz; and d) 3110~$\mu$Hz  $\leq \nu <~$ 3790~$\mu$Hz. The 10.7-cm radio flux, $F_{10.7}$, smoothed in the same way and scaled to match the rising phase and maximum of Cycle 23, is shown as a proxy of the solar surface activity (solid line).}
\end{figure*}

Figure~\ref{fig:figappendixB1} shows the temporal variations of the frequency shifts, $\langle  \Delta\nu_{n,l=0} \rangle$ , $\langle  \Delta\nu_{n,l=1} \rangle$ , $\langle  \Delta\nu_{n,l=2} \rangle$, at each individual angular degree $l$. The shifts were calculated over four frequency ranges:  1800~$\mu$Hz  $\leq \nu <~$ 3790~$\mu$Hz; 1800~$\mu$Hz  $\leq \nu <~$2450~$\mu$Hz (the low-frequency range); 2450~$\mu$Hz  $\leq \nu <~$3110~$\mu$Hz (the mid-frequency range); and 3110~$\mu$Hz  $\leq \nu <~$ 3790~$\mu$Hz (the high-frequency range). The QBO's signature was removed by applying a proper smoothing of 2.5 years as prescribed in \citet{fletcher10} for the four frequency ranges and for the 10.7-cm radio flux.  Note that in order to avoid misinterpretation due to possible border effects introduced by the smoothing, we do not consider the extreme points corresponding to the first three and the last three points of the smoothed data. 
Each frequency range is sensitive to different depths in the sub-surface layers of the Sun, as calculated by \citet{basu12}. The averaged kernels of the mid- and the high-frequency ranges have their largest sensitivities at $0.9981~R_{\sun}$ (i.e., 1300~km) and $0.9989~R_{\sun}$ (i.e., 760~km) respectively, while the low-frequency modes peak deeper in the solar interior at $0.9965~R_{\sun}$ (i.e., 2400~km). See Section~3 for results about the frequency shifts averaged over the modes $l=0$, 1, and 2,  $\langle  \Delta\nu_{n,l=0,1,2} \rangle$.

While the temporal variations of the frequency shifts at $l=0$, 1, and 2 show a good agreement with the 10.7-cm radio flux for the mid- and high-frequency modes, clear differences between the two quantities can be seen for the low-frequency modes. A residual after removing the QBO is still present in the $l=0$ and $l=2$ modes. The longer period observed in the low-frequency modes starting around 2006 and mentioned in Section~3 appears in the $l=0$ and $l=2$ modes only. In the other hand, the low-frequency $l=1$ has a very different behavior. After a long period between 2004 and 2010 of practically no changes one the frequency shifts, a sharp increase is measured starting from 2011, which is actually not present in the other angular degrees. 

\begin{figure*}[tbp]
\begin{center} 
\includegraphics[width=0.3\textwidth]{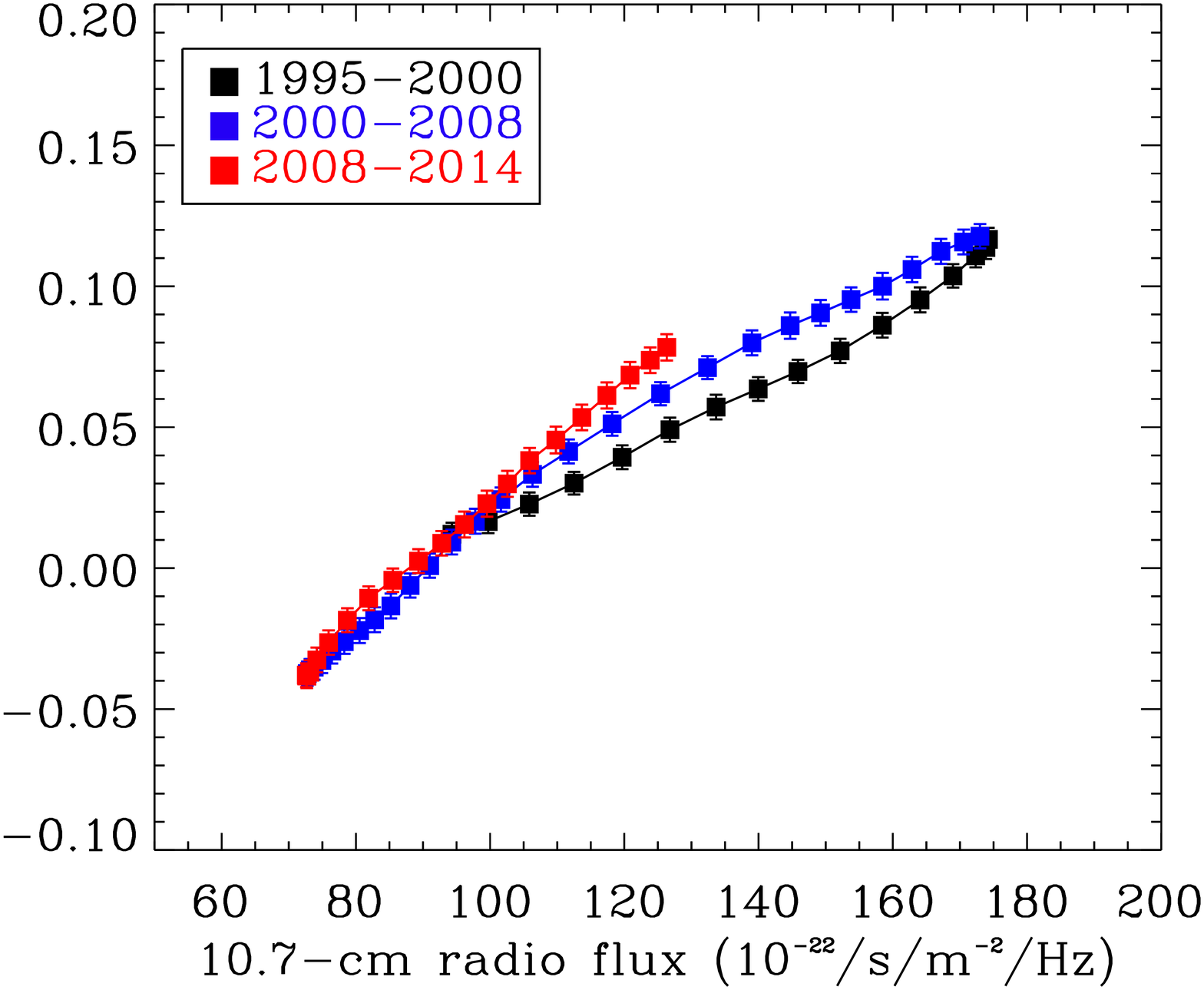}
\includegraphics[width=0.3\textwidth]{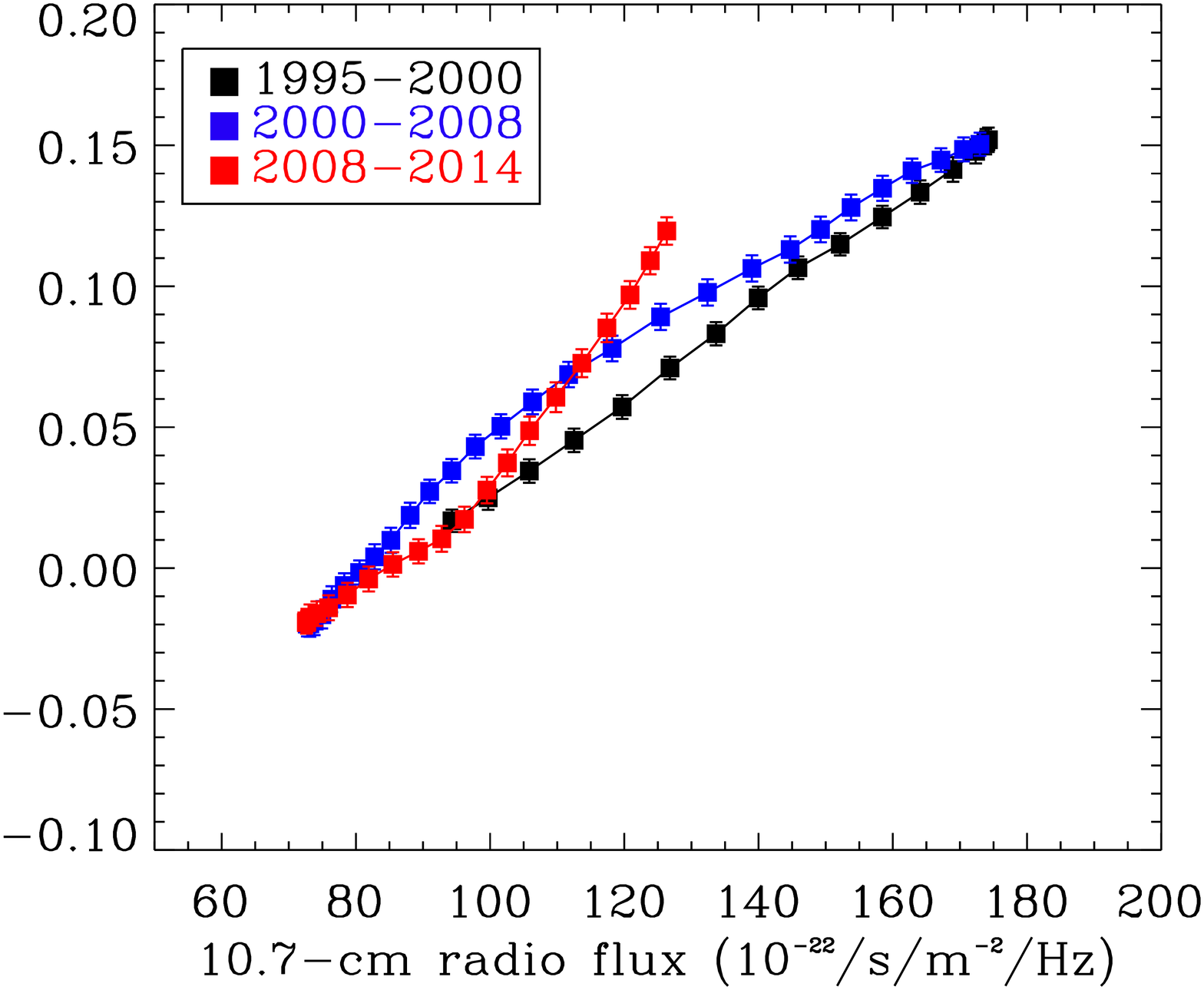}
\includegraphics[width=0.3\textwidth]{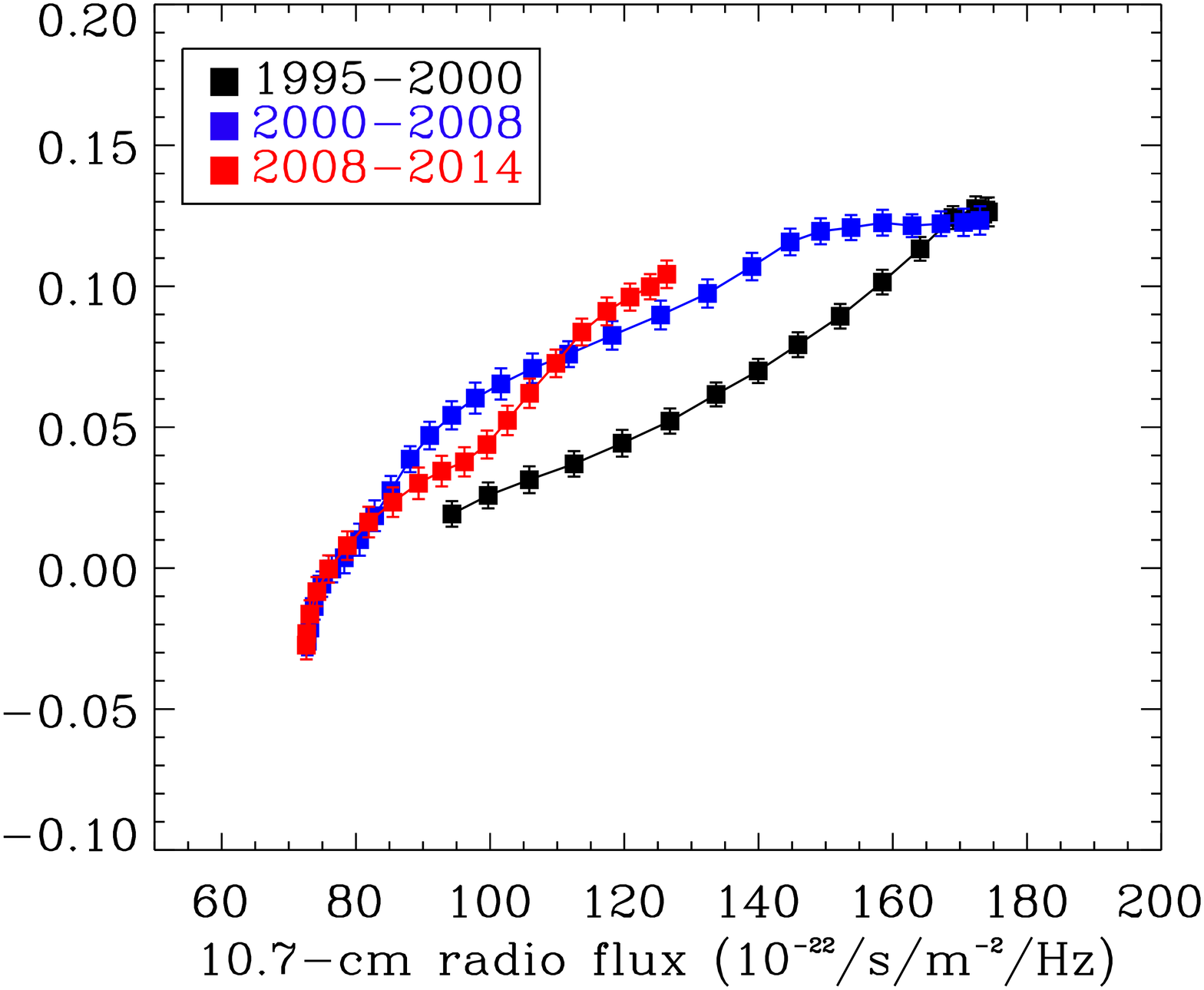}

\includegraphics[width=0.3\textwidth]{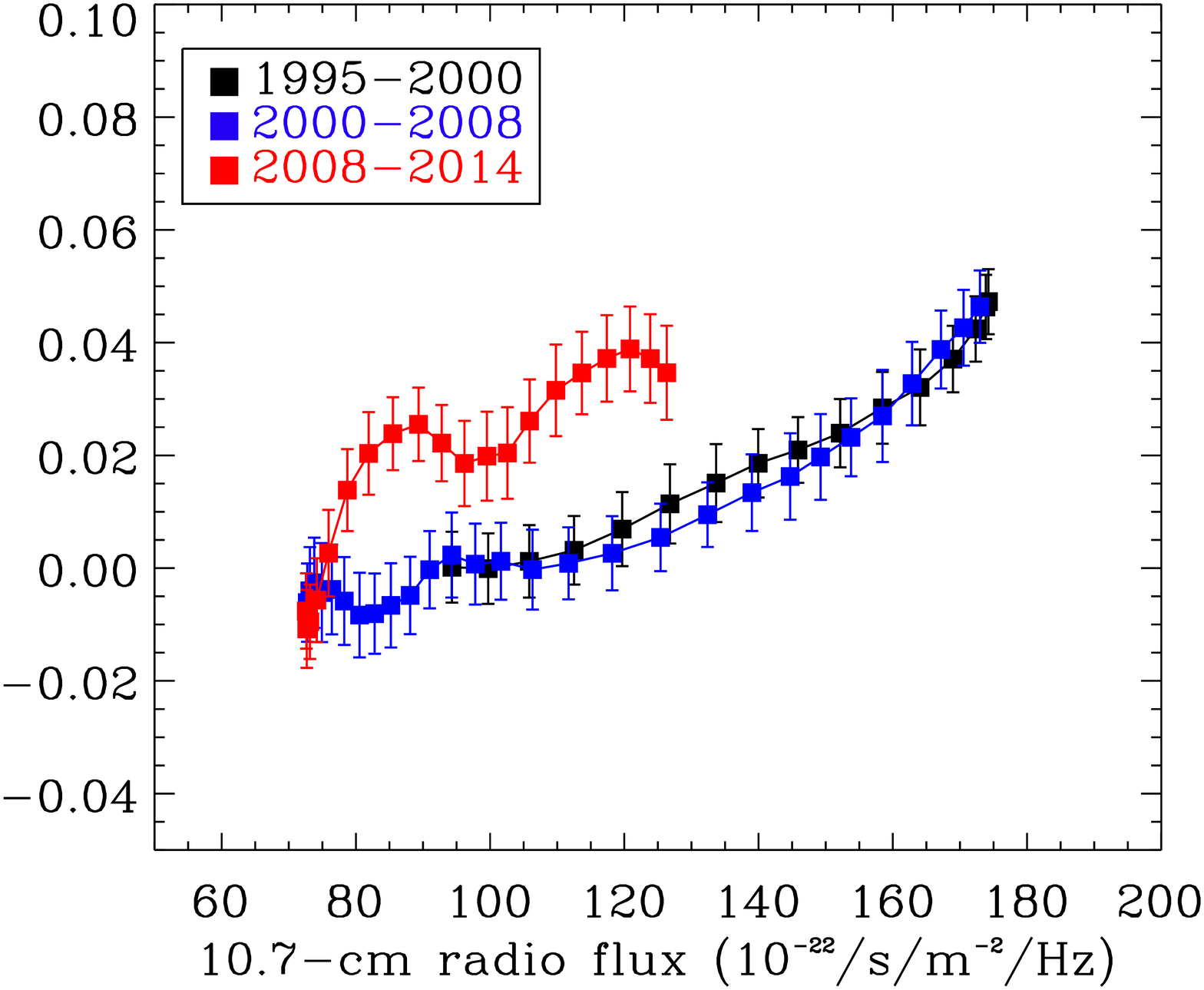}
\includegraphics[width=0.3\textwidth]{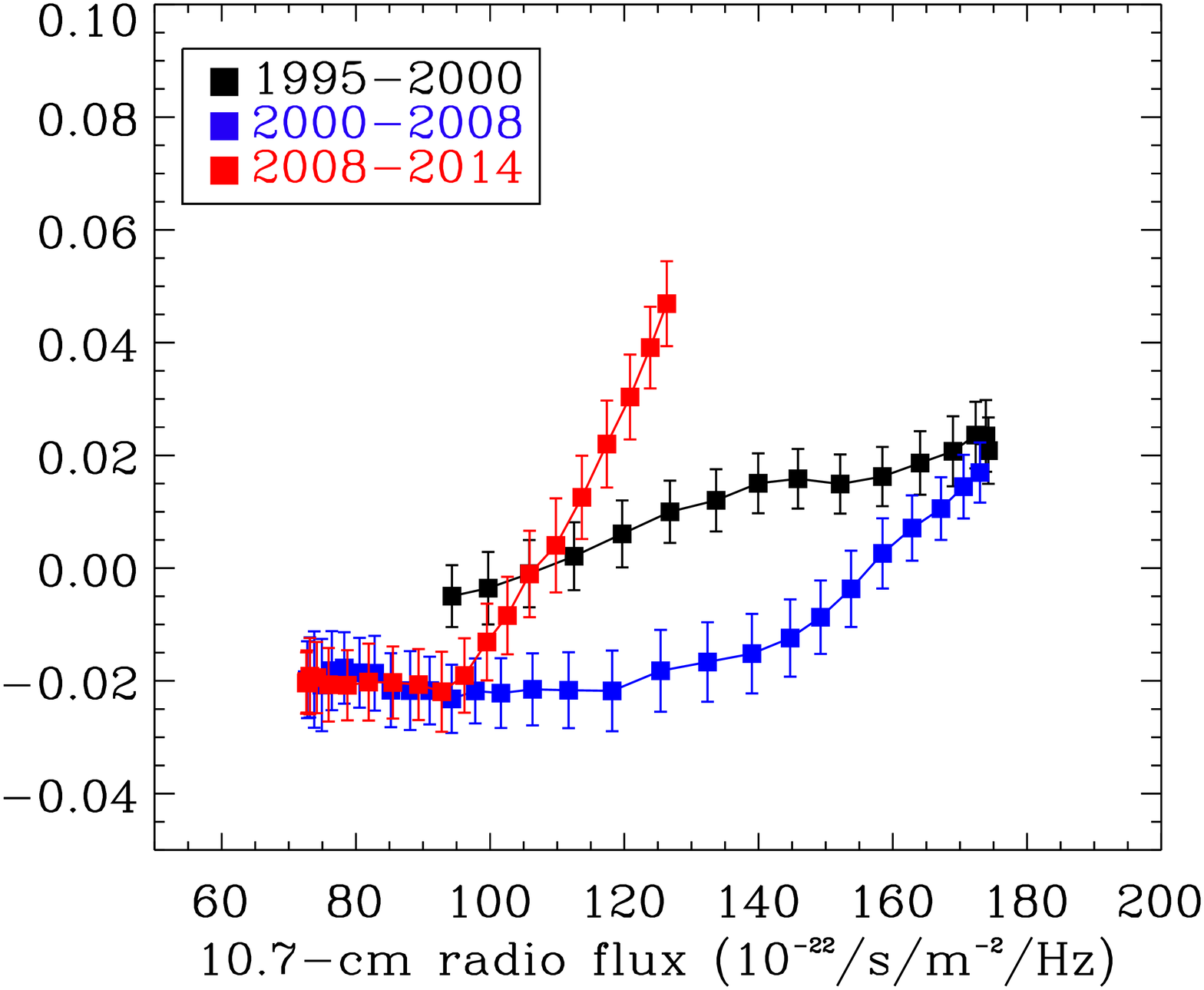}
\includegraphics[width=0.3\textwidth]{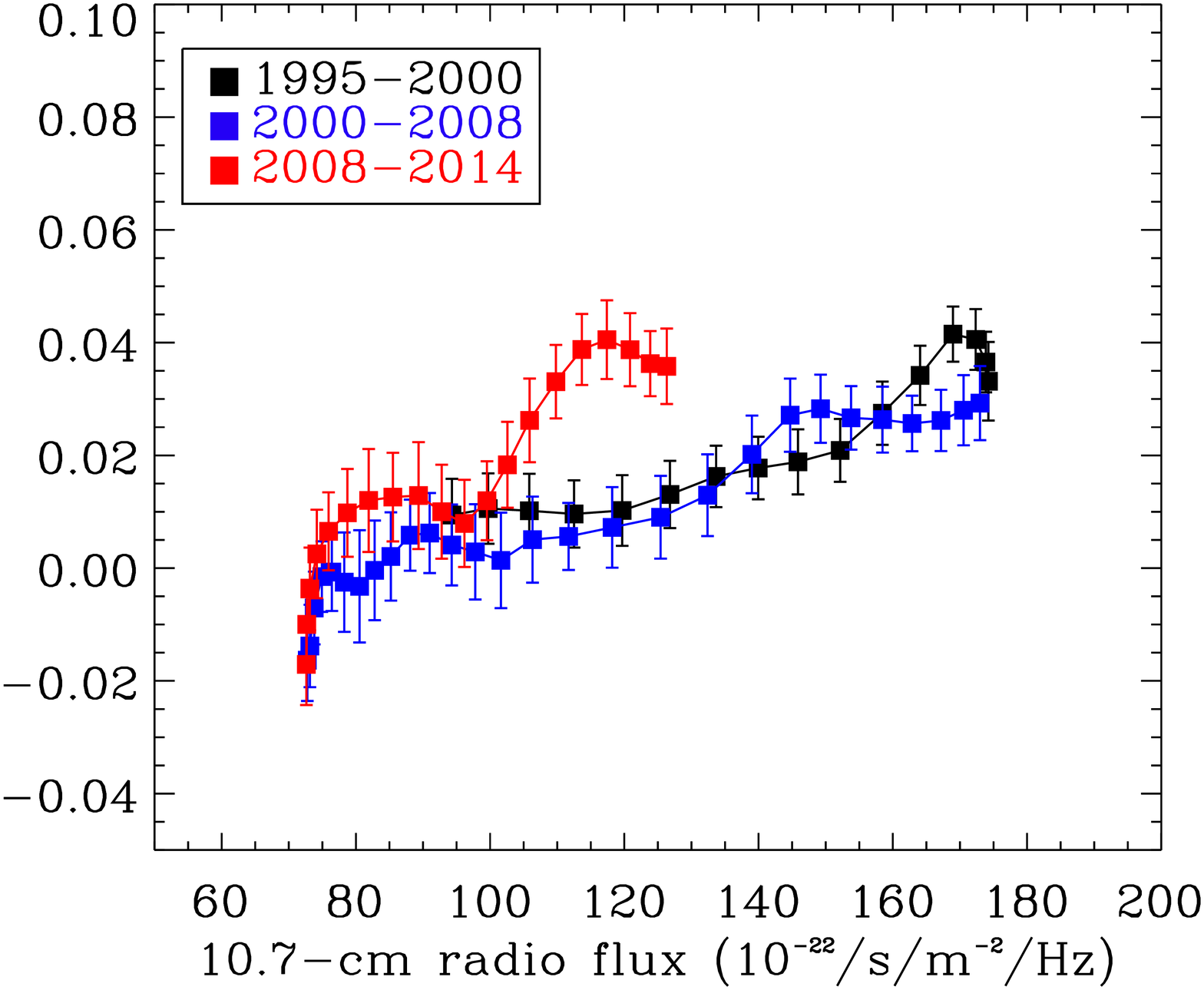}

\includegraphics[width=0.3\textwidth]{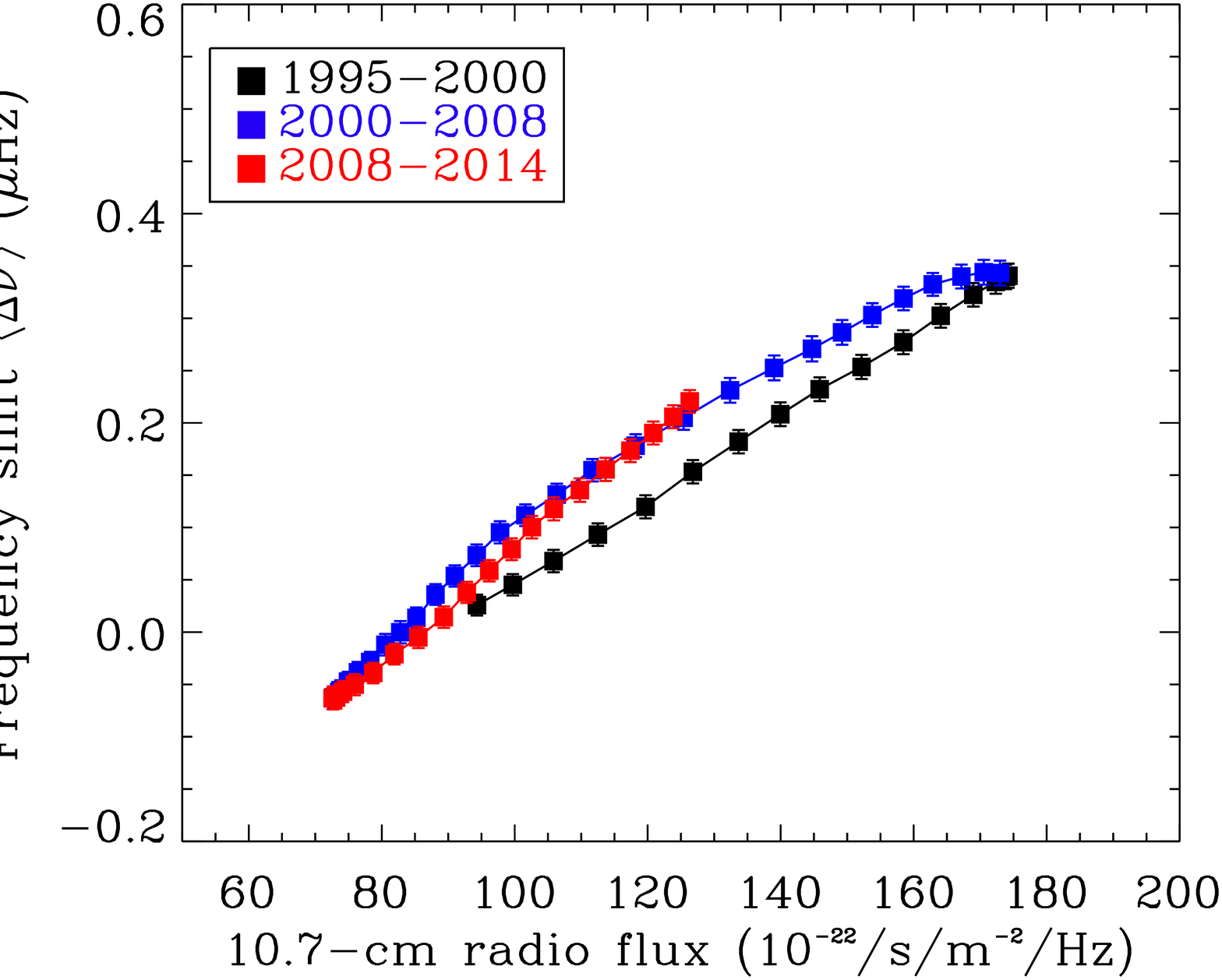}
\includegraphics[width=0.3\textwidth]{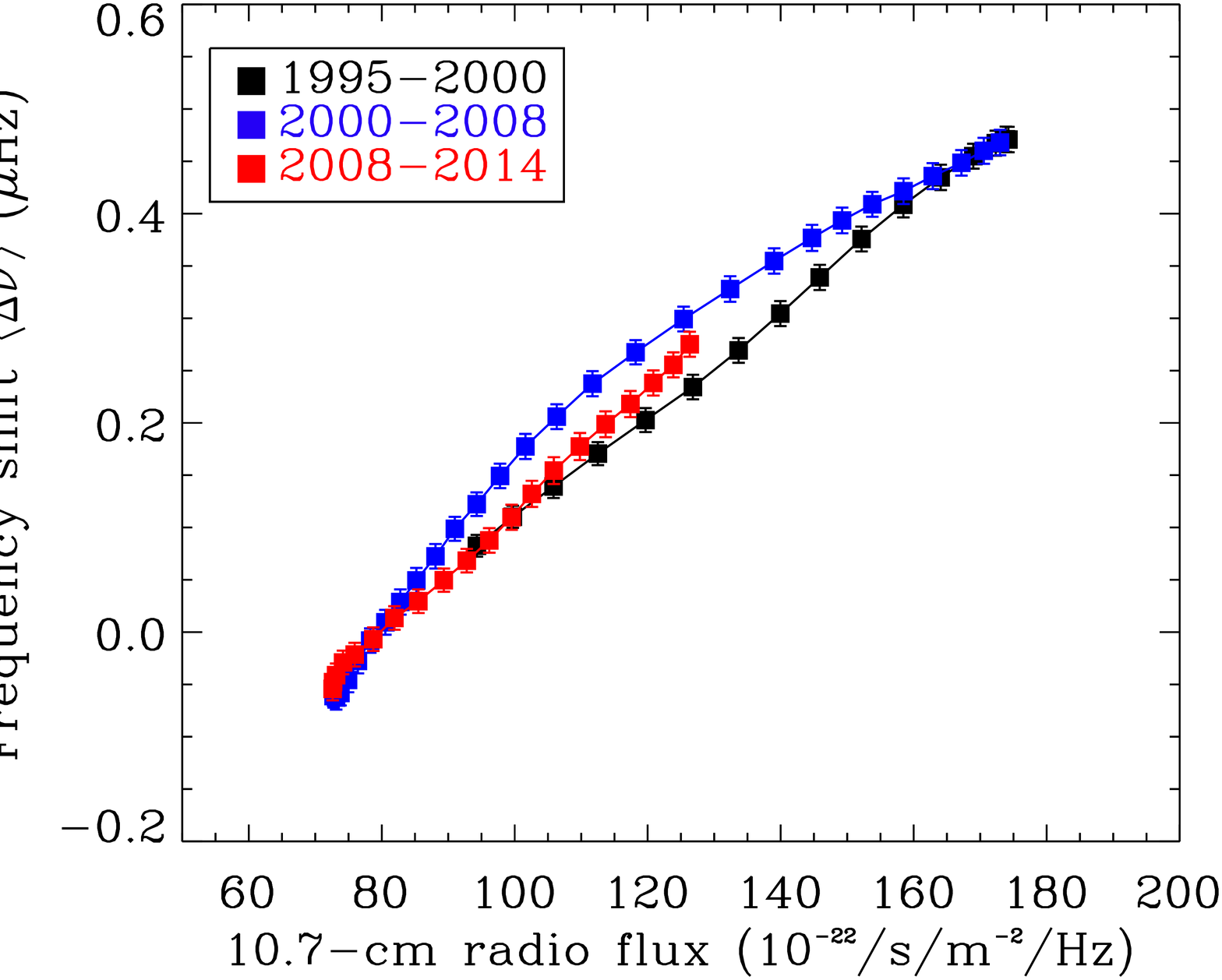}
\includegraphics[width=0.3\textwidth]{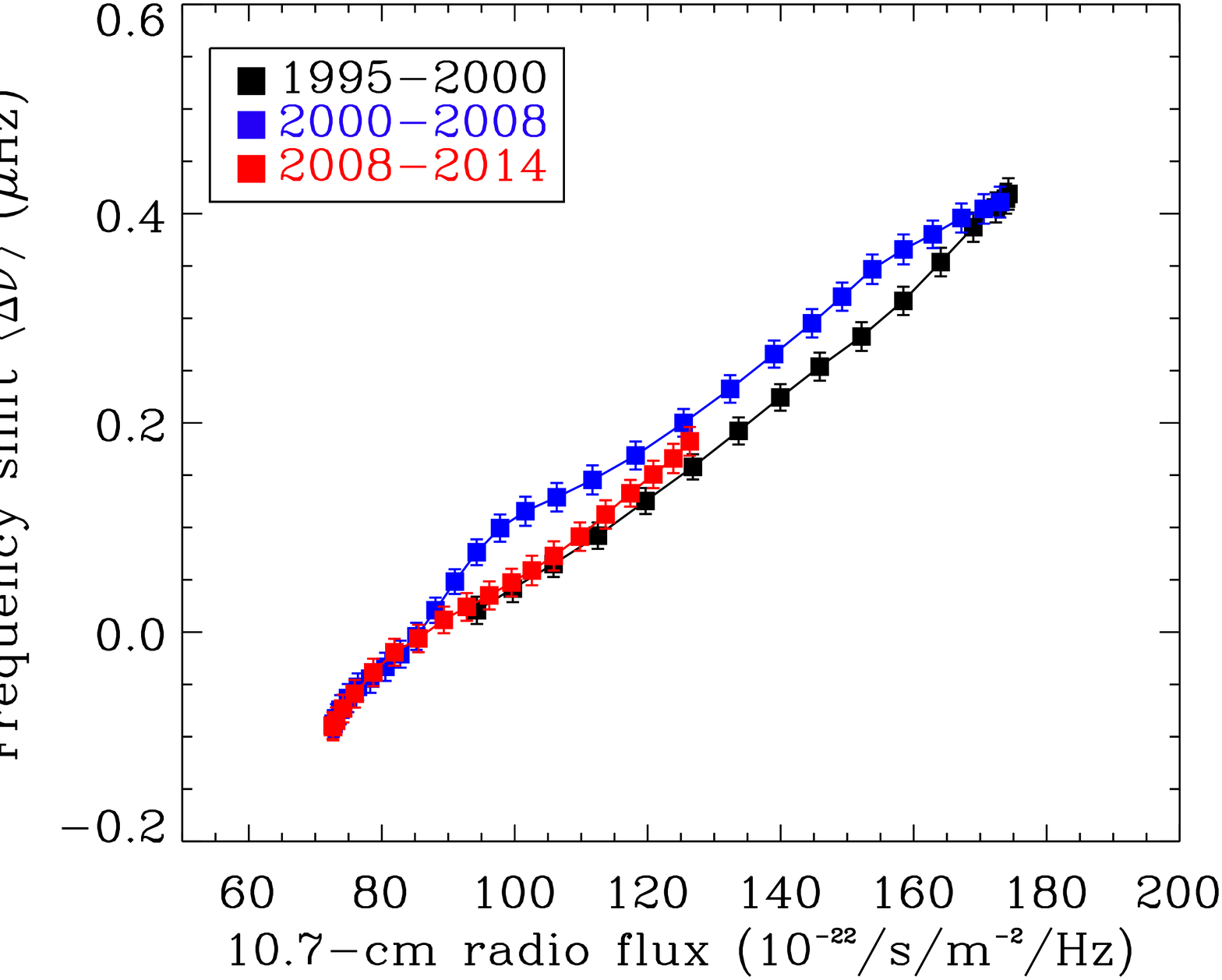}
\end{center} 
\caption{\label{fig:figappendixB2} Frequency shifts, $\langle  \Delta\nu_{n,l=0} \rangle$ , $\langle  \Delta\nu_{n,l=1} \rangle$ , $\langle  \Delta\nu_{n,l=2} \rangle$, at each individual angular degree $l=0$, 1, and 2 (from left to right),  as a function of the corresponding 10.7-cm radio flux, $F_{10.7}$, and calculated for three different frequency ranges once the QBO's signature was removed.  From top to bottom, the frequency ranges are the following: a) 1800~$\mu$Hz  $\leq \nu <~$3790~$\mu$Hz; b) 1800~$\mu$Hz  $\leq \nu <~$2450~$\mu$Hz (the low-frequency range); and c) 3110~$\mu$Hz  $\leq \nu <~$ 3790~$\mu$Hz (the high-frequency range). The rising (black dots) and declining (blue dots) phases of Cycle 23, and the rising phase (red dots) of Cycle 24 are indicated by different colors.}
\end{figure*}

Figure~\ref{fig:figappendixB2} shows the same frequency shifts, $\langle  \Delta\nu_{n,l=0} \rangle$ , $\langle  \Delta\nu_{n,l=1} \rangle$ , $\langle  \Delta\nu_{n,l=2} \rangle$, at each individual angular degree $l$, but as a function of the corresponding 10.7-cm radio flux, $F_{10.7}$. Both quantities were properly smoothed as in \citet{fletcher10}. The shifts obtained over three frequency ranges are represented: 1800~$\mu$Hz  $\leq \nu <~$3790~$\mu$Hz; 1800~$\mu$Hz  $\leq \nu <~$2450~$\mu$Hz (the low-frequency range); and c) 3110~$\mu$Hz  $\leq \nu <~$ 3790~$\mu$Hz (the high-frequency range). Different colors are used  for three different phases of solar activity: a) the rising phase of Cycle 23 going from April 1996 to October 200; b) the declining phase of Cycle 23 going from October 2001 to January 2009; and c) the rising phase of Cycle 24 going from January 2009 to April 2014 (see Section~3 for more details). Again, the  low-frequency modes show a very different behavior compared to higher-frequency modes. Moreover, a longer period than the quasi-biennial oscillation is observed for the radial and quadrupole modes during the rise of Cycle~24, while the dipole modes show a very distinct behavior.

\end{appendix}

\end{document}